\documentclass[twocolumn]{article}

\usepackage[fontsize=9pt]{fontsize}

\usepackage{cite} 		  
\usepackage{xcolor}       
\usepackage{hyperref}	  
\hypersetup{colorlinks=true,linkcolor=blue,citecolor=blue,breaklinks={true}} 
\usepackage[normalem]{ulem}

\usepackage{graphicx}
\usepackage{sidecap}
\usepackage[font=footnotesize,labelfont=bf,justification=justified]{caption}
\graphicspath{{./FIG/}}
\DeclareGraphicsExtensions{.eps,.png,.jpg,.jpeg,.bmp,.gif,.pdf}

\usepackage[left=45pt,%
                right=45pt,%
                top=54pt,
                bottom=54pt,%
                headheight=0pt,
                headsep=10pt,%
                footskip=25pt,
                marginparwidth=38pt]{geometry}

\usepackage{tabularx,booktabs}	
\usepackage{multirow}	

\usepackage{amsmath}      		
\usepackage{amssymb}	  		
\usepackage{bm}		      		
\usepackage{physics}	  		
\usepackage{mathtools}			
\usepackage{dsfont}		  		
\usepackage{soul}				
\usepackage{relsize}			

\newcommand{\N}{\mathbb{N}}		
\newcommand{\R}{\mathbb{R}}		
\newcommand{\transp}{\mathsf{T}}					

\DeclareMathOperator*{\argmin}{arg\,min}

\usepackage{amsthm}

\theoremstyle{definition}

\newcommand{\rev}[1]{#1}

\usepackage[switch]{lineno}         
\usepackage[resetlabels]{multibib}

\usepackage{authblk}                

\usepackage{titlesec}               
\titleformat{\section}{\large\bfseries}{\thesection.}{1em}{}
\titleformat{\subsection}{\normalsize\bfseries}{\thesubsection.}{1em}{}
\titlespacing*{\section}
{0pt}{1em}{0.2em} 
\titlespacing*{\subsection}
{0pt}{1em}{0em} 
\title{\huge\textbf{Optimal flock formation induced by \rev{agent} heterogeneity}}

\author[a,b,1,2]{Arthur~N.~Montanari}
\author[a,b,1]{Ana Elisa D. Barioni}
\author[c]{Chao Duan}
\author[a,b,d,e]{Adilson E. Motter}

\affil[a]{Center for Network Dynamics, Northwestern University, Evanston, IL 60208, USA}
\affil[b]{Department of Physics and Astronomy, Northwestern University, Evanston, IL 60208, USA}
\affil[c]{School of Electrical Engineering, Xi'an Jiaotong University, Xi'an 710049, China}
\affil[d]{Department of Engineering Sciences and Applied Mathematics, Northwestern University, Evanston, IL, 60208, USA}
\affil[e]{Northwestern Institute on Complex Systems, Northwestern University, Evanston, IL 60208, USA}

\date{}

\begin{document}

\twocolumn[
    \maketitle
    \vspace{-1em} 
    \begin{center}
        \begin{minipage}{1\textwidth}
            \begin{abstract}
            \vspace{0cm}

            The study of flocking in biological systems has identified conditions for self-organized collective behavior, inspiring the development of decentralized strategies to coordinate the dynamics of swarms of drones and other autonomous vehicles. Previous research has focused primarily on the role of the time-varying interaction network among agents while assuming that the agents themselves are identical or nearly identical. Here, we depart from this conventional assumption to investigate how inter-individual differences between agents affect the stability and convergence in flocking dynamics. We show that flocks of agents with optimally assigned heterogeneous parameters significantly outperform their homogeneous counterparts, achieving 20-40\% faster convergence to desired formations across various control tasks. These tasks include target tracking, flock formation, and obstacle maneuvering. In systems with communication delays, heterogeneity can enable convergence even when flocking is unstable for identical agents. Our results challenge existing paradigms in multi-agent control and establish system disorder as \rev{an adaptive, distributed} mechanism to promote collective behavior in flocking dynamics.

\medskip\noindent
\textbf{Keywords:} consensus dynamics, 
dynamical networks, multi-agent systems, nonlinear dynamics, flocking,  
autonomous vehicles
\end{abstract}
        \end{minipage}
    \end{center}
    \vspace{2em} 
]

\let\thefootnote\relax\footnotetext{$^1$These authors contributed equally to this work.}
\let\thefootnote\relax\footnotetext{$^2$Corresponding author: arthur.montanari@northwestern.edu.}


\noindent
In 1987, Reynolds introduced three basic
rules to emulate the flocking behavior of animals \cite{reynolds1987flocks}: 1)~agents must avoid collisions with nearby flock mates (\textit{separation}), 2)~agents must match their velocity with nearby agents (\textit{alignment}), and 3)~agents must move towards the center of mass of the local group of flock mates (\textit{cohesion}). 
Models based on Reynold's rules, known as boids, remain a standard solution in computer graphics for animating group behavior \cite{silva2010boids}. 
Beyond computer graphics, these rules have also found interdisciplinary applications in the modeling of sociobiological systems, stimulating research on the conditions required for the emergence of self-organization 
\cite{vicsek1995novel,helbing2000simulating,gazi2004stability,couzin2009collective,katz2011inferring,marras2012fish,pearce2014role,gomez2022intermittent,sinha2023optimal,sar2023flocking}. 
The underlying distributed decision-making strategies observed in animal flocks, which are governed mostly by local interactions, have inspired the design of multi-agent engineering systems \cite{xiao2024perception}, such as swarms of unmanned aerial vehicles (UAVs). 
Swarms of small vehicles offer a cost-effective alternative to large vehicles in a wide variety of applications, ranging from surveillance and reconnaissance \cite{wang2022coverage} to target tracking \cite{bertuccelli2009real}, operation management \cite{balazs2024decentralized}, and transportation \cite{nguyen2021swarm}. However, the deployment of these technologies faces  fundamental challenges associated with controlling a large number of agents \cite{chen2019control,beaver2021overview,leonard2024fast}. Overcoming these challenges requires the discovery of scalable decision-making mechanisms that can adapt to dynamic environments, operate under data communication constraints, and coordinate hundreds of agents.

The analysis of flocking dynamics is often formulated in the context of multi-agent consensus problems. Agents are said to achieve consensus if they all eventually agree on a common state or behavior (e.g., a specified formation) despite operating only with local information on the state of the flock. Lack of consensus can lead to group fragmentation in the presence of stochastic disturbances, physical obstacles, and loss of communication \cite{olfati2006flocking}. Previous studies have focused mainly on the role of the interaction (communication) network \cite{ren2007formation,nagy2010hierarchical,baronchelli2012consensus,griparic2022consensus}, including the impact of the time-varying topology \cite{cucker2007emergent,valcher2017consensus,mikaberidze2024consensus,amichay2024revealing} and data communication constraints \cite{OlfatiSaber2004,blondel2005convergence,ren2008consensus,yu2010some,zhang2017sliding}. 
In this context, Lyapunov stability has been a major tool for deriving the conditions for flock formation in numerous control tasks 
\cite{olfati2006flocking,ogren2004cooperative,beaver2021overview}.

A common implicit assumption in multi-agent studies is that consensus is facilitated when agents are identical or nearly identical. 
Still, empirical research on animal behavior, such as fish schooling \cite{jolles2017consistent,niizato2024information} and ant synchronization \cite{doering2022noise}, 
has identified scenarios in which inter-individual differences can facilitate coordination \cite{jolles2020role}.  
In the study of network synchronization, disorder in the parameters of the oscillators has been shown to improve synchronization in various systems \cite{nishikawa2016symmetric,molnar2020network}, including power grids \cite{molnar2021asymmetry}, electronic circuits \cite{mallada2015distributed,sugitani2021synchronizing}, coupled lasers \cite{nair2021using,cao2022harnessing}, neuronal oscillators \cite{gast2024neural}, and chemical oscillators \cite{zhang2021random}. 
 Experimental studies have explored similar effects in self-organization and pattern formation \cite{teng2022heterogeneity,yang2022emergent,nicolaou2021heterogeneity,ceron2023programmable}. 
%
Despite these advances and the connections between synchronization and consensus \cite{keeffe2017oscillators,ghosh2022synchronized},  heterogeneity among agents has yet to be explored as a potential framework to promote flocking.

In this paper, we investigate the impact of optimizing inter-individual differences in real-time as an adaptive mechanism to enhance flocking behavior.
We show that the stability and convergence rate of the collective dynamics substantially improve for suitable heterogeneous parameters when compared to their homogeneous counterparts. 
%
%
Despite the (possibly nonlinear) time-varying dynamics, Lyapunov stability analysis shows that this optimization is tightly bounded by the minimization of the largest Lyapunov exponent of the system. 
Our formulation highlights the dependence of the flocking dynamics on the interplay between the parameters of the agents, the flock formation, and the underlying communication network.
The results are established for several control tasks and flocking models, with increasing degrees of complexity. We first consider a system designed for target tracking and formation keeping. 
We then generalize the results to a time-delay consensus model and a gradient-based flocking model. The latter accounts for sparse communication networks, emergent formations, and obstacle avoidance. 
In all scenarios, we show evidence that heterogeneous parameter optimization can improve the flock convergence rate by 20-40\% relative to homogeneous parameter optimization under the same constraints. We further show that consensus can be achieved over a larger range of communication delays for heterogeneous systems than for homogeneous ones.


\section*{Pre-Assigned Flock Formation}

\subsection*{Flocking model}
\begin{figure*}[t]
	\centering
	\includegraphics[width=0.85\textwidth]{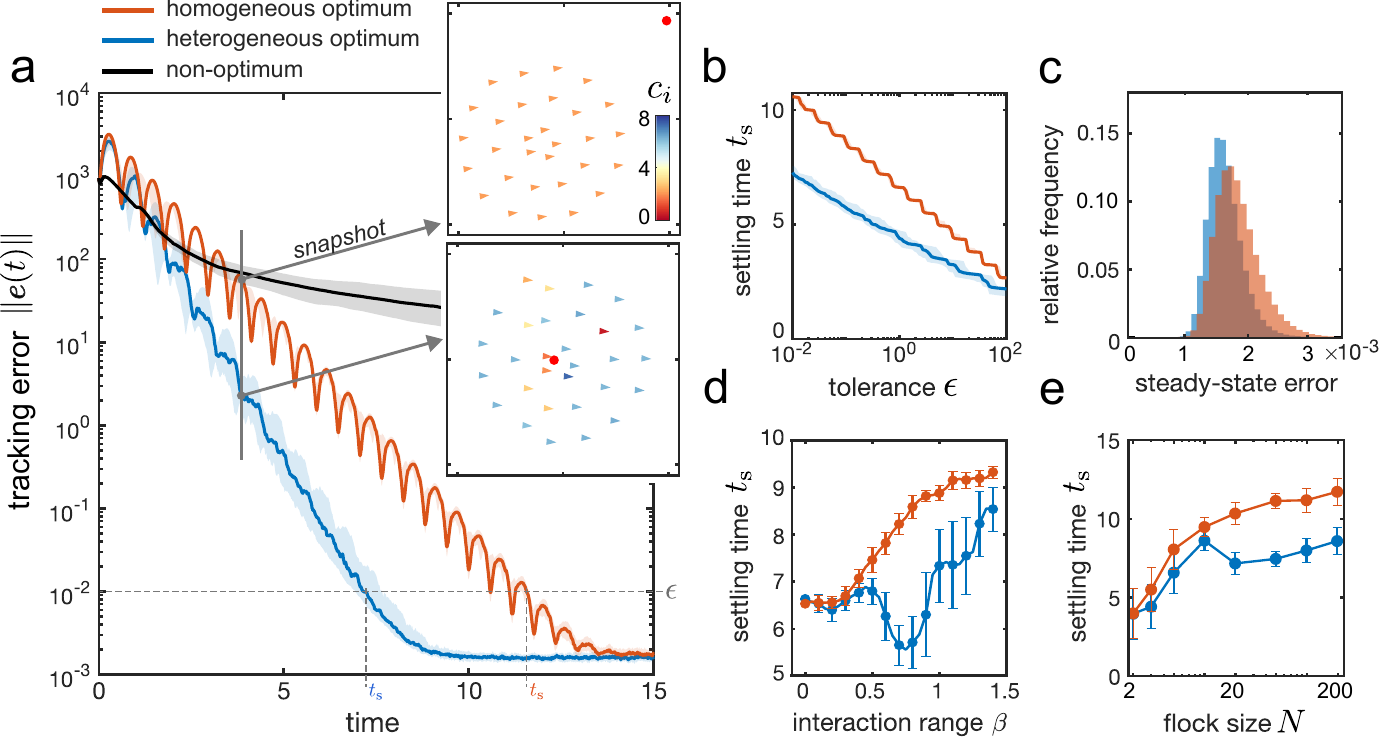}
	\caption{\textbf{Optimal formation in flocks of heterogeneous and homogeneous agents.}
            (\textbf{a})~Tracking error as a function of time for an optimal flock of $N=30$ agents in the 2D space. The blue and orange lines represent flocks of heterogeneous and homogeneous agents, respectively, in which feedback gains are optimized in real time at every time step $w = T = 0.1$. For reference, the black line represents the non-optimal case of randomly assigned time-independent feedback gains. The solid lines represent the median over 100 realizations with different initial conditions (and parameters in the non-optimal case), while the shaded areas indicate the first and third quartile. The insets show a snapshot of the agents position at $t=4$ for the homogeneous flock (top inset) and heterogeneous flock (bottom inset). Agents are color coded by their velocity feedback gain $\bm c_i$. In this simulation, the agents start at a random stationary positions around the origin ($\bm q_i(0)\sim \mathcal U[-2,2]^2$) and are tasked to track a virtual target (red dot) that starts far away from the agents ($\bm q_{\rm t}(0) = [100,\, 100]$) and moves with constant velocity ($\bm p_{\rm t}(t) = [100,\, 0]$, $\forall t\geq 0$). 
            (\textbf{b})~Settling time $t_{\rm s}$ as a function of the tolerance $\epsilon$, 
            where the relationship $t_{\rm s}$ vs. $\epsilon$ is illustrated by the dashed lines in panel \textbf{a}.
            (\textbf{c})~Histogram of the steady-state error $\norm{\bm e(t)}$ of the heterogeneous (blue) and homogeneous (orange) flock across all realizations in panel \textbf{a} for $20\leq t\leq 30$.
            (\textbf{d},\textbf{e})~Settling time $t_{\rm s}$ as a function of the  interaction range $\beta$ (\textbf{d}) and the  number of agents $N$ (\textbf{e}) for heterogeneous (blue) and homogeneous (orange) flocks. The dots and error bars represent the average and one standard deviation over 100 realizations, respectively.
            All simulations implement Eq.~\eqref{eq.flockmodel} with additive Gaussian noise to probe the robustness to small perturbations.
            See Methods for details on the simulation parameters and Supplementary Movie 1 for an animation of the  dynamics.
        }
	\label{fig.converror}
    \vspace{-0.1cm}
\end{figure*}

To describe the dynamics of a flock of $N$ agents in an $m$-dimensional Euclidean space, we represent each agent $i=1,\ldots,N$ as a state variable $\bm x_i = [\bm q_i, \,\, \bm p_i]$, where $\bm q_i\in\R^m$ and $\bm p_i\in\R^m$ are respectively the position and momentum of agent $i$. \rev{The model assumes unit mass and all quantities expressed in dimensionless form; time, position, and velocity are scaled relative to a chosen reference.} The full-system state is denoted as $\bm x = [\bm x_1, \ldots, \bm x_N]\in\R^n$, where $n = 2Nm$. Here, the flock is tasked to maintain a pre-specified formation and follow a (physical or virtual) target moving in space, which is represented by the state vector $\bm x_{\rm t}(t) = [\bm q_{\rm t}(t), \,\, \bm p_{\rm t}(t)]$. The target may represent a moving vehicle/animal or a pre-programmed trajectory. To perform this task, we consider the following multi-agent model \cite{ren2007formation,ren2007consensus}:
\begin{equation}
    \begin{aligned}
        \dot{\bm q}_i &= \bm p_i, \\
        \dot{\bm p}_i &= \dot{\bm p}_{\rm t} - b_{i}\left(\bm q_i - \bm q_{\rm t} - \bm r_i\right) - \gamma c_{i}\left(\bm p_i - \bm p_{\rm t}\right) \\
        &\,\,\,\, +\sum_{j = 1}^N A_{ij}(t)\Big[ (\bm q_j - \bm r_j) - (\bm q_i - \bm r_i) + \gamma(\bm p_j - \bm p_i) \Big],
    \end{aligned}
\label{eq.flockmodel}
\end{equation}

\noindent
where 
$\bm r_i\in\R^m$ indicates the intended position of agent $i$ within the desired formation relative to the target position $\bm q_{\rm t}$.
The parameters $\bm b = [b_1,\ldots,b_N]$ and $\bm c = [c_1,\ldots,c_N]$ are the controller gains associated with position and velocity feedback, respectively. This feedback control law ensures that the agents achieve target tracking: $\bm q_i(t) \rightarrow \bm q_{\rm t}(t) + \bm r_i$ and $\bm p_i(t)\rightarrow \bm p_{\rm t}(t)$ as $t\rightarrow \infty$, $\forall i$.
The adjacency matrix $A \in\R^{N\times N}$ encodes the pairwise coupling between agents, guaranteeing the formation keeping and velocity matching (per Reynolds' rules \#1 and \#2, respectively).
The damping parameter $\gamma$ parameterizes the velocity feedback gain relative to the position gain. Fig.~\ref{fig.converror}{a} (inset) illustrates the model for a flock of agents sustaining a circular formation centered at the virtual target.

In flocking dynamics, both biological and artificial agents typically interact more strongly with nearby agents, either due to sensing and information-processing constraints \cite{ren2008distributed,pearce2014role} or as a mechanism to ensure alignment and avoid collisions \cite{olfati2006flocking,cucker2007emergent}. Accordingly, we define the adjacency matrix as a time-dependent matrix\cite{cucker2007emergent} $\tilde A_{ij}(t) = K(\rho^2 + \norm{\bm q_i(t) - \bm q_j(t)}^2)^{-\beta}$, where $\rho = 0.1$, \rev{$K> 0$} is the coupling strength, and $\beta\geq 0$ represents the interaction range (larger $\beta$ corresponds to weaker interaction \rev{at long inter-agent distances, as illustrated in Fig.~\ref{fig.sm.beta}}). 
To account for communication constraints in swarms of UAVs, we assume that positional and velocity data are exchanged periodically among agents \cite{ren2008distributed}. Thus, the entries of the adjacency matrix are modeled as piecewise-constant functions \cite{su2011stability}, that is, $A_{ij}(t) = \tilde A_{ij}(t_k)$, for $t\in[t_k, t_k+T]$, where $T$ is the time interval between communication events and $t_k = kT$, for $k\in\N$, is the update time instant. Note that, even though agents operate with information about the network structure updated at discrete times, the dynamics of system \eqref{eq.flockmodel} are still continuous. Denoting the corresponding graph of the adjacency matrix $A(t_k)$ as $\mathcal G(A(t_k))$, it follows from Ref.~\cite[Theorem 2.31]{ren2008distributed} that, since $\bigcup_k \mathcal G(A(t_k))$ is undirected and connected, 
consensus is guaranteed to be achieved asymptotically: $(\bm q_i(t) + \bm r_i) - (\bm q_j(t) + \bm r_j) \rightarrow 0$ and $\bm p_j(t) - \bm p_i(t) \rightarrow 0$ as $t\rightarrow \infty$ for all pairs $(i,j)$.

\subsection*{Optimal flock formation for target tracking}
Given that consensus is guaranteed theoretically, our goal is to optimize the controller parameters $b_i$ and $c_i$ to maximize the convergence rate towards the intended formation (centered at the target). To this end, we define the tracking error of each agent $i$ as $\bm e_{i} = [\bm e_{q,i}, \,\, \bm e_{p,i}] = [\bm q_i - (\bm q_{\rm t} + \bm r_i), \,\, \bm p_i - \bm p_{\rm t}]$, in which $\bm e_q = [\bm e_{q,1},\ldots,\bm e_{q,N}]$ and $\bm e_p = [\bm e_{p,1},\ldots,\bm e_{p,N}]$. From Eq.~\eqref{eq.flockmodel}, the tracking error dynamics are given by
\begin{equation}
    \begin{bmatrix}
        \dot{\bm e}_{q} \\ \dot{\bm e}_{p}
    \end{bmatrix}
    =
    \underbrace{
    \begin{bmatrix}
        0_{Nm} & I_{Nm} \\
        - J_1(t)  \otimes I_m & - J_2(t) \otimes  I_m
    \end{bmatrix}
    }_{J(t)}
    \underbrace{
    \begin{bmatrix}
        {\bm e}_{q} \\ {\bm e}_{p}
    \end{bmatrix}
    }_{\bm e},
\label{eq.ltv}
\end{equation}

\vspace{-0.1cm}
\noindent
where $0_{Nm}$ is an $Nm \times Nm$ zero matrix, $I_{Nm}$ is an identity matrix of size $Nm$, and $\otimes$ denotes the Kronecker product. 
It follows that $J_1(t) = B+L(t)$ and $J_2(t) = \gamma (C+L(t))$, where
$B = \operatorname{diag}(\bm b)$ is the position feedback matrix, $C = \operatorname{diag}(\bm c)$ is the velocity feedback matrix, and $\operatorname{diag}(\cdot)$ denotes a diagonal matrix with the respective input vector elements along its diagonal.
The Laplacian matrix is given by 
$L(t) = D(t) - A(t)$, where $D(t) = \operatorname{diag}(\sum_j A_{1j}(t), \ldots, \sum_j A_{Nj}(t))$.

Eq.~\eqref{eq.ltv} is a linear time-varying (LTV) system whose solution is given by $\bm e(t) = \bm\Phi(t,0)\bm e(0)$, where $\bm e(0)$ is the initial condition and $\bm\Phi(t,0)$ is the state-transition matrix. Since $A(t)$, and hence $J(t)$, are piecewise-constant matrices, it follows that $\bm\Phi(t,0) = \prod_{k=0}^{t/T} \bm\Phi(t_k+T,t_k) = \prod_{k=0}^{t/T} e^{J(t_k)T}$ (we assume for simplicity that $t/T\in\N$). Therefore, we have
\vspace{-0.1cm}
\begin{equation}
\begin{aligned}
    \norm{\bm e(t)} &\leq \prod_{k=0}^{t/T} 
    \norm{e^{J(t_k)T}} \norm{\bm e(0)} \\
    &\leq 
    \eta \exp\Big\{{\sum\limits_{k=0}^{t/T} \Lambda_{\rm max}(J(t_k))T}\Big\}  \norm{\bm e(0)},
\end{aligned}
\label{eq.upperboundlambda}
\end{equation}

\vspace{-0.1cm}
\noindent
where $\eta = \prod_k \norm{U_k}\norm{U_k^{-1}}$ and $U_k$ is the transformation matrix in the Jordan decomposition $J(t_k) = U_k^{-1} \tilde J_k U_k$ ($\tilde J_k$ is the corresponding Jordan matrix). 
In this case, the convergence rate of $\bm e(t)$ is characterized by the spectral properties of $J(t_k)$, $\forall k$, and is upper bounded by the largest Lyapunov exponent $\Lambda_{\rm max}(J(t_k))=\operatorname{max}_i \operatorname{Re}\{\lambda_i(J(t_k))\}$, where $\lambda_i(J(t_k))$ is the $i$th eigenvalue of $J(t_k)$.
To maximize the convergence time of the tracking error $\bm e(t)$, we formulate the optimization problem as
\begin{equation}
    \begin{aligned}
        \min_{\bm b,\bm c} \quad &\Lambda_{\rm max}(J(t_k)), \\
        \text{s.t.} \quad & 0 < \bm b \leq b_{\rm max}, \\
                    & 0 < \bm c \leq c_{\rm max},
    \end{aligned}
    \label{eq.optimization}
\end{equation}

\noindent
for each time step $t_k$, where the inequality applies element-wise.
Since $\mathcal G(A(t))$ is strongly connected for all $t$, $L(t)$ has only one null eigenvalue and hence, by the Gershgorin’s disc theorem \cite{horn2012matrix}, the eigenvalues of $J_1(t)$ and $J_2(t)$ have strictly negative real parts if the feedback gains satisfy the lower bound $b_i, c_i > 0$, $\forall i$. The upper bounds
$b_{\rm max}$ and $c_{\rm max}$ represent physical limitations in the controller actuation. 

At each time instant $t_k$, $\bm b^{(k)}$ and $\bm c^{(k)}$ denote the optimal feedback gains given by the solution of Eq. \eqref{eq.optimization}. These optimal gains, which depend on the agents' positions through $J(t_k)$, are set constant within each interval $[t_k, t_k+w]$, where $w$ is the optimization window size. They are then recurrently reoptimized for subsequent time windows. The optimization window size $w$ is assumed to be synchronous with the interval $T$ between communication events such that $w=\kappa T$, where $\kappa \in \N$. Except when noted otherwise, we set both windows to have the same size (i.e., $\kappa = 1$).
In what follows, we implement the real-time optimization procedure for two scenarios:
\begin{enumerate}
\itemsep0em 
    \item optimal flocks of \textit{homogeneous} agents, where parameters are optimized subject to the constraint that all agents have identical gains, i.e., $\bm b^{(k)} = [b^{(k)},\ldots,b^{(k)}]$ and $\bm c^{(k)} = [c^{(k)},\ldots,c^{(k)}]$;
    \item optimal flocks of \textit{heterogeneous} agents, where gains are optimized independently for each agent, i.e., $\bm b^{(k)} = [b_1^{(k)},\ldots,b_N^{(k)}]$ and $\bm c^{(k)} = [c_1^{(k)},\ldots,c_N^{(k)}]$.
\end{enumerate}

\noindent
Thus, the feedback matrices are also piecewise-constant functions: $B(t) = \operatorname{diag}(\bm b^{(k)})$ and $C(t) = \operatorname{diag}(\bm c^{(k)}),\forall t\in[t_k,t_k+w]$. The procedure to solve the optimization problem \eqref{eq.optimization} is discussed later in this section.

\begin{figure*}
	\centering
	\includegraphics[width=0.9\textwidth]{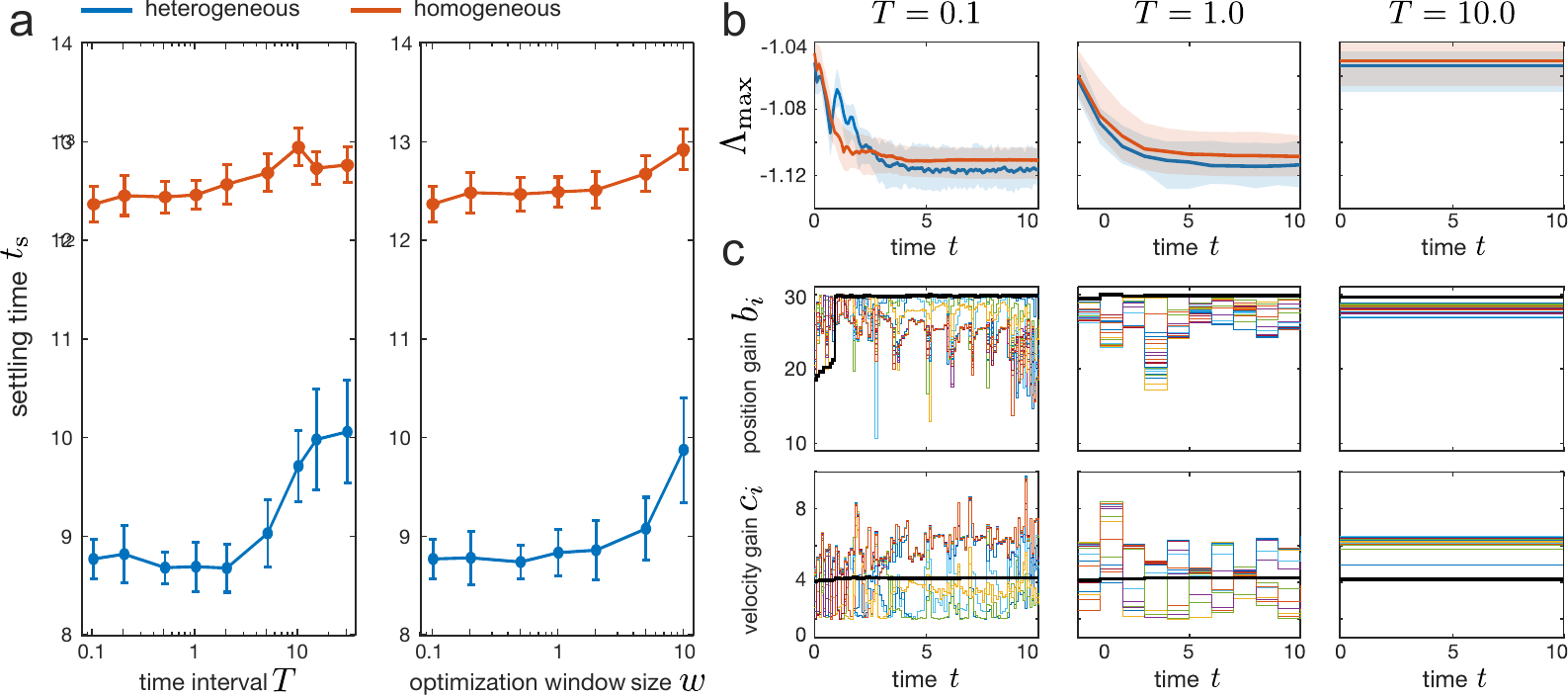}
	\caption{\textbf{Real-time optimization performance.}
		(\textbf{a})~Settling times $t_{\rm s}$ as functions of the time interval $T$ between communication events (left for $w=T, \epsilon = 0.01$) and the optimization window size $w$ (right for $T = 0.1, \epsilon = 0.01$). The data points represent an average over 100 independent realizations of the initial conditions, and the error bars indicate one standard deviation. 
		(\textbf{b})~Lyapunov exponent $\Lambda_{\rm max}$ of the optimal flock as a function of time for different parameter choices: $T=0.1$ (left), $T=1$ (middle), and $T=10$ (right); $w=T$ in all cases. The color scheme is the same as in panel \textbf{a}, with the blue and orange lines representing optimal flocks of heterogeneous and homogeneous agents, respectively. The solid lines represent the median over 100 realizations and the shaded areas indicate the first and third quartiles.
            (\textbf{c})~Optimal position (top) and velocity (bottom) feedback gains as functions of time in a representative realization in a heterogeneous flock (a colored line for each agent) and a homogeneous flock (a common black line for all agents). The different columns correspond to the choices of $T$ and $w$ in panel \textbf{b}.
            In all cases, the number of agents is $N=30$ and the simulation time is 30 (panels \textbf{b} and \textbf{c} are zoomed in to facilitate visualization). 
            See Methods for details on the system parameters.
        }
	\label{fig.realtimeopt}
    \vspace{-0.3cm}
\end{figure*}

\subsection*{Heterogeneous versus homogeneous flocks}
Fig.~\ref{fig.converror} compares the performance in the target tracking task for optimized flocks of heterogeneous and homogeneous agents by first considering a target that moves with constant speed. Fig.~\ref{fig.converror}{a} shows that the convergence of a flock to the desired formation centered at the virtual target is substantially faster for the optimal flocks when compared to a flock of agents with randomly assigned parameters (black line). However, the heterogeneous flock exhibits a noticeably faster convergence than the homogeneous one. As illustrated for the specified tolerance $\epsilon = 10^{-2}$, the heterogeneous flock converges on average within time $t_{\rm s}=7.16$ while the homogeneous flock takes $t_{\rm s} = 11.62$, where the settling time $t_{\rm s}$ is defined such that $\norm{\bm e(t)} < \epsilon$ for $t\geq t_{\rm s}$. 
\rev{The optimized homogeneous flocks shown in Fig.~\ref{fig.converror}a exhibit an underdamped response,
as evidenced in the oscillations of the tracking error $\norm{\bm e(t)}$; in contrast, the heterogeneous flocks display a strongly damped response, with minimal oscillations around the target.
}
Fig.~\ref{fig.converror}{b} shows the dependence of the settling time on the specified tolerance. The slopes $\alpha = \Delta t_{\rm s}/\Delta\log_{10}\epsilon$ for the heterogeneous and homogeneous flocks are respectively $\alpha_{\rm het} \approx -1.32$ and $\alpha_{\rm hom} \approx -2.06$, indicating that the optimal heterogeneous flocks converge on average 36\% faster than their homogeneous counterparts for any threshold within the depicted range. 

Fig.~\ref{fig.converror}{c} shows that, along with the improvement in the convergence rate, optimizing agent heterogeneity also enhances the robustness of  flock formation against noise. (To comprehensively account for the effect of noise, additive Gaussian noise is incorporated to all simulations in the paper; see Methods for details.)
Indeed, the tracking error at steady state is on average 13\% smaller for heterogeneous flocks. An improvement is indeed expected given that the optimization problem \eqref{eq.optimization} is also related to the stability of the flocking model \eqref{eq.flockmodel} against small disturbances (in the linear regime); note that, for $t\rightarrow\infty$, the time-varying matrix $J(t)$ coincides with the Jacobian matrix of Eq.~\eqref{eq.flockmodel} evaluated at the equilibrium point $\bm e=0$. This analysis demonstrates that heterogeneity can also confer improved robustness in noisy environments. 

The performance improvement  promoted by heterogeneity depends on the choice of the network parameters, including the interaction range $\beta$ and the number of agents $N$. Fig.~\ref{fig.converror}{d} shows that for large $\beta$ the settling times $t_{\rm s}$ of homogeneous and heterogeneous flocks increase and approach each other. \rev{This occurs because, at long inter-agent distances $\norm{\bm q_i(t) - \bm q_j(t)}>1$, a higher $\beta$ reduces the coupling $A_{ij}(t)$, leading to $L(t)\approx 0$ and, consequently, $J_1(t)\approx B(t)$ and $J_2(t)\approx \gamma C(t)$.} Thus, in this case, assigning larger gains $b_i$ and $c_i$ for all $i$ directly minimizes $\Lambda_{\rm max}(J(t))$. An analogous analysis can be conducted for sufficiently small $\beta$ such that $A_{ij}(t)\approx K$ for all pairs $(i,j)$ and hence the Laplacian matrix $L$ has all nonzero eigenvalues equal to $KN$.
In summary, if $J_1$ is dominated by $L$ or $B$ (and $J_2$ is dominated by $L$ or $C$), the optimal parameters $\bm b$ and $\bm c$ that minimize $\Lambda_{\rm max}(J)$ are given by a homogeneous solution. 
Therefore, it is the interplay between the network structure, encoded by $L(t)$, and the nodal dynamics, encoded by the feedback gains $B(t)$ and $C(t)$, that enables heterogeneity to promote optimal stability and optimal convergence rate in the flocking model. 
Importantly, this mechanism is scalable in the sense that heterogeneous flocks exhibit a higher convergence rate even when the number of agents increases (Fig.~\ref{fig.converror}{e}). 
%

Performance analyses for different types of spatial formations, target trajectories (deterministic and stochastic), and \rev{optimization constraints} are reported in the Supplementary Information (SI), Section~\ref{sec.targettrack}. The results confirm that heterogeneous flocks also attain faster convergence in different scenarios.

\subsection*{Real-time parameter optimization}
We now investigate the reliance of the results presented thus far on the real-time optimization procedure, specifically the choice of the time intervals $T$ and $w$. Fig.~\ref{fig.realtimeopt}{a} shows that, on average, the settling time increases with larger $T$ and larger $w$.
The decline in performance results from an increased lag between the continuous changes in the agents' positions and the discrete updates of the network structure (determined by $T$) and the agents' parameters (determined by $w$).
In particular, the settling time of heterogeneous flocks is more sensitive to the choice of $T$ and $w$, which leads to a decrease in its relative improvement with respect to the homogeneous flock from 31\% (for $T=w=0.1$) to 21\% (for $T=w=30$; left panel) and 23\% (for $w=10$, $T=0.1$; right panel).
As shown in Eq.~\eqref{eq.upperboundlambda}, improving the flock's convergence time is tied to the minimization of the Lyapunov exponent $\Lambda_{\rm max}(J(t_k))$.
Fig.~\ref{fig.realtimeopt}{b} shows that $\Lambda_{\rm max}(J(t_k))$ is generally smaller for heterogeneous flocks, even though the relative improvement may be smaller (or negative during short transients) depending on the choice of $T$.
\rev{Nonetheless, the heterogeneous flock retains superior performance even when the optimization procedure is computed every $\kappa = 100$ rounds of communication events (e.g., $w=10$ and $T=0.1$), demonstrating the robustness of the approach with respect to the optimization window.} 

The higher sensitivity of heterogeneous flocks to parameter choices can be observed in Fig.~\ref{fig.realtimeopt}{c}: the optimal gains $\bm b^{(k)}$ and $\bm c^{(k)}$ change non-trivially across time windows, whereas the optimal homogeneous gains remain roughly the same as the agents approach the desired formation.
This suggests that changes in the agents' relative positions impact the optimal assignment of gains more strongly in heterogeneous systems than in homogeneous ones.
\rev{In the SI, Section~\ref{sec.sm.symmetries}, we also demonstrate that the optimization procedure is robust for networks with varying levels of connectivity, and that there is no simple relation between the agent's optimal gains and its structural properties within the flock (e.g., distance to target, node in-degree, and network symmetries). As shown next, the optimal gains are instead strongly determined by the spectral properties of the Jacobian matrix $J$.}

\begin{figure}
    \centering
    \includegraphics[width=\columnwidth]{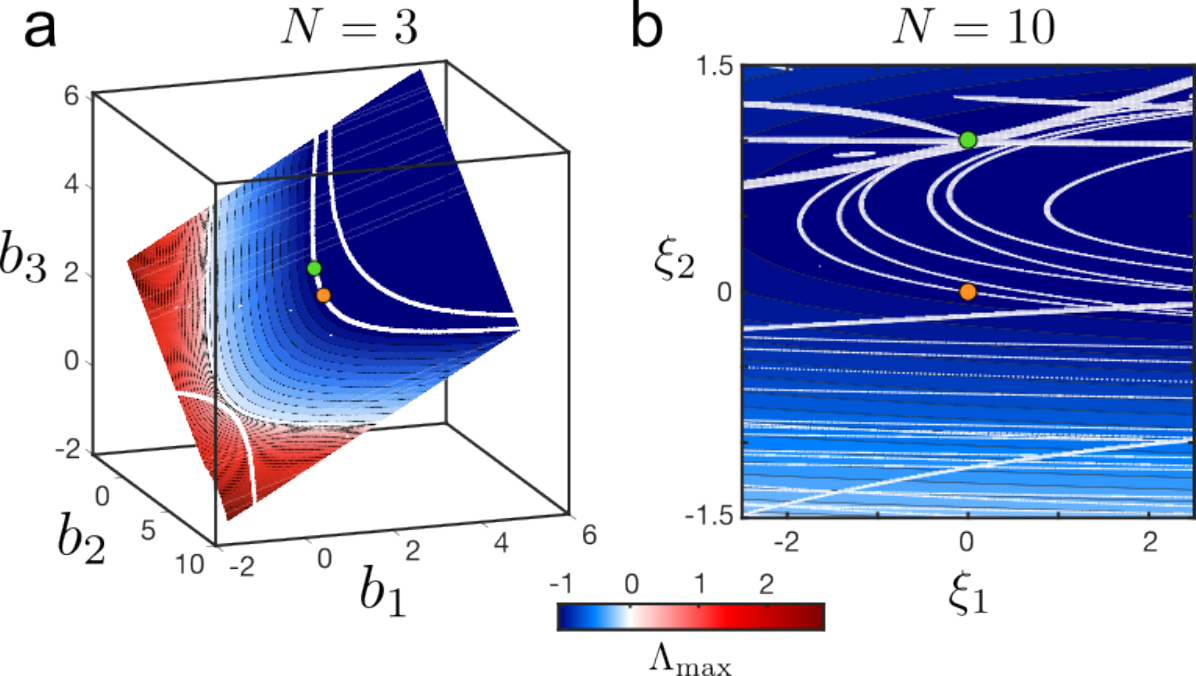}
    \caption{\textbf{Stability landscape for the pre-assigned flocking model.}
    (\textbf{a})~Lyapunov exponent $\Lambda_{\rm max}(J')$ as a function of the feedback gains ($b_1, b_2, b_3$) for $N=3$ agents. The color-coded section shows the stability landscape on a plane containing both the optimal homogeneous gain $\bm b^*_{\rm hom}$ (orange dot) and the optimal heterogeneous gain $\bm b^*_{\rm het}$ (green dot). The flock formation is stable (unstable) for $\Lambda_{\rm max}<0$ ($\Lambda_{\rm max}>0$).
    (\textbf{b})~Lyapunov exponent $\Lambda_{\rm max}(J')$ for $N=10$ on a plane ($\bm \xi_1,\bm \xi_2$)  containing $\bm b^*_{\rm hom}$ and $\bm b^*_{\rm het}$.
    In both panels, the white curves indicate the cross-sections of a  hypersurface (codimension 1) corresponding to single degeneracy of the real parts of the eigenvalues of the Jacobian $J'$.
    }
    \label{fig.eigenvalue}
    \vspace{-0.3cm}
\end{figure}

\smallskip
\subsection*{Lyapunov exponent minimization}
To explicitly characterize the landscape of the optimization problem \eqref{eq.optimization}, let us consider a simplified setup in which the position gains and velocity gains are equal for each agent (i.e., $b_i=c_i$, $\forall i$). For a fixed, time-independent Laplacian matrix $L$, this leads to the following Jacobian matrix
\begin{equation}
    J'(\bm b) = \begin{bmatrix}
        0_{Nm} & I_{Nm} \\
        -(B + L) \otimes I_m & - \gamma (B + L) \otimes I_m 
    \end{bmatrix},
\label{eq.constrainedjacobian}
\end{equation}

\noindent
where $B = \operatorname{diag}(\bm b)$. 
In the case of homogeneous gain ($B = bI_N$), Ref.~\cite[Theorem 5.8]{ren2008distributed} establishes that all eigenvalues of $J'$ have negative real part if $\gamma > \bar\gamma$, where
\begin{equation} \label{eq.gammamin}
    \bar\gamma = 
    \begin{cases}
        0, \,\,\, \text{if all nonzero eigenvalues of $L$ are positive},\\ 
        \max\limits_{\substack{\Re\{\nu_i\}<0\\ \Im\{\nu_i\}>0}} \,\, \sqrt{\frac{2}{|\nu_i|\cos\left(\operatorname{arctan}\left(-\frac{\Im\{\nu_i\}}{\Re\{\nu_i\}}\right)\right)}} \,\, \text{otherwise},
    \end{cases}
\end{equation}
\noindent
and $\nu_i$ is the $i$th eigenvalue of $-(bI_N + L)$. 
Now, let us denote the \textit{optimal homogeneous gain} as the $N$-dimensional vector $\bm b^*_{\rm hom} = [b^*_{\rm hom},\ldots,b^*_{\rm hom}] = \argmin_{\bm b}\Lambda_{\rm max}(J'(\bm b))$ subject to $ \, b_i = b$, $\forall i$.
Based on Eq.~\eqref{eq.gammamin}, we can show that the optimal homogeneous gain is given by
\begin{equation}
b_{\rm hom}^* = \frac{2}{\gamma^2} - \ell_N + \sqrt{(\ell_N-\ell_1)^2+ \frac{4}{\gamma^4}},
\label{eq.homopt}
\end{equation}

\noindent
where $\ell_1<\ldots<\ell_N$ are the eigenvalues of $L$ (see SI, Section~\ref{sec.optimalhom}, for a derivation). 
Eq.~\eqref{eq.homopt} draws a direct link between the network structure and the agents' parameters\textemdash a relation analogous to results previously established for the synchronization of coupled oscillators \cite{Pecora1998,nishikawa2006synchronization} and power grids \cite{Motter2013,Dorfler2013}. 

Fig.~\ref{fig.eigenvalue} illustrates the stability landscape, characterized by $\Lambda_{\rm max}(J')$, for flocks of different sizes. At the homogeneous optimum $\bm b^*_{\rm hom}$, $\Lambda_{\rm max}(J'(\bm b^*_{\rm hom}))$ is non-differentiable and has positive directional derivative $\frac{{\rm d}\Lambda_{\rm max}(J')}{{\rm d}\bm b }|_{\bm b^*_{\rm hom}}>0$ along any vector $\bm b\in\R^N$ (SI, Section~\ref{sec.optimalhom}).
Yet, 
the homogeneous gain $\bm b^*_{\rm hom}$ is not the \textit{best} solution for $\min_{\bm b}\Lambda_{\rm max}(J'(\bm b))$ when $\bm b$ is unconstrained. Although it may seem impossible to further minimize $\Lambda_{\rm max}(J'(\bm b^*_{\rm hom}))$ locally, Ref.~\cite{molnar2021asymmetry} has shown the existence of curved paths out of $\bm b^*_{\rm hom}$ along which $\Lambda_{\rm max}$ locally decreases in the particular case of power-grid networks. Crucially, following these paths, the largest Lyapunov exponent reaches a minimum at some point $\bm b^*_{\rm het}$ corresponding to a heterogeneous choice of parameters. These results can also be extended to the multi-agent consensus model considered here (SI, Section~\ref{sec.stabilitylandscape}). As illustrated in Fig.~\ref{fig.eigenvalue}, such curved paths follow the surfaces of codimension one in the stability landscape where the Jacobian matrix $J'$ is degenerate in the sense of having at least two eigenvalues with identical real parts. 
The fact that the paths connecting $\bm b^*_{\rm hom}$ and $\bm b^*_{\rm het}$ are locally curved hinder the effectiveness of first-order methods (e.g., gradient descent) in solving the optimization problem \eqref{eq.optimization}. To circumvent this issue, we employ solvers that incorporate higher-order approximations of the objective function $\Lambda_{\rm max}$ (e.g., by estimating the Hessian matrix), such as the interior-point method or quasi-Newton methods \cite{nocedal1999numerical}.
\rev{See Methods for computational details on the optimization solver used in the simulations.}

\smallskip
\subsection*{\rev{Distributed optimization}}
\rev{The multi-agent system \eqref{eq.flockmodel} is decentralized as agents rely primarily on local information from nearby peers, especially when $\beta$ is large or $A$ is sparse (see Fig.~\ref{fig.sm.directned} for performance analysis on sparse networks). Yet, the optimization problem \eqref{eq.optimization} has thus far been formulated in a centralized form that requires global knowledge of the adjacency matrix $A(t)$ and hence the full vector of agent positions $\bm q(t)$. When $w$ is large, the optimization operates on slow timescales, allowing enough time for decentralized agent-to-agent communication to gather state information and distribute optimized parameters. In contrast, for small $w$, the flocking dynamics may outpace the computational time required for data collection, optimization, and distribution. To address this challenge, we propose a \textit{distributed} optimization variant of the approach that enables each agent to solve the optimization based on local information.}

\begin{figure}[t!]
    \centering
    \includegraphics[width=\columnwidth]{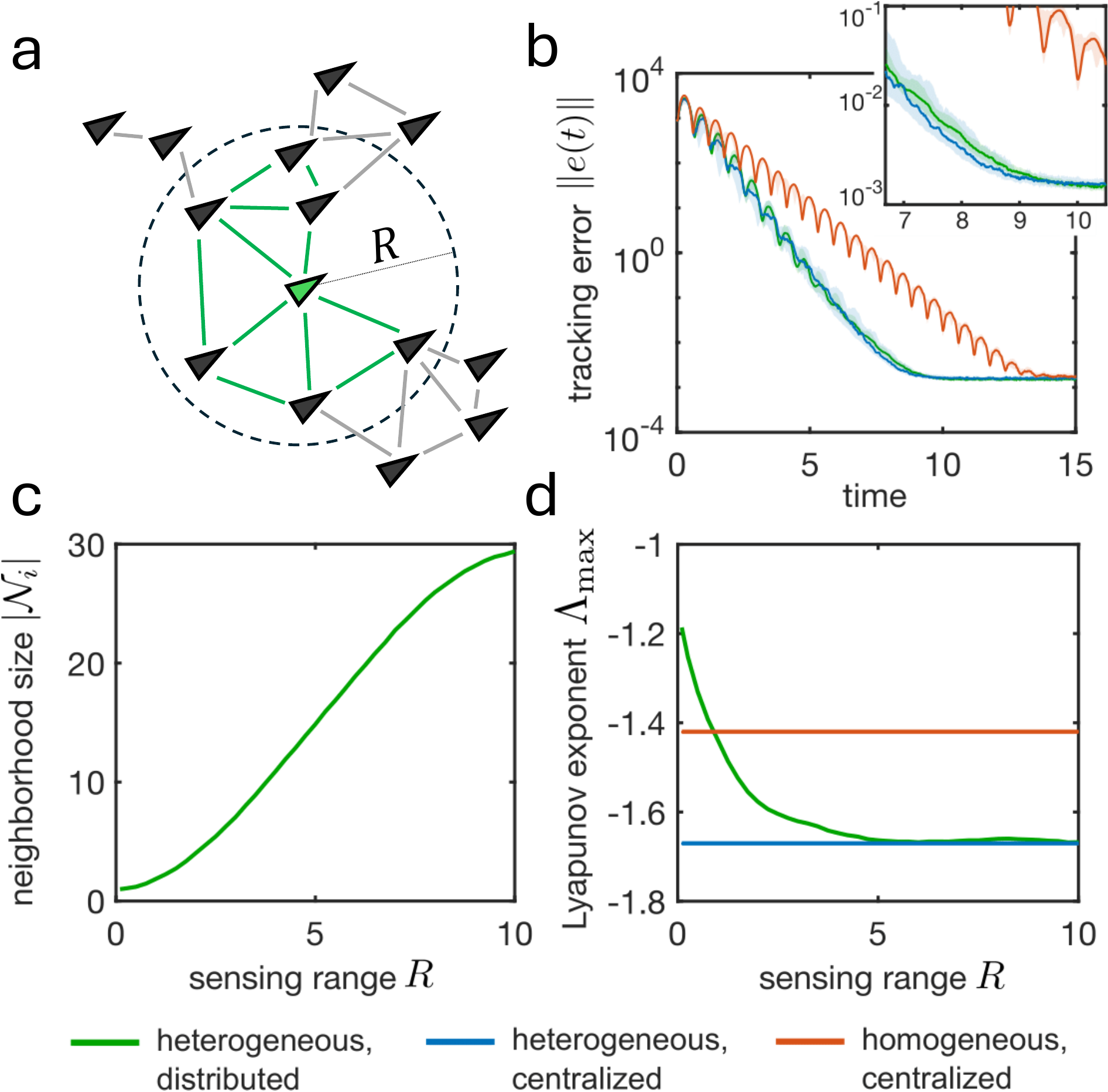}
    \caption{\rev{\textbf{Distributed versus centralized optimization.}
    (\textbf{a}) Schematic diagram of the distributed optimization approach, where each agent has access only to local information within a specified range $R$ and the associated subnetwork $\mathcal G_i$ (green edges).
    (\textbf{b}) Tracking error over time for a flock of $N= 30$ agents using heterogeneous distributed optimization with $R=2$ (green), heterogeneous centralized optimization (blue), and homogeneous centralized optimization (orange).
    (\textbf{c, d}) Average neighborhood size $|\mathcal N_i|$ (c) and optimal Lyapunov exponent $\Lambda_{\rm max}$ (d) as functions of $R$ for the distributed method (green curve). In panel d, the blue and orange lines show the optimal values obtained by the centralized heterogeneous and homogeneous approaches, respectively.
    }}
    \label{fig.distributed}
    \vspace{-0.3cm}
\end{figure}

\rev{Fig.~\ref{fig.distributed}a illustrates the distributed approach. Each agent $i$ is assumed to only access state information of the agents within a spatial neighborhood $\mathcal N_i(\bm q) = \{j \, : \, \norm{\bm q_i - \bm q_j} \leq R\}$, where $R$ is the sensing range. This partial information defines a subgraph $\mathcal G_i\subseteq\mathcal G$ of the communication network by the network of agents within $\mathcal N_i$. At each optimization time $t_k$, every agent $i$ independently performs the following steps: i) retrieves the positions of neighboring agents within $\mathcal N_i$, ii) constructs the corresponding subgraph $\mathcal G_i$, and iii) solves a local, lower-dimensional optimization problem associated with subgraph $\mathcal G_i$ to determine its optimal gains $b_i^{(k)}$ and $c_i^{(k)}$ over the interval $[t_k,t_k+w]$ (see Methods for details). Thus, each agent adapts its parameter based both on local information and local computation, reducing computational burden and enabling paralellization. 
}

\rev{We compare the flock convergence under three optimization scenarios: i)  the distributed formulation, ii) the centralized formulation, and iii) the centralized formulation in which the agent gains are constrained to be homogeneous. 
%
Fig.~\ref{fig.distributed}b shows that flocks optimized with the distributed method perform comparably to those optimized by the centralized (heterogeneous) method for a sensing range $R=2$, while also converging 34\% faster than their homogeneous counterparts (under the same simulation conditions as Fig.~\ref{fig.converror}). In this context, $\mathcal N_i$ contains on average only 4.1 agents for $R=2$ (Fig~\ref{fig.distributed}c), substantially reducing the dimension of the local optimization problem and its computational burden. Fig.~\ref{fig.distributed}d also shows that, as $R$ increases, the optimal Lyapunov exponent obtained by the distributed approach converges to global optimum obtained by the centralized approach, as expected. Notably, for $R>1.5$ ($|\mathcal N_i|=3.7$ on average), the distributed (heterogeneous) approach already outperforms the centralized homogeneous approach. 
}

\section*{Extension to Time-Delay Systems}

\begin{figure}[t!]
    \centering
    \includegraphics[width=0.95\columnwidth]{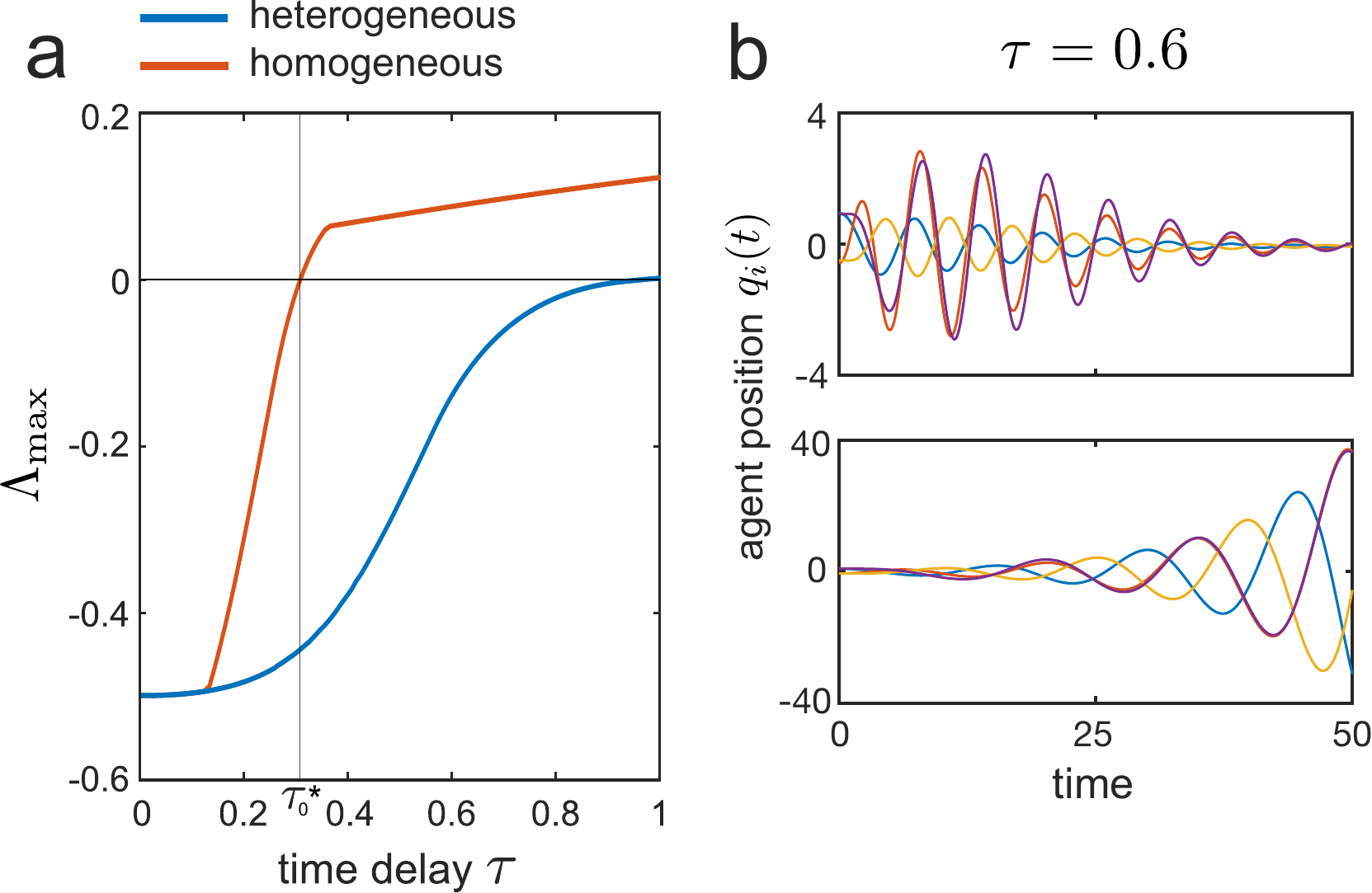}
    \caption{\textbf{Heterogeneity-induced stability in consensus dynamics with time delay.}
    (\textbf{a}) Lyapunov exponent $\Lambda_{\rm max}$ as a function of the time delay $\tau$ for an optimal flock of heterogeneous (blue) and homogeneous (orange) agents for $N=4$. Consensus is stable (unstable) for all initial conditions $\bm x(0)$ if $\Lambda_{\rm max} > 0$ ($\Lambda_{\rm max} > 0$). A flock is said to be optimal if the choice of parameters $k_i$ minimizes $\Lambda_{\rm max}$.
    (\textbf{b})~Dynamical evolution of the agents' positions $\bm q_i(t)$ for an optimal flock of heterogeneous (top) and homogeneous (bottom) agents. The time delay is set as $\tau = 0.6$. See SI, Section~\ref{sec.dde}, for details on the optimization problem. 
    }
    \label{fig.timedelay}
    \vspace{-0.3cm}
\end{figure}

Having established that agent heterogeneity can improve \textit{convergence}, we now show that it can improve \textit{stability} and lead to stable consensus even when a homogeneous flock is necessarily unstable.
%
%
Consider the  second-order consensus model with time delay \cite{yu2010some}:
\begin{equation}
    \begin{aligned}
        \dot{\bm q}_i(t) &= \bm p_i(t), \\
        \dot{\bm p}_i(t) &= - k_i \left(\sum_{j=1}^N L_{ij}\bm q_j(t-\tau) + \sum_{j=1}^N L_{ij}\bm p_j(t-\tau) \right),
    \end{aligned}
\label{eq.timedelayconsensus}
\end{equation}

\noindent
where $L\in\R^{N\times N}$ is a (time-invariant) Laplacian matrix, $\tau$ is the time delay modeling the communication lag between agents, and $k_i$ is the coupling gain of agent $i$. 
Consensus is achieved if $\|\bm x_i(t)-\bm x_j(t)\|\rightarrow 0$ as $t\rightarrow\infty$ for all pairs $(i,j)$ 
(illustrated in Fig.~\ref{fig.timedelay}{b}, top). 
The system of delay differential equations (DDE) \eqref{eq.timedelayconsensus} reaches consensus, or is said to be asymptotically stable, if and only if all ``eigenvalues'' have negative real part (see SI, Section~\ref{sec.dde}, for details). Lack of consensus leads to irregular fragmentation, a common pitfall where the flock breaks into subgroups of agents that diverge from each other in space.

It follows from Ref.~\cite[Theorem 2]{yu2010some} that, for a \textit{fixed} homogeneous choice of coupling gain $k_i = \bar k > 0$, $\forall i$, consensus can be achieved if and only if $\tau < \tau_0$ (where $\tau_0$ depends explicitly on $\bar k$ and the eigenvalues of $L$; see Eq. \eqref{eq.tau0bound} in the SI). Based on this analytical relation, we can show that there exists a maximum delay $\tau_0^* = \max_{\bar k} \tau_0$, subject to the constraint $\bar k\leq k_{\rm max}$, for which consensus can be achieved using some homogeneous parameter assignment (see Fig.~\ref{fig.tau0} depicting $\tau_0$ as a function of $\bar k$).
Indeed, Fig.~\ref{fig.timedelay}{a} shows that for $N=4$ agents constrained by $k_{\rm max}=1$, there exists a flock of homogeneous agents that can achieve consensus if and only if $\tau < \tau_0^* = 0.306$ (corresponding to $\bar{k}^* = 0.724$). In principle, this sets an upper bound on the largest communication delay for which consensus is possible. 
However, this limitation can be circumvented by optimizing the agents' coupling gain in an heterogeneous manner (also subject to the constraint $k_i \leq k_{\rm max}$). Fig. \ref{fig.timedelay}{a} shows that consensus can be achieved for flocks with much larger communication delay (up to $\tau = 0.98$). 
For $\tau = 0.6$, Fig.~\ref{fig.timedelay}{b} confirms that the optimal heterogeneous flock reaches consensus whereas the optimal homogeneous flock irregularly fragments into three separate groups. 

\section*{Extension to Free-Flocking Systems}

Thus far, we have analyzed the convergence and stability of flocking models encompassing two of Reynolds' rules. The multi-agent model \eqref{eq.flockmodel} adheres to rules \#1 and \#2, but excludes rule \#3 as each agent is assigned a fixed position within the flock formation. The consensus model \eqref{eq.timedelayconsensus} implements rules \#2 and \#3 instead, but lacks a mechanism to avoid collisions (note that $\bm q_i(t)\rightarrow \bm q_j(t)$ as $t\rightarrow\infty$ if the system reaches consensus). We now show that heterogeneity can also promote optimal flocking in models that account for all three of Reynolds' rules. This section extends our methodology to more complex dynamics.

\subsection*{Optimal free flocking}
\begin{SCfigure*}[\sidecaptionrelwidth]
    \centering
    \includegraphics[width=0.68\textwidth]{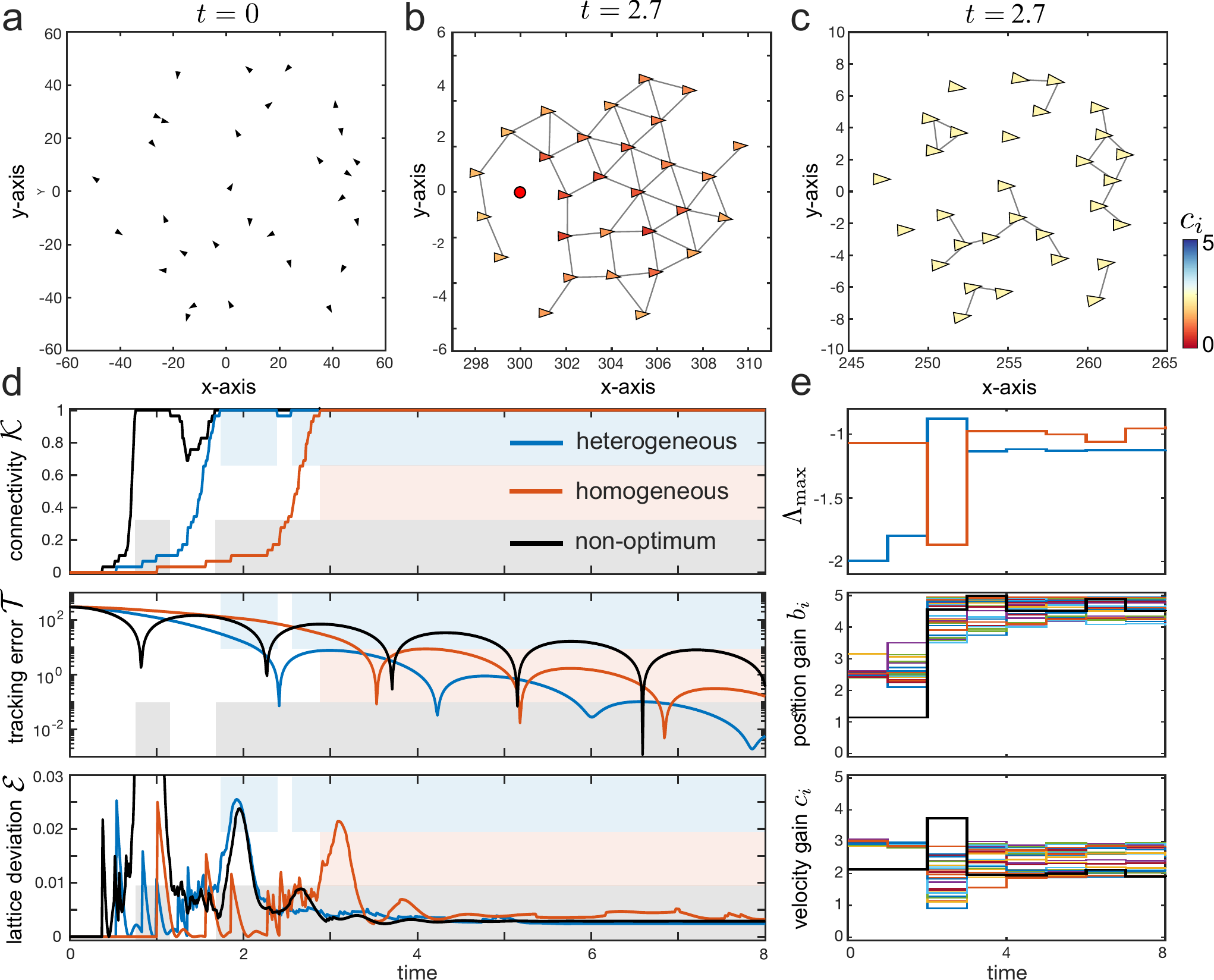}
    \hspace{-0.3cm}
    \caption{\textbf{Optimal free flocking.}
    (\textbf{a})~Initial positions of a group of 30 agents in the 2D Euclidean space.
    (\textbf{b},\textbf{c})~Snapshot of the agents' positions at $t=2.7$ for an optimal flock of heterogeneous (\textbf{b}) and homogeneous (\textbf{c}) agents. The agents are color coded by the feedback gain $c_i$, and the edges indicate the underlying communication network. The target (red dot) is stationary at $\bm q_{\rm t} = [300,\,\, 0]$ (note that the target is distant from the neighborhood of the flock plotted in panel \textbf{c}).
    (\textbf{d}) Performance metrics of the flock convergence as functions of time for a group of heterogeneous (blue) and homogeneous (orange) agents; the performance of a non-optimal flock (black) with constant gains $b_i=b_{\rm max}$ and $c_i=c_{\rm max}$, $\forall i$, is shown as a reference. The shaded areas indicate time intervals in which the network is fully connected ($\mathcal K = 1$) for the respective cases.
    (\textbf{e})~Lyapunov exponent $\Lambda_{\rm max}(J(t_k))$, optimal position gain $b_i^{(k)}$, and optimal velocity gain $c_i^{(k)}$ solved at each time interval $t\in[t_k,t_k+w]$ for $w=1$. {Top}: Heterogeneous (blue) and homogeneous (orange) flocks. {Middle and bottom}: Heterogeneous flock (a colored line for each agent) and homogeneous flock (a common black line for all agents).
    See Methods for details on the system parameters and  Supplementary Movie 2 for an animation of the dynamics.
    }
    \label{fig.freeflock}
    \vspace{-0.5cm}
\end{SCfigure*}

%
We consider the following flocking model proposed by Olfati-Saber \cite{olfati2006flocking}:
\begin{equation}
    \begin{aligned}
        \dot{\bm q}_i &= \bm p_i, \\
        \dot{\bm p}_i &=
        \bm u_i^\alpha + \bm u_i^\gamma + \bm u_i^\beta,
    \end{aligned}
\label{eq.osflock}
\end{equation}

\noindent
for $i=1,\ldots,N$, where each term $\{\bm u_i^\alpha, \bm u_i^\gamma, \bm u_i^\beta\}$ describes a different control objective between agent $i$ and its environment. The agent-agent interaction is given by
\begin{equation}
    \bm u_i^\alpha =  -k_1^\alpha \gradient_{\bm q_i}V(\bm q) + k_2^\alpha \sum_{j\in\mathcal N_i(\bm q)}A_{ij}(\bm q)(\bm p_j-\bm p_i),
\label{eq.ualpha}
\end{equation}

\noindent
for some gains $k_1^\alpha, k_2^\alpha > 0$. In Eq.~\eqref{eq.ualpha}, the second term enforces the velocity consensus among agents (Reynolds' rule~\#2), which is governed by a time-dependent adjacency matrix whose entries are inversely proportional to the agents' relative distance: $A_{ij}(\bm q)\propto\norm{\bm q_i-\bm q_j}^{-1}$. Furthermore, each agent $i$ only interacts with agents located within the spatial neighborhood $\mathcal N_i(\bm q) = \{j  :  \norm{\bm q_i - \bm q_j} < R, \, j\neq i\}$; hence, $A_{ij} = 0$ if $j\notin\mathcal N_i$ and the underlying communication network $\mathcal G(A(\bm q))$ is sparse and possibly disconnected in this model.
The gradient term in Eq.~\eqref{eq.ualpha} involves a smooth collective potential $V(\bm q)$ whose local minima $\bm q^*$ form \textit{lattices} \cite[Lemma 3]{olfati2006flocking}. That is, $\bm q^*$ corresponds to a configuration of agent positions satisfying $\norm{\bm q_i - \bm q_j} = d$ for all $j\in\mathcal N_i(\bm q)$ and some pre-specified distance $0<d<R$ (ensuring no collisions according to Reynold's rule \#1). The precise definition of $A(\bm q)$ and $V(\bm q)$ follows Ref.~\cite{olfati2006flocking} and is reported in Methods. 
Since no formation is pre-assigned and any lattice is an admissible solution, Eq.~\eqref{eq.osflock} defines a free-flocking model. Fig.~\ref{fig.freeflock}{a,b} illustrates the agents converging to a lattice configuration.

The control terms $\bm u_i^\gamma$ and $\bm u_i^\beta$ encode the agent-target and agent-obstacle interactions, respectively. The mathematical modeling of $\bm u_i^\beta$ follows a structure similar to Eq.~\eqref{eq.ualpha}, in which the boundaries of the obstacles are represented by additional virtual agents (see Methods). For simplicity, we first present our results for scenarios with no obstacles ($\bm u_i^\beta = 0$, $\forall i$), also known as flocking in free space.
We define the feedback term modeling the agent-target interaction:
\begin{equation}
    \bm u_i^\gamma = - b_i(\bm q_i - \bm q_{\rm t}) - c_i(\bm p_i - \bm p_{\rm t}),
\label{eq.ugamma}
\end{equation}    

\noindent
where $\bm x_{\rm t}(t) = [\bm q_{\rm t}(t), \,\, \bm p_{\rm t}(t)]$ represents the state of a target moving across space. As before, $b_i, c_i > 0$ are feedback gains sought to be optimized in a homogeneous or heterogeneous manner.
Note that, unlike model \eqref{eq.flockmodel}, Eq.~\eqref{eq.ugamma} does not specify the desired agents' position relative to the target. Instead, each agent seeks to minimize its distance to the target. As a consequence, agents navigate towards the flock's center of mass in accordance to Reynolds' rule \#3.
Following the convergence to a final formation, the position $\bm q_{\rm c} = \frac{1}{N}\sum_{i}\bm q_i$ and momentum $\bm p_{\rm c} = \frac{1}{N}\sum_{i}\bm p_i$ of the flock's center of mass matches the target's position and momentum.

To optimize the flock convergence, we measure the centering deviation $[\bm e_{q,i}, \,\, \bm e_{p,i}] = [\bm q_i - \bm q_c, \,\, \bm p_i - \bm p_c]$ of each agent $i$ with respect to the flock's center of mass. 
Accordingly, we define $\bm e_q = [\bm e_{q,1},\ldots,\bm e_{q,N}]$, $\bm e_p = [\bm e_{p,1},\ldots,\bm e_{p,N}]$, and $\bm e = [\bm e_q, \bm e_p]$. 
In the reference frame of the center of mass, the flock formation at steady-state is given by $\bm e^* = [\bm e_q^*, 0]$, where the agents form a lattice (satisfying $\gradient V(\bm q^*) = \gradient V(\bm e_q^*) = 0$) and match the velocity of the target ($\bm p_i = \bm p_{\rm t}$, $\forall i$). 
Using Lyapunov stability analysis, we prove that \rev{the centering deviation} to a desired formation is upper bounded as 
\begin{equation}
    \norm{{\bm e}(t) - \bm e^*} \leq \eta \exp\left\{\frac{\eta_k}{2\alpha_2}\Lambda_{\rm max}(J(t_k)) T\right\}\norm{{\bm e}(t_k) - \bm e^*}
\label{eq.upperboundosmodel}
\end{equation}

\noindent
for each interval $t\in[t_k,t_k+T]$ and  constants $\eta,\eta_k,\alpha_2, T > 0$, where $t_k = kT$, $k\in\N$. The matrix $J(t_k)$ has a Jacobian-like structure and is defined as
\begin{equation}
    J(t_k) = 
    \begin{bmatrix}
        0_{Nm} & I_{Nm} \\
        -B(t_k) \otimes I_m  & - \left(C(t_k) + L(\bm q(t_k))\right) \otimes I_m
    \end{bmatrix},
    \label{eq.osjacobian}
\end{equation}

\noindent
where $B(t_k) = \operatorname{diag}(\bm b^{(k)})$, $C(t_k) = \operatorname{diag}(\bm c^{(k)})$, and $L(\bm q)$ is the Laplacian matrix associated with $A(\bm q)$. In contrast to model \eqref{eq.flockmodel}, $A(\bm q)$ is \textit{not} a piecewise-constant function, but rather a continuous function. Nonetheless, due to the timescale separation between changes in the network structure and the motion of agents in space, we can approximate $A(\bm q)$  by a piecewise-constant function within each interval $t\in[t_k,t_k + T]$, leading to Eq.~\eqref{eq.osjacobian}. This approximation was used to derive the upper bound \eqref{eq.upperboundosmodel} (SI, Section~\ref{sec.stabilityfreeflock}).

Once again, by solving Eq.~\eqref{eq.optimization} at each interval $[t_k,t_k+w]$ (where $J$ is now given by Eq.~\eqref{eq.osjacobian}), we can optimize the convergence rate of the flock to a lattice  formation. This procedure determines the optimal gains $\bm b^{(k)}$ and $\bm c^{(k)}$ for each instant $t_k$.

\subsection*{Free-flocking performance}
Fig.~\ref{fig.freeflock} compares the free-flocking performance between optimal flocks of heterogeneous and homogeneous agents. Starting from the same initial condition $\bm q_i(0)\sim\mathcal U[-60,60]^2$ (Fig. \ref{fig.freeflock}{a}), the heterogeneous flock forms a connected, lattice-like formation (Fig.~\ref{fig.freeflock}{b}), whereas the homogeneous flock remains largely disconnected within the same convergence time (Fig.~\ref{fig.freeflock}{c}). To measure this heterogeneity-promoted improvement, we evaluate the flock convergence using the following three metrics reported in Fig.~\ref{fig.freeflock}{d}:
\begin{enumerate} \itemsep0em 
    \item the relative connectivity of the agents' communication network, $\mathcal K(t) = \frac{1}{N-1}\rank(L(\bm q(t))$, where 0 corresponds to a fully disconnected network and 1 to a network with a single connected component;
    \item the tracking error of the center of mass with respect to the target position, $\mathcal T(t) = \norm{\bm q_{\rm t}(t) - \bm q_{\rm c}(t)}$, where $\mathcal T = 0$ corresponds to full convergence; 
    \item the formation deviation from a perfect lattice configuration, $\mathcal E(t) = \frac{1}{N_{\rm e} + 1}\sum_{i=1}^N\sum_{j\in\mathcal N_i}\psi(\norm{\bm q_j(t) - \bm q_i(t)}-d)$, where $N_{\rm e}$ is the number of edges in $\mathcal G(A(t))$ and the function $\psi(\cdot)$ is defined in Methods. 
\end{enumerate}

\noindent
Note that $\mathcal E$ only measures the lattice deviation within connected components, and thus this quantity is most useful when the network comprises a single connected component ($\mathcal K = 1$). 

The simulations show that allowing parameter heterogeneity in the optimization procedure can enhance the convergence rate of flocks by 36\%. For the tolerance  $\epsilon = 1$, we observe a settling time of $t_{\rm s} = 4.01$ in the tracking task for heterogeneous flocks, which contrasts with $t_{\rm s}=6.27$ for homogeneous flocks (here, $t_{\rm s}$ is defined such  that $\mathcal T(t) \leq \epsilon$, $\forall t\geq t_{\rm s}$). This result aligns with our expectations given that, following Eq.~\eqref{eq.upperboundosmodel}, $\mathcal T(t)$ is directly related to the optimized cost function $\Lambda_{\rm max}(J(t_k))$. The improvement in  $\Lambda_{\rm max}$ is illustrated in Fig.~\ref{fig.freeflock}{e}; for instance, at $t=0$ (initial formation) and $t=8$ (final formation), $\Lambda_{\rm max}$ is respectively 97\% and 19\% smaller for the heterogeneous flock. Again, the gains change nontrivially over time, despite the tendency of $\bm b^{(k)}$ to increase as the flock formation converges.


It is instructive to further examine the performance metrics in Fig.~\ref{fig.freeflock}{d}. As a result of the faster decay in $\mathcal T(t)$, the heterogeneous agents form a single connected component 38\% faster than their homogeneous counterparts. However, there is a trade-off between the flock's convergence to a connected formation and its tracking error. For example, trivially setting $(b_i,c_i) = (b_{\rm max},c_{\rm max})$, $\forall i$, leads to faster convergence to a fully connected formation than either of the optimized flocks (Fig.~\ref{fig.freeflock}{d}, black line). Yet, this improvement comes at the expense of large underdamped oscillations of the flock's center of mass around the target, yielding an overall slower decay of $\mathcal T(t)$.

The flock convergence can also be quantified using the lattice deviation $\mathcal{E}$.
A constant value of $\mathcal E(t)$ over time indicates that the flock converged to a steady-state formation in which the relative motion between agents is negligible.
As observed in the other performance measures, $\mathcal{E}(t)$ stabilizes more rapidly for heterogeneous flocks compared to homogeneous ones. 
Specifically, $\mathcal E(t)$ converges to a fixed value (within 10\% deviation) at $t = 5.25$ and $t = 7.81$ for the optimal heterogeneous and homogeneous flocks, respectively. 
The spikes in $\mathcal{E}(t)$ are due to discontinuities in network connectivity.
For all cases shown in Fig.~\ref{fig.freeflock}{d}, the steady-state values of $\mathcal{E}(t)$ are small and comparable. This suggests that the resulting formations are consistently well-structured.


\begin{SCfigure*}[\sidecaptionrelwidth]
    \centering
    \includegraphics[width=0.62\textwidth]{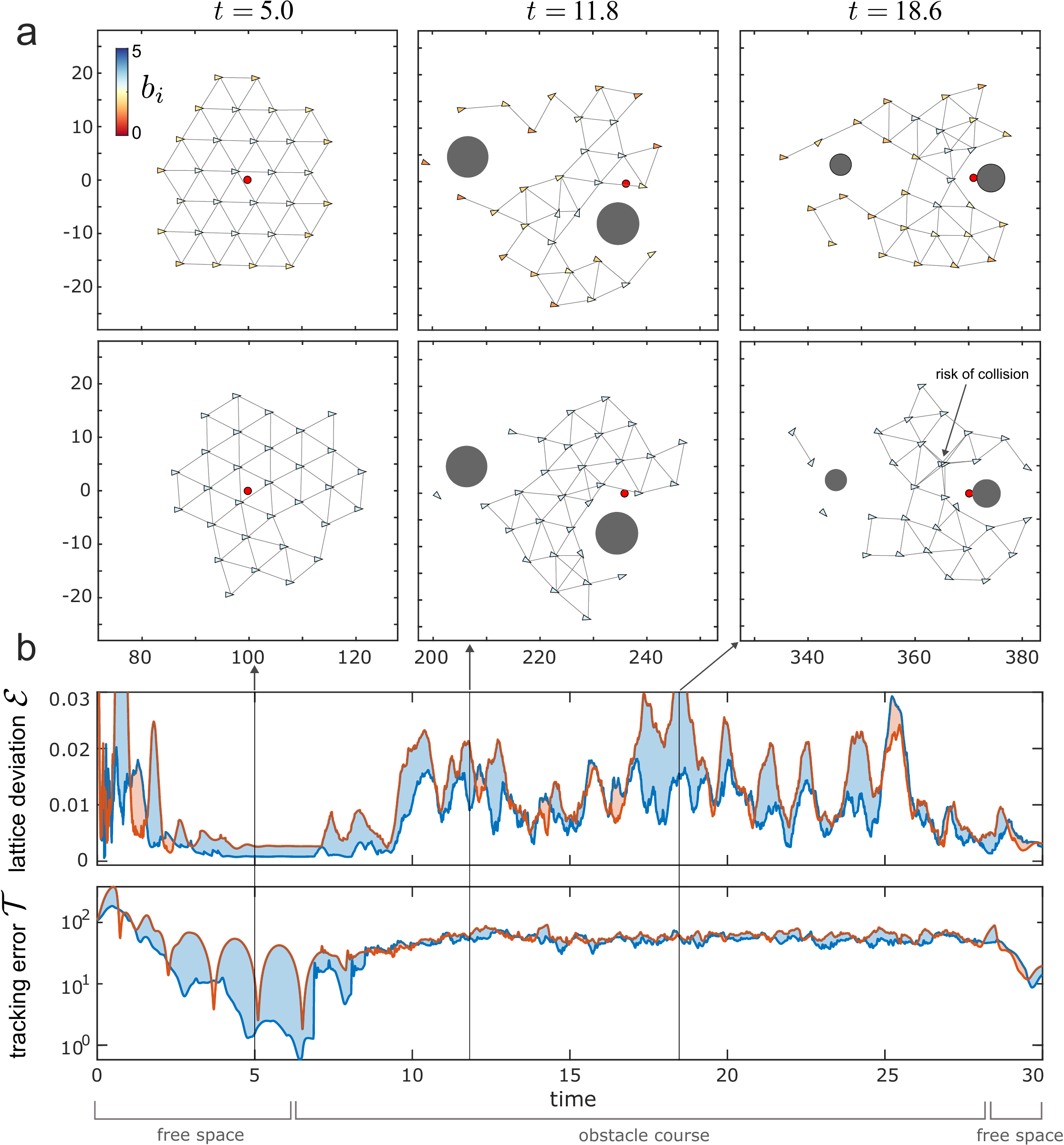}
    \caption{\textbf{Optimized maneuvering of obstacles.}
    (\textbf{a})~Snapshot of the agents' positions at different time instants for optimal flocks of heterogeneous (top) and homogeneous (bottom) agents. The agents are color coded by the feedback gain $b_i$, and the edges indicate the underlying communication network. The obstacles are represented by gray circles. The virtual target (red dot) moves with constant speed along the x-axis ($\bm p_{\rm t}(t) = [20,\,\, 0]$, $\forall t\geq 0$, and $\bm q_{\rm t}(0) = [0,0]$).
    (\textbf{b}) Performance metrics of the flock convergence as functions of time for optimal heterogeneous (blue) and homogeneous (orange) flocks. If the heterogeneous (homogeneous) flock has a superior performance, then the area between the curves is colored in blue (orange).
    The obstacle course contains 15 circular objects, each with radius $r_k\sim\mathcal U[1,4]$, randomly distributed in the 2D-space $[150,-20]\times[600,20]$. The segments where flocks navigate in free space or in an environment with obstacles is marked in panel \textbf{b}. The agents start from the same random initial condition $\bm q_i(0)\sim\mathcal U[-60,60]^2$. See Methods for details on the system parameters.
    }
    \label{fig.obstacle}
    \vspace{-0.3cm}
\end{SCfigure*}

\subsection*{Obstacle maneuvering}
Fig.~\ref{fig.obstacle} compares the flocking of optimized heterogeneous and homogeneous agents navigating through an obstacle course.
In this application, agents must track a virtual target moving with constant velocity along the x-axis while maneuvering around 15 static obstacles randomly placed along the course. The optimization procedure is given by Eq.~\eqref{eq.optimization}, where $J(t_k)$ is a slightly modified version of Eq.~\eqref{eq.osjacobian} to handle the presence of obstacles (SI, Section~\ref{sec.stabilityfreeflock}). The time evolution is  subdivided in three segments in which agents first navigate in free space ($t< 6$), then maneuver around obstacles ($6< t< 28$), and finally leave the obstacle course ($t>28)$. 
This setup is such that the agents form a cohesive flock before encountering the first obstacle. See Supplementary Movie 3 for an animation of the flocking dynamics.

The superior performance of heterogeneous flocks is characterized by an overall smaller lattice deviation $\mathcal E(t)$ and tracking error $\mathcal T(t)$. 
While flocking in free space, the heterogeneous agents converge faster to the lattice formation, which is also more symmetric, as seen at $t=5$. The homogeneous flock  formation, on the other hand, exhibits gaps in the lattice structure. These differences in the lattice structures are captured by the deviation measure $\mathcal E$. Overall, the performance metrics in the interval $t\in[0,6]$ are consistent with the behavior observed in Fig.~\ref{fig.freeflock}, where the agents also flock in free space. 

Along the segment with obstacles, the tracking error $\mathcal T$ is slightly smaller for the heterogeneous flock, with an average improvement of $14\%$. Crucially, $\mathcal E$ is consistently smaller for the heterogeneous flock throughout most of this segment, yielding an improvement of $30\%$ on average. This improvement is illustrated in the snapshots at $t=11.8$ and $t=18.6$, showing that the homogeneous flock have agents undesirably close to each other (at a distance $\norm{\bm q_j-\bm q_i}$ much smaller than the specified lattice spacing $d=7$), as indicated by a larger $\mathcal E$. The difference between $\mathcal E(t)$ for heterogeneous and homogeneous flocks is  largest at $t= 18.6$, where the homogeneous flock is most susceptible to collisions among agents. The heterogeneous agents, on the other hand, sustain a more stable formation with less variability in $\mathcal E$.
These results suggest that the proposed methodology can be applied to optimize consensus protocols for drone swarms operating in \textit{unmapped} environments since in our simulations agents only detect obstacles in real time. 


\smallskip
\section*{Discussion}
\label{sec.discussion}

Our results show that heterogeneity, when appropriately designed, can improve the stability of flocking dynamics beyond what can be achieved in homogeneous systems. 
We have focused our analysis on three different consensus models as they account for a variety of control tasks and  levels of complexity. 
The control tasks include trajectory tracking of static and dynamic targets, setpoint tracking and spatial formation, rendezvous operation, emergent flocking in free space, and obstacle maneuvering in unmapped terrain.
%
The model complexity accounts for the sparsity and time-dependence of the communication networks, the presence of communication delays, the structure of the flock formation, and the Reynold's rules incorporated by each model. These features are summarized in Table \ref{tab.complexity}.
%
%


This is not the first study to model consensus in heterogeneous multi-agent systems \cite{chen2024cooperative,lee2020tool,zheng2011consensus}. \rev{However, the literature has focused primarily on stability conditions to address the \textit{detrimental} rather than the \textit{beneficial} effect of heterogeneity considered here.
Indeed, previous work generally assumes that heterogeneity inhibits consensus and thus seeks stability conditions characterizing the maximum level of heterogeneity under which consensus is still possible. The conclusion that heterogeneity is detrimental is reached because 1) the stability conditions are usually sufficient rather than necessary, and 2) the heterogeneity tends to be modeled as small and/or random deviations from a homogeneous baseline rather than as a design parameter that can be strategically optimized.
That is,} the main goal has been to determine when consensus can be achieved \textit{despite} heterogeneity, whereas here we identify scenarios in which consensus can be achieved and enhanced \textit{because of} heterogeneity.




\begin{table*}
    \centering
\begin{small}
    \begin{tabular}{llll}
    \toprule[1.5pt]
        \textbf{Feature} & \textbf{Pre-assigned flocking model \eqref{eq.flockmodel}} & \textbf{Time-delay model \eqref{eq.timedelayconsensus}} & \textbf{Free-flocking model \eqref{eq.osflock}} \\
    \toprule[1.5pt]
Reynold's rules                                            & 1, 2                   & 2, 3                               & 1, 2, 3              \\
\midrule[0.5pt]
\multicolumn{1}{c}{\multirow{3}{*}{communication network}} & piecewise constant     & time invariant                     & continuous           \\
\multicolumn{1}{c}{}                                       & undirected, weighted   & directed, binary, signed           & undirected, weighted \\
\multicolumn{1}{c}{}                                       & all-to-all             & sparse                             & sparse               \\
\midrule[0.5pt]
communication delay                                        & no                     & yes                                & no                   \\
\midrule[0.5pt]
flock structure                                            & pre-assigned positions & fixed point                        & lattice (emergent)   \\
\midrule[0.5pt]
feasible control task                                      & flock formation        & position consensus                 & flock formation      \\
                                                           & velocity consensus     & velocity consensus                 & velocity consensus   \\
                                                           & target tracking        &                                    & target tracking      \\
                                                           &                        &                                    & obstacle avoidance  \\
\midrule[0.5pt]
related figures                                            & Figs.~\ref{fig.converror}--\ref{fig.eigenvalue}, \ref{fig.targettrack}--\ref{fig.sm.symmetry} & Figs.~\ref{fig.timedelay}, \ref{fig.tau0} & Figs.~\ref{fig.freeflock}, \ref{fig.obstacle}, \ref{fig.timescale} \\
\toprule[1.5pt]
\end{tabular}
\end{small}
    \vspace{-0.3cm}
    \caption{Complexity features per flocking model.}
    \label{tab.complexity}
\vspace{-0.4cm}
\end{table*}

This work leads to fundamental questions worth pursuing in future research. 
It is well known in control theory that the dynamical characteristics of a system, such as its settling time and overshoot, depend not only on the largest eigenvalue but also on the placement of \textit{all} eigenvalues in the complex plane. Thus, while here we focused on the largest Lyapunov exponent, our formulation can be recasted as an optimal control problem in the context of model predictive control \cite{zhan2012flocking,nascimento2023nmpc}, potentially leading to a tighter bound on the tracking error (see SI, Section~\ref{sec.optimalcontrol}, for a discussion of the challenges involved in this approach).
\rev{To further enhance flock convergence, we could jointly optimize additional parameters of the model\textemdash such as the coupling strength $K$ and exponent $\beta$\textemdash alongside the  gains $\bm{b}$ and $ \bm{c}$.}
\rev{Another promising direction for future work is to extend our approach to multi-agent systems with leader-follower roles \cite{gomez2022intermittent} and/or predator behavior \cite{sar2025dynamics}. These scenarios introduce additional features, such as asymmetries in the interaction network, that could be exploited through our adaptive mechanisms for flocking control.}

It is instructive to note that not all systems are expected to benefit from agent heterogeneity. Consider, for example, the first-order consensus model $\dot{\bm x}=-K L\bm x$, where $K = \operatorname{diag}(k_1,\ldots,k_N)$ represents the individual gains of each agent, satisfying $k_i \leq k_{\rm max}$, $\forall i$. The convergence rate of this model, characterized by its largest (transversal) Lyapunov exponent, is globally optimized by homogeneously setting $k_i =  k_{\rm max}$, $\forall i$ (a result that follows from Gershgorin's disc theorem). 
Thus, homogeneity is better than heterogeneity for achieving consensus in this case, even though we have shown that heterogeneity can enhance consensus in the more complex models considered in this paper. 
\rev{
Other models used to describe spin alignment, such as the XY model and the Vicsek model \cite{ginelli2016physics}, reduce to the first-order consensus model when linearized around the equilibrium, and thus are also consistent with the conclusion that homogeneity is preferable. 
The key factor enabling heterogeneity to promote consensus in our study is the second-order nature of flocking dynamics, which incorporates agent inertia as dictated by Newton’s second law.
}

Given the generality of our results, we suggest that the approach will find applications in the optimization of a broad range of other consensus problems in networks, including mobile sensor networks \cite{leonard2007collective,shi2019fast}, distributed state estimation \cite{battistelli2016stability,soatti2016consensus,montanari2022functional}, opinion dynamics in social networks \cite{meng2018opinion,redner2019reality,bernardo2021achieving,crabtree2024influential}, and energy management for the coordinated charging of electrical vehicles and other Internet-of-Things devices \cite{wang2019distributed,yi2024optimal}. For applications in swarms of UAVs, the approach can also be extended to account for practical challenges, such as asynchronous communication \cite{cao2008agreeing}, 
data packet loss \cite{zhang2016sampled}, actuator saturation \cite{wang2016global}, and cyber-attacks \cite{pasqualetti2011consensus}.

\section*{Methods}

\begin{small}

\noindent
\textbf{Parameters of the pre-assigned formation model.}
We report the parameters used in simulations of the multi-agent system \eqref{eq.flockmodel}. Unless specified otherwise, all parameters are set as follows. The agents are constrained to move in the $m=2$ dimensional space and 
the damping coefficient is set as $\gamma = 1$ (in Figs.~\ref{fig.converror}, \ref{fig.realtimeopt}, \ref{fig.targettrack}) or $\gamma = 3$ (in Fig.~\ref{fig.flock}). The weights of the adjacency matrix $\tilde A$ are computed for $\rho = 0.1$, $\beta = 0.8$, and $K=2$ (in Figs.~\ref{fig.converror}, \ref{fig.realtimeopt}, \ref{fig.targettrack}) or $K=5$ (in Fig.~\ref{fig.flock}). 
The noise is added to the acceleration equation of each agent as $\dot{\bm p_i} = \bm f(\bm q,\bm p) + \bm\nu_i$, where $\bm f(\bm q,\bm p)$ represents the corresponding right-hand side in Eq.~\eqref{eq.flockmodel} and  $\bm\nu_i(t)\sim\mathcal N(\mu,\sigma^2)^m$ is a random variable drawn from an $m$-dimensional Gaussian distribution with mean $\mu$ and standard deviation $\sigma$. We report simulations for $(\mu,\sigma) = (0,0.1)$, but we note that the results in Fig.~\ref{fig.converror}{c} remain qualitatively consistent for the range $\sigma\in[10^{-4},10^0]$.

The initial conditions in the independent realizations are set as $\bm p_i(0) = 0$ and $\bm q_i(0)\sim\mathcal U\big[-\sqrt{N/7.5},\sqrt{N/7.5}\big]^2$ (in Figs.~\ref{fig.converror},~\ref{fig.realtimeopt},~\ref{fig.targettrack}) or $\bm q_i(0)\sim\mathcal U[0,1000]^2$ (in Fig.~\ref{fig.flock}), where $\mathcal U[a,b]^m$ denotes an $m$-dimensional uniform distribution within the interval $[a,b]$. Since our goal in Figs. \ref{fig.converror}, \ref{fig.realtimeopt}, and  \ref{fig.targettrack} is to statistically evaluate the performance of flocks operating under different conditions, we also set the desired formation to be random according to $\bm r_i\sim\mathcal U\big[-\sqrt{N/1.2},\sqrt{N/1.2}\big]^2$; the only exception is the inset of Fig.~\ref{fig.converror}{a} (and  corresponding Supplementary Movie 1), where we adopted an ordered circular pattern for illustration purposes.
Note that the size of the intervals containing the initial conditions $\bm q_i(0)$ and relative positions $\bm r_i$ are scaled to equalize the flock density for any number of agents $N$ (Fig.~\ref{fig.converror}{e}).

\medskip\noindent
\textbf{Free-flocking model.}
We define each of the terms in the free-flocking model \eqref{eq.osflock}. Starting with the agent-agent interaction in Eq.~\eqref{eq.ualpha}, the adjacency matrix is defined as
\begin{equation}
    A_{ij}(\bm q) = \rho_h(\norm{\bm q_j - \bm q_i}_\sigma/\norm{R}_\sigma) \in [0,1],
\end{equation}

\noindent
where the $\sigma$-norm is defined as $\norm{z}_\sigma = \frac{1}{\varepsilon}\big(\sqrt{1+\varepsilon\norm{z}^2} - 1\big)$ and the bump function $\rho_h$ is a scalar function given by
\begin{equation}
    \rho_h(z) = \begin{cases}
        1,& z\in[0,h), \\
        \frac{1}{2}\left(1+\cos(\frac{z-h}{1-h})\right),& z\in[h,1], \\
        0,& \text{otherwise},
    \end{cases}
\end{equation}

\noindent
where $0< h < 1$. We set $\varepsilon = 0.1$ and $h = 0.2$ in the simulations. The interaction range is set to $R = 1.2d$, where $d = 2$ (in Fig. \ref{fig.freeflock}) or 7 (in Fig. \ref{fig.obstacle}) is the constrained distance between agents in the lattice structure.

The collective potential is $V(\bm q) = \frac{1}{2}\sum_i\sum_{j\neq i} \psi_\alpha (\norm{\bm q_j- \bm q_i}_\sigma)$. Let $\psi_\alpha(z) = \int_{\norm{d}_\sigma}^z \phi_\alpha (s){\rm d}s$, where $\phi_{\alpha}(z) = \rho_h(z/\norm{R}_\sigma)\phi(z-\norm{d}_\sigma)$ and $\phi(z) = \frac{1}{2}[(a+b)(z+c)/\sqrt{1+(z+c)^2}+(a-b)]$. It follows that
\begin{equation}
    -\gradient_{\bm q_i}V(\bm q) = \sum_{j\in\mathcal N_i(\bm q)}\phi_\alpha(\norm{\bm q_j-\bm q_i}_\sigma)\frac{\bm q_j - \bm q_i}{\sqrt{1 + \varepsilon\norm{\bm q_j - \bm q_i}^2}}.
\end{equation}

\noindent
We set $a = b = 5$, $c = |a - b|/\sqrt{4ab}$,  $k_1^\alpha = 30$, and $k_2^\alpha = 2\sqrt{k_1^\alpha}$ in the simulations.

We now define the agent-obstacle interaction term:
\begin{equation}
    \begin{aligned}
        \bm u_i^\beta &= - k_1^\beta \gradient_{\bm q_i}V_\beta(\bm q) + k_2^\beta \sum_{k\in\mathcal N_i^\beta} A^\beta_{ik}(\bm q)(\hat{\bm p}_{i,k} - \bm p_i),
    \end{aligned}
\label{eq.ubeta}
\end{equation}
\noindent
where the gradient of the collective potential between the agents and obstacles is given explicitly by
\begin{equation}
    -\gradient_{\bm q_i}V_\beta (\bm q) = \sum_{k\in\mathcal N_i^\beta(\bm q)} \phi_\beta \left(\norm{\hat{\bm q}_{i,k} - \bm q_i}_\sigma\right) \frac{\hat{\bm q}_{i,k} - \bm q_i}{\sqrt{1 + \varepsilon\norm{\hat{\bm q}_{i,k} - \bm q_i}^2}} .
\end{equation}

\noindent
For each obstacle $k=1,\ldots,N_{\rm obs}$ and agent $i=1,\ldots,N$, there is a virtual agent with position $\hat{\bm q}_{i,k}$ and momentum $\hat{\bm p}_{i,k}$, where $N_{\rm obs}$ is the number of obstacles. Note that an agent $i$ can only perceive obstacles within its spatial neighborhood $\mathcal N_i^\beta = \{1,\ldots,N_{\rm obs} :  \norm{\hat{\bm q}_{i,k} - \bm q_i}<R'\}$, where $R'$ is an interaction range. 
    Following Ref.~\cite[Lemma 4]{olfati2006flocking}, spherical obstacles of radius $r_k$ and centered at $\bm y_k\in\R^m$ are represented by $\hat{\bm q}_{i,k} =\mu_{i,k}\bm q_i + (1-\mu_{i,k})\bm y_k$ and $\hat{\bm p}_{i,k} = \mu_{i,k} P_{i,k} \bm p_i$, where $\mu_{i,k} = r_k/\norm{\bm q_i- \bm y_k}$, $\bm \eta_{i,k} = (\bm q_i - \bm y_k)/\norm{\bm q_i - \bm y_k}$, and $P_{i,k} = I_m - \bm\eta_{i,k} \bm\eta_{i,k}^\transp$. The gradient potential and adjacency matrix are respectively given by $\phi_\beta(z) = \rho_{h'}(z/\norm{d'}_\sigma)((z - \norm{d'}_\sigma)/\sqrt{1 + (z - \norm{d'}_{\sigma})^2} - 1)$ and $A^\beta_{ij}(\bm q) = \rho_{h'}(\| \hat{\bm q}_{i,k}-\bm q_i \|_\sigma/\norm{d'}_\sigma)$. We set $h' = 0.9$, $d' = 0.6d$, $R' = 1.2d'$, $k_1^\beta = 300$, and $k_2^\beta = 2\sqrt{k_1^\beta}$. We assign $k_1^\beta\gg k_1^\alpha$ so that agents prioritize collision avoidance with obstacles over retaining formation.

A MATLAB implementation of the free-flocking model \eqref{eq.osflock} is provided in GitHub (see Data availability).

\medskip\noindent
\textbf{Solving the optimization problem.}
To solve the constrained optimization problem \eqref{eq.optimization}, we employ the interior-point method, as implemented by the function \texttt{fmincon} in MATLAB.
At each time interval $[t_k, t_k + w]$, we solve the optimization problem \eqref{eq.optimization} for  10 random initial conditions $b_i\sim\mathcal N(b_{\rm max}/2, 0.01)$ and $c_i\sim\mathcal N(c_{\rm max}/2, 0.01)$, and then select the best solution.
The upper bounds on the feedback gains are set as $b_{\rm max} = c_{\rm max} = 30$ (in Figs.~\ref{fig.converror}, \ref{fig.realtimeopt},  \ref{fig.targettrack}), $10$ (in Fig.~\ref{fig.flock}), or $5$ (in Figs.~\ref{fig.freeflock}, \ref{fig.obstacle}). For the optimization of the flocking models \eqref{eq.flockmodel} and \eqref{eq.osflock}, the Jacobian matrix $J$ is defined in Eqs.~\eqref{eq.ltv} and \eqref{eq.osjacobian}, respectively.
The optimization problem for the time-delay system \eqref{eq.timedelayconsensus} is described in SI, Section \ref{sec.dde}.

\rev{We note that the dimension $m$ of the physical space of the agents does not impact the optimization time of the parameters $\bm{b}$ and $\bm{c}$. Because of the Kronecker product structure of $J$, the set of eigenvalues of the Jacobian matrix $J$, denoted by the operator $\operatorname{spec}(\cdot)$, is given by
\begin{equation}
    \operatorname{spec}(J) = \bigcup_{i=1}^m \operatorname{spec}(J^{(1)}), \quad \text{where} \,\, J^{(1)} = \begin{bmatrix}
        0_{N} & I_{N} \\ -J_1 & J_2
    \end{bmatrix}.
\end{equation}
\noindent
That is, for a specific dimension $m$, the spectrum of $J$ consists of $m$ repeated sets of eigenvalues of the Jacobian matrix $J^{(1)}$. Thus, to determine the largest Lyapunov exponent $\Lambda_{\rm max}(J)$, it suffices to calculate the Lyapunov exponents of the $2N\times 2N$ matrix $J^{(1)}$.}

\medskip\noindent
\textbf{\rev{Distributed optimization formulation.}}
    \rev{In this approach, each agent $i$ computes its optimal gains $b_i$ and $c_i$ using only local information determined by its neighborhood $\mathcal N_i$. Thus, each agent has access to a subgraph $\mathcal G_i\subseteq\mathcal G$, where a node $j$ belongs to $\mathcal G_i$ if $j\in\mathcal N_i$ and an edge $(j,k)$ exists in $\mathcal G_i$ if both $j,k\in\mathcal N_i$. Let $A_i = A[\mathcal N_i]$ denote the submatrix of $A$ formed by selecting rows and columns indexed by $j\in\mathcal N_i$. Since $A$ is the $N\times N$ adjacency matrix of $\mathcal G$, it follows that $A_i$ is the $|\mathcal N_i|\times |\mathcal N_i|$ adjacency matrix of $\mathcal G_i$. The neighborhoods $\mathcal N_i$ generally change over time, and the agent positions $\bm q_j(t_k)$, for $j\in\mathcal N_i$, is sufficient to construct $A_i(t_k)$ at any time $t_k$.}

\rev{For each agent $i$, we define the \textit{local} Jacobian matrix:}
\begin{equation}
    \rev{
    J_i = \begin{bmatrix}
        0_{|\mathcal N_i|m} & I_{|\mathcal N_i|m} \\
        -(B_i + L_i) \otimes I_m & - \gamma (C_i + L_i) \otimes I_m 
    \end{bmatrix},
    }
\label{eq.distributedeig}
\end{equation}

\noindent
\rev{where $B_i = B[\mathcal N_i]$, $C_i = C[\mathcal N_i]$, and $L_i$ is the Laplacian matrix associated with $A_i$. 
At time step $t_k$, each agent $i$  solves\textemdash independently and in parallel\textemdash the following low-dimensional optimization problem:}
\begin{equation}
    \rev{
    \begin{aligned}
        \min_{\bm b^{(i,k)},\bm c^{(i,k)}} \quad &\Lambda_{\rm max}(J_i(t_k)),  \\
        \text{s.t.} \quad & 0 < \bm b^{(i,k)} \leq b_{\rm max},  \\
                    & 0 < \bm c^{(i,k)} \leq c_{\rm max},
    \end{aligned}
    }
    \label{eq.distributedopt}
\end{equation}

\noindent
\rev{where $\bm b^{(i,k)}$ and $\bm c^{(i,k)}$ represent the local $|\mathcal N_i|$-dimensional optimization variables at time $t_k$. Accordingly, the local feedback matrices take the form $B_i(t_k) = \operatorname{diag}(\bm b^{(i,k)})$ and $C_i(t_k) = \operatorname{diag}(\bm c^{(i,k)})$. After solving the problem, each agent $i$ extracts the entries of $\bm b^{(i,k)}$ and $\bm c^{(i,k)}$ corresponding to itself (i.e., the index $j\in\mathcal N_i$ such that $j=i$) and assign them as the optimal gain of agent $i$ for the time interval $[t_k, t_k+w]$.}

\medskip\noindent
\textbf{Data availability.}
Codes and data are available through our GitHub repository: \href{https://github.com/montanariarthur/OptFlock}{https://github.com/montanariarthur/OptFlock}.

\medskip\noindent
\textbf{Code availability.}
The GitHub repository contains the codes used to simulate and optimize the flocking dynamics in all models considered in this study. Both the centralized and  distributed formulations are included.

\medskip\noindent
\textbf{Acknowledgements.}
The authors thank Pietro Zanin for insightful discussions on the eigenstructure of the Jacobian matrix. This work is supported by the U.S. Army Research Office (Grant No.\ W911NF-23-1-0102), National Science Foundation (Grant No.\ DMS-2308341), and Office of Naval Research (Grant No.\ N00014-22-1-2200). The authors also acknowledge support from the National Institute for Theory and Mathematics in Biology (NSF Grant No.\ DMS-2235451 and Simons Foundation Grant No.\ MP-TMPS-00005320) and the use of Quest High-Performance Computing Cluster at Northwestern University.

\medskip\noindent
\textbf{Author contributions statement.}
A.N.M., C.D., and A.E.M. designed the research; A.N.M. and A.E.D.B. developed the theory; A.N.M. performed the numerical simulations; A.N.M. and A.E.D.B. analyzed the data; A.N.M. led the writing of the manuscript; all authors contributed to the interpretation of the results and editing of the final version of the paper.

\medskip\noindent
\textbf{Competing interests.}
The authors declare no competing interests.

\end{small}

\begin{footnotesize}

\end{footnotesize}

\onecolumn
\section*{\LARGE Supplementary Information: 
\\ ``Optimal flock formation induced by agent heterogeneity''}

\vspace{10pt}

{\large Arthur N. Montanari, Ana Elisa D. Barioni, Chao Duan, and Adilson E. Motter}

\vspace{5pt}

\noindent
{\normalsize E-mail: arthur.montanari@northwestern.edu}

\setcounter{equation}{0}
\setcounter{figure}{0}
\setcounter{table}{0}
\setcounter{page}{1}
\setcounter{section}{0}
\makeatletter
\renewcommand{\theequation}{S\arabic{equation}}
\renewcommand{\thefigure}{S\arabic{figure}}
\renewcommand{\thesection}{S\arabic{section}}

\section{Additional examples of target tracking, flock formation, and parameter optimization}
\label{sec.targettrack}

\begin{SCfigure*}[\sidecaptionrelwidth][b!]
	\centering
	\includegraphics[width=0.6\textwidth]{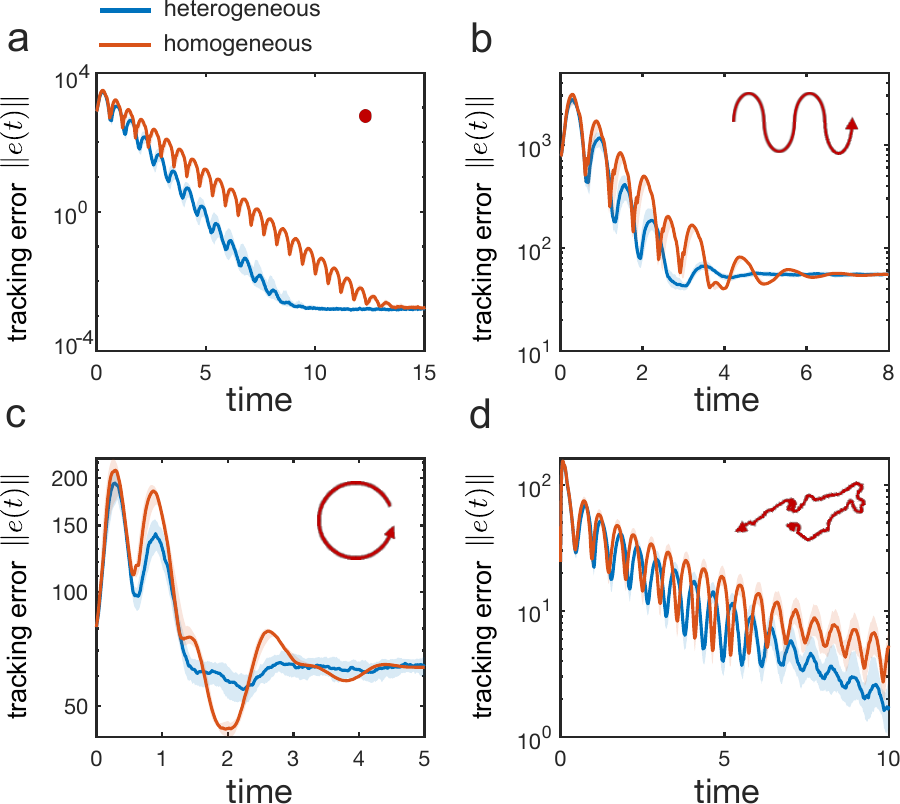}
	\caption{\textbf{Performance of the flock formation for different target trajectories.}
            Each panel shows the tracking error as a function of time for an optimal flock of heterogeneous (blue) and homogeneous (orange) agents. The target trajectories are sketched as insets:
            (\textbf{a})~stationary; (\textbf{b})~sinusoidal; (\textbf{c})~circular; and (\textbf{d})~Brownian.
            The solid lines represent the median over 100 realizations with different initial conditions, while the shaded areas indicate the first and third quartile. The other parameters are set as in Fig.~\ref{fig.converror}. See Supplementary Movie~1 for an animation of the flocking dynamics.
        }
	\label{fig.targettrack}
\end{SCfigure*}

\noindent
\textbf{Target tracking.} Following the analysis reported in Fig.~\ref{fig.converror} for a target moving with constant speed, Fig. \ref{fig.targettrack} presents a performance comparison of the convergence time of heterogeneous and homogeneous flocks for other four types of target trajectories. The initial conditions and desired flock formation of the agents are set as reported in Methods: $\bm q_i(0)\sim\mathcal U[-2,2]^2$, $\bm p_i(0) = 0$, and $\bm r_i\sim\mathcal U[-5,5]^2$ (for $N=30$ agents). The considered target trajectories are defined as follows.
\begin{enumerate}
    \item stationary target: $\bm q_{\rm t}(t) = [100, 100]$ and $\bm p_{\rm t}(t) = [0, 0]$, $\forall t\geq 0$;
    
    \item sinusoidal trajectory with constant velocity along x-axis: $\dot{\bm q}_{\rm t}(t) = [\bm p_{\rm t,1}, 10\sin(t)]$, $\dot{\bm p}_{\rm t}(t) = [0, 10\cos(t)]$, $\bm q_{\rm t}(0) = [100,100]$, and $\bm p_{\rm t}(0) = [10,10]$;
    
    \item circular trajectory: $\dot{\bm q}_{\rm t} = [- \Omega\bm q_{\rm t,2} - (\bm q_{\rm t,1}^2 + \bm q_{\rm t,2}^2 - R^2)\bm q_{\rm t,1}, + \Omega\bm q_{\rm t,1} -(\bm q_{\rm t,1}^2 + \bm q_{\rm t,2}^2 - R^2)\bm q_{\rm t,2}]$, $\dot{\bm p}_{\rm t} = \ddot{\bm q}_{\rm t}(t)$, $\bm q_{\rm t}(0) = [10,0]$, $\bm p_{\rm t}(0) = [0,10]$, and $(\Omega,R) = (2,10)$;

    \item underdamped Brownian motion: $\dot{\bm q}_{\rm t} = \bm p_{\rm t}$, $0.1\dot{\bm p}_{\rm t} = - 0.1\bm p_{\rm t} + \tilde{\nu}_i$, $\tilde{\nu}_i\sim\mathcal N(0,900)^2$, and $[\bm q_{\rm t}(0),\bm p_{\rm t}(0)] = 0$.
\end{enumerate}

\begin{figure*}[t]
	\centering
	\includegraphics[width=\textwidth]{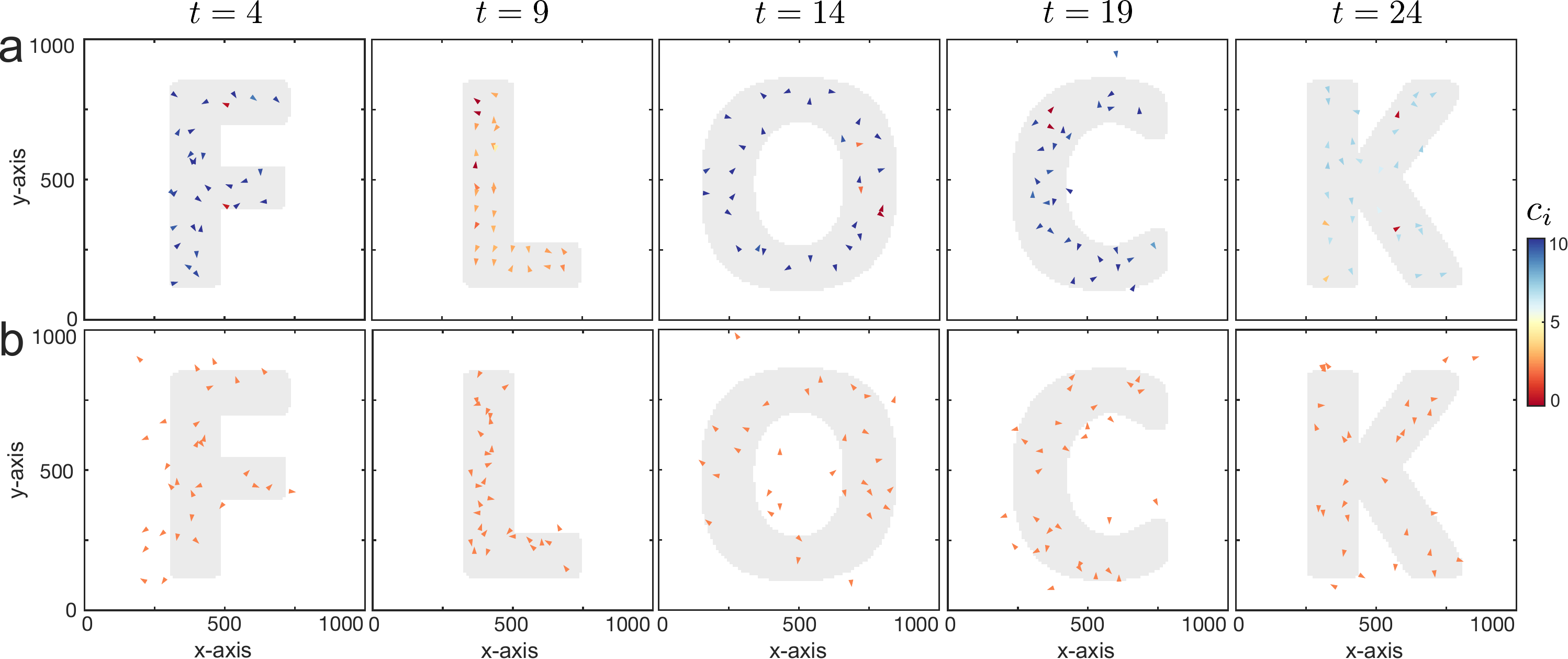}
	\caption{\textbf{Setpoint tracking for a flock animation.}
            Snapshots of the agents' position and orientation on the 2D Euclidean space for: (\textbf{a})~heterogeneous and (\textbf{b})~homogeneous flocks. The colorbar indicates the agents' velocity feedback gain. In this simulation, agents are tasked to move to pre-specified positions within the assigned letter region (highlighted in gray). The switching time between letters is 5 time units; the snapshots are shown 1 time unit before each letter switches to the next iteration. 
            For each subsequent letter, the agents' initial condition corresponds to their final state at the previous letter iteration.
            See Methods for details on the system parameters and Supplementary Movie 4 for an animation of the dynamics.
        }
	\label{fig.flock}
\end{figure*}

\begin{figure}[t!]
	\centering
	\includegraphics[width=0.77\textwidth]{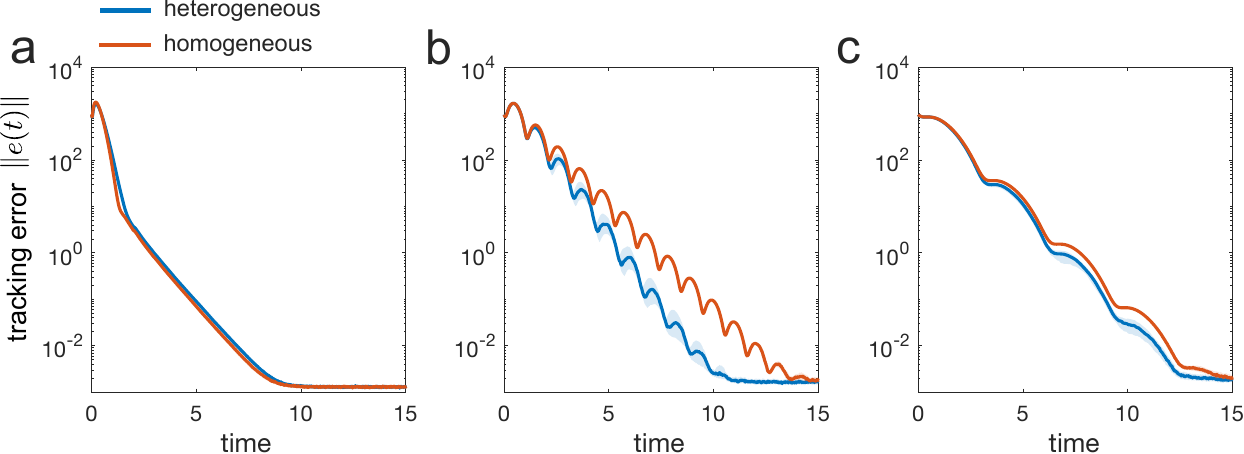}
	\caption{\rev{\textbf{Performance of the optimal flocks for different parameter constraints.}
            Tracking error as a function of time for an optimal flock of heterogeneous (blue) and homogeneous (orange) agents. The optimized parameters in each panel are: (\textbf{a}) position feedback $b_i$ (where $c_i=10$, $\forall t$), (\textbf{b}) velocity feedback $c_i$ (where $b_i=10$, $\forall t$), and (\textbf{c}) position and velocity feedback  constrained to be equal (i.e., $b_i=c_i$, $\forall i$).
            The solid lines represent the median over 100 realizations with different initial conditions, while the shaded areas indicate the first and third quartile. The other parameters are set as in Fig.~\ref{fig.converror}.
        }}
	\label{fig.sm.differentBs}
\end{figure}

\medskip\noindent
\textbf{Setpoint tracking.} Inspired by stop-motion animations performed by drones forming static shapes on the sky, we consider the problem of optimizing the convergence time of a flock of agents towards distinct spatial formations that change at regular time intervals. This problem is known in the control literature as setpoint tracking, since each agent must move towards a specified point in space.
Fig.~\ref{fig.flock} compares the performance of heterogeneous and homogeneous flocks in this task, where setpoints are programmed to display the letters F, L, O, C, and K sequentially. 
The snapshots show that the heterogeneous flock is capable of converging to the desired setpoint substantially faster than the homogeneous flock. 
The heterogeneity in the controller gains enables individual agents to simultaneously compensate for large and small positional errors relative to the desired setpoints, yielding a faster settling time. In contrast, the homogeneous flock's performance is limited by the uniform response of all agents, which can lead to suboptimal, underdamped maneuvers and slower convergence to the intended formation, as observed by the relatively large number of agents outside the boundaries of the specified letters.

\newpage\noindent
\rev{\textbf{Parameter optimization.} The optimization problem \eqref{eq.optimization} proposed in the main paper is based on simultaneously tuning the position feedback gains $b_i$ and the velocity feedback gains $c_i$. Here, we evaluate system performance when optimization is constrained to a single parameter in the following scenarios: 
\begin{enumerate}
    \item $\min_{\bm b} \Lambda_{\rm max}(J(t_k))$, s.t. $0\leq \bm b\leq b_{\rm max}$;
    \item $\min_{\bm c} \Lambda_{\rm max}(J(t_k))$, s.t. $0\leq \bm c\leq c_{\rm max}$;
    \item $\min_{\bm b} \Lambda_{\rm max}(J(t_k))$, s.t. $b_i = c_i$, $\forall i$, and $0\leq \bm b\leq b_{\rm max}$.
\end{enumerate}
\noindent
Fig.~\ref{fig.sm.differentBs} shows that, in all cases, heterogeneity improves the flock convergence time. However, this improvement is marginal when optimizing only the position feedback $\bm b$. Such result is expected given the integrator dynamics in model \eqref{eq.flockmodel}: while the feedback gain $c_i$ has a direct impact on the velocity mismatch $(\bm p_i-\bm p_{\rm t})$ of an agent, the feedback gain $b_i$ must be first integrated before correcting the position error $(\bm q_i-\bm q_{\rm t} - \bm r_i)$. As a result, changes in $\bm b$ influence the system more gradually, leading to an overdamped convergence to steady state. In contrast, optimizing solely $\bm c$ leads to a substantial improvement in the convergence time of heterogeneous flocks, although the settling times for both heterogeneous ($t_{\rm s}=8.73$) and homogeneous ($t_{\rm s}=12.16$) flocks are still slower than the values reported in the main paper when both  gains $\bm b$ and $\bm c$ are jointly optimized (respectively, $t_{\rm s} = 7.16$ and $11.62$ for the same tolerance $\epsilon = 10^{-2}$).}

\begin{SCfigure*}[\sidecaptionrelwidth][t]
	\centering
	\includegraphics[width=0.5\textwidth]{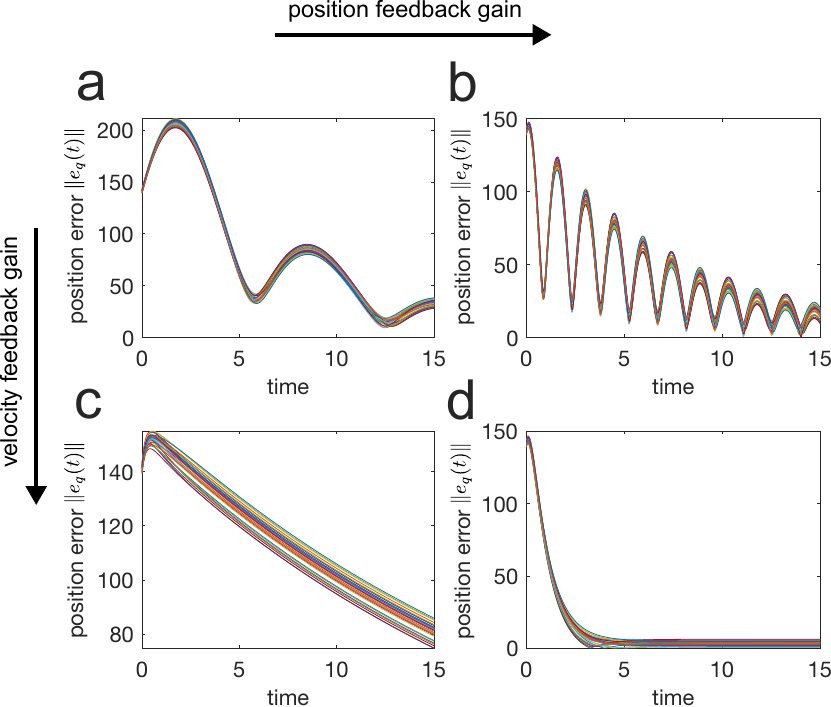}
	\caption{\rev{\textbf{Underdamped and overdamped regimes in the flocking model.}
            Position error $\norm{\bm e_q}$ as a function of time in a non-optimal homogeneous flock with constant feedback gains. Each colored line represents a different agent. The parameters are set as follows: (\textbf{a}) $b = c = 0.5$, (\textbf{b}) $b=10$ and $c=0.5$, (\textbf{c}) $b = 0.5$ and $c=10$, and (\textbf{d}) $b=c=10$. The initial conditions are the same in all panels, and the other parameters are set as in Fig.~\ref{fig.converror}.
        }}
	\label{fig.sm.damping}
\end{SCfigure*}

\rev{To further understand the influence of parameters $b_i$ and $c_i$ on the flocking dynamics, Fig.~\ref{fig.sm.damping} shows the response of a (non-optimal) flock of homogeneous agents with constant feedback gains  (i.e., $b_i=b$ and $c_i=c$, $\forall i$). The four parameter combinations illustrate that the velocity gain $c$ has a strong influence on damping the oscillations of the system. 
Particularly, when $b=0.5$ and $c=10$, the system response is completely overdamped, with no oscillations around the equilibrium but slow convergence. The best performance among the non-optimal cases is achieved for $b=c=10$, which exhibits a response close to a critically damped regime, with fast convergence and minimal oscillations. For the tolerance $\epsilon=10^{-2}$, this parameter choice yields a settling time of $t_{\rm s}=14.4$, which is  slower than the convergence of the optimal flocks. (Note that Fig.~\ref{fig.sm.damping} shows the \textit{position} error $\norm{\bm e_{q,i}}$ of each agent, whereas Fig.~\ref{fig.converror} shows the \textit{tracking} error $\norm{\bm e}$ of all agents.) Indeed, as we explore throughout the paper, the optimal response\textemdash which minimizes both the convergence time and the oscillatory behavior of a flock towards a desired formation\textemdash is given by a heterogeneous combination of parameters.} 

\section{\rev{Optimal parameter set, flock formation, and network symmetries}}
\label{sec.sm.symmetries}

\begin{figure}
	\centering
	\includegraphics[width=0.83\textwidth]{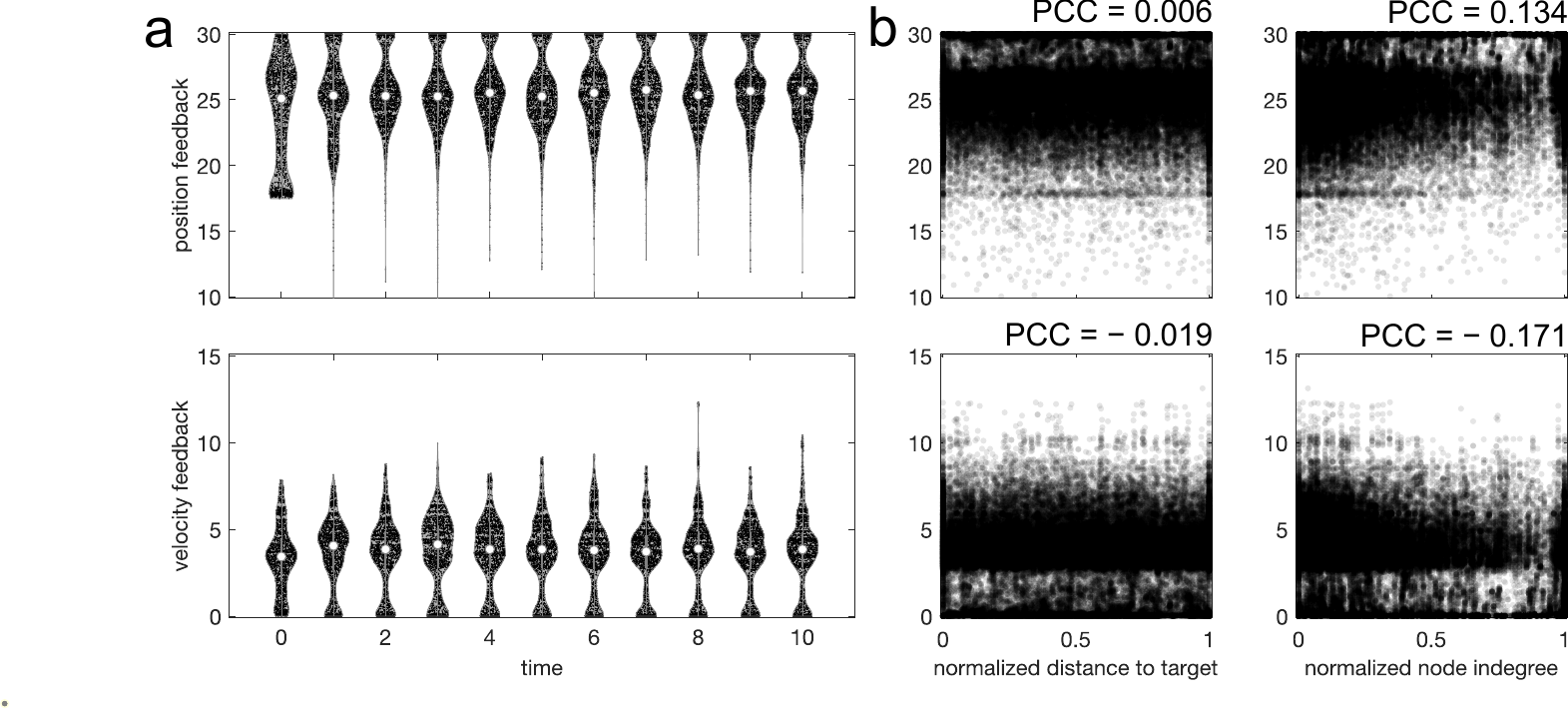}\vspace{-0.3cm}
	\caption{\rev{\textbf{Relationship between optimal feedback gains and flock formation.}
            (\textbf{a}) Violin plots showing the distribution of optimal gains $b_i$ (top) and $c_i$ (bottom) as functions of time. 
            (\textbf{b}) Scatter plots of the  optimal gains $b_i$ (top row) and $c_i$ (bottom row) versus the corresponding agent's normalized distance to target $\bm e_{q,i}$ (left column) and node in-degree $\sum_j A_{ij}$ (right column). The corresponding Pearson correlation coefficient (PCC) is reported on top of each panel.
            In all panels, the results are shown for $N=30$ agents across 100 realizations with random initial conditions, in which parameters are optimized every $w = T = 1$ time units over $t\in[0,10]$.
            Each data point represents the corresponding gain of an agent  at a particular time instant.
            The other parameters are set as in Fig.~\ref{fig.converror}.
        }}
	\label{fig.sm.paramdistrib}
\end{figure}

\rev{
As illustrated in Fig. \ref{fig.realtimeopt}c for a representative flock simulation, the assigned optimal gains exhibit significant heterogeneity both across agents and over time. This heterogeneity depends on the network structure, which in turn is defined by the time-varying flock formation. Thus, to better understand this behavior, we examine the relationship between the optimal gains and the flock formation at each time instant by using statistical results comprising 100 independent realizations. Fig.~\ref{fig.sm.paramdistrib}a shows the probability distribution functions (PDFs) of optimal gains over time, revealing broad distributions with high variability for both types of feedback gains. However, as the flock formation evolves over time, we also observe that the PDFs become increasingly consistent, suggesting an emergent regularity in the optimization process as the agents settle converge towards the desired formation. Despite this trend, Fig.~\ref{fig.sm.paramdistrib}b indicates that there is no systematic correlation between an agent's optimal gain and its structural properties within the flock (specifically, distance to target or node in-degree).
}

\begin{figure}
    \centering
    \includegraphics[width=0.5\linewidth]{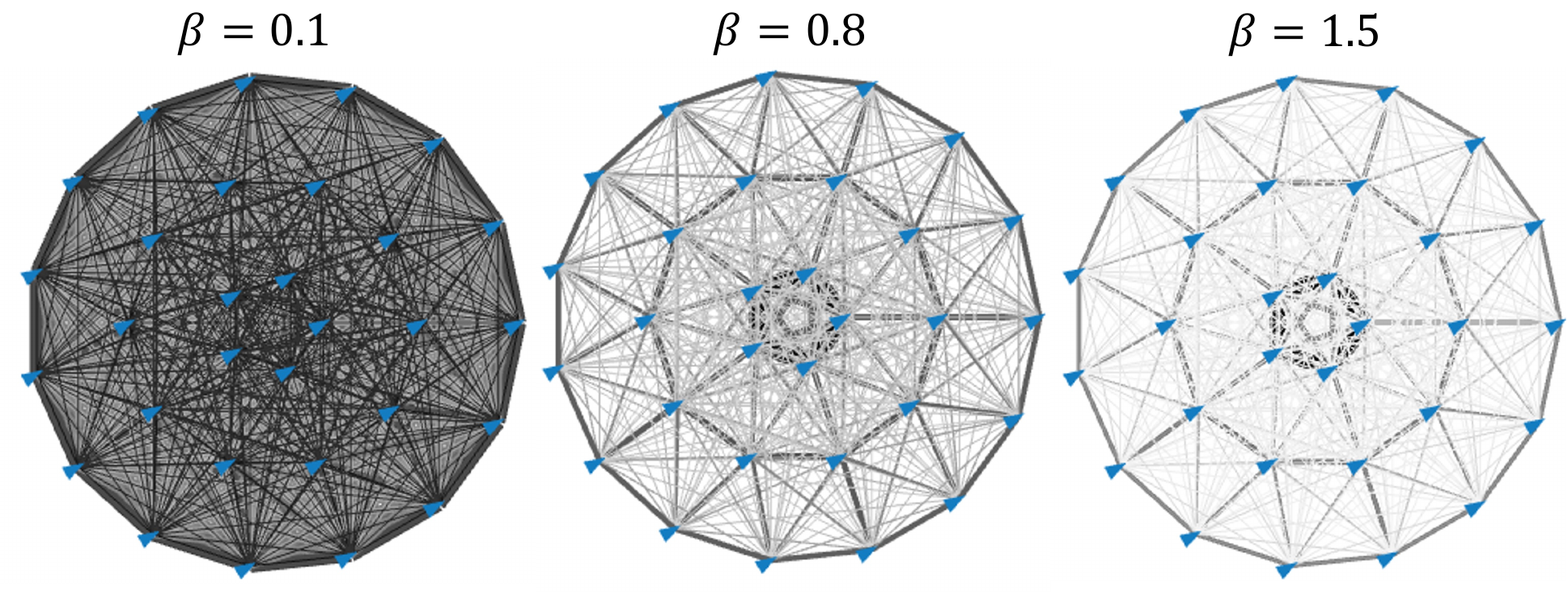}
    \caption{\rev{\textbf{Network structures for varying interaction ranges $\beta$.} The networks depict $N=30$ agents arranged in concentric circular formation (Methods), where the adjacency matrix parameters are $\sigma = 0.1$ and $K=2$. The edge thickness is proportional to $A_{ij}$.}}
    \label{fig.sm.beta}
\end{figure}

\begin{figure}
	\centering
	\includegraphics[width=0.6\textwidth]{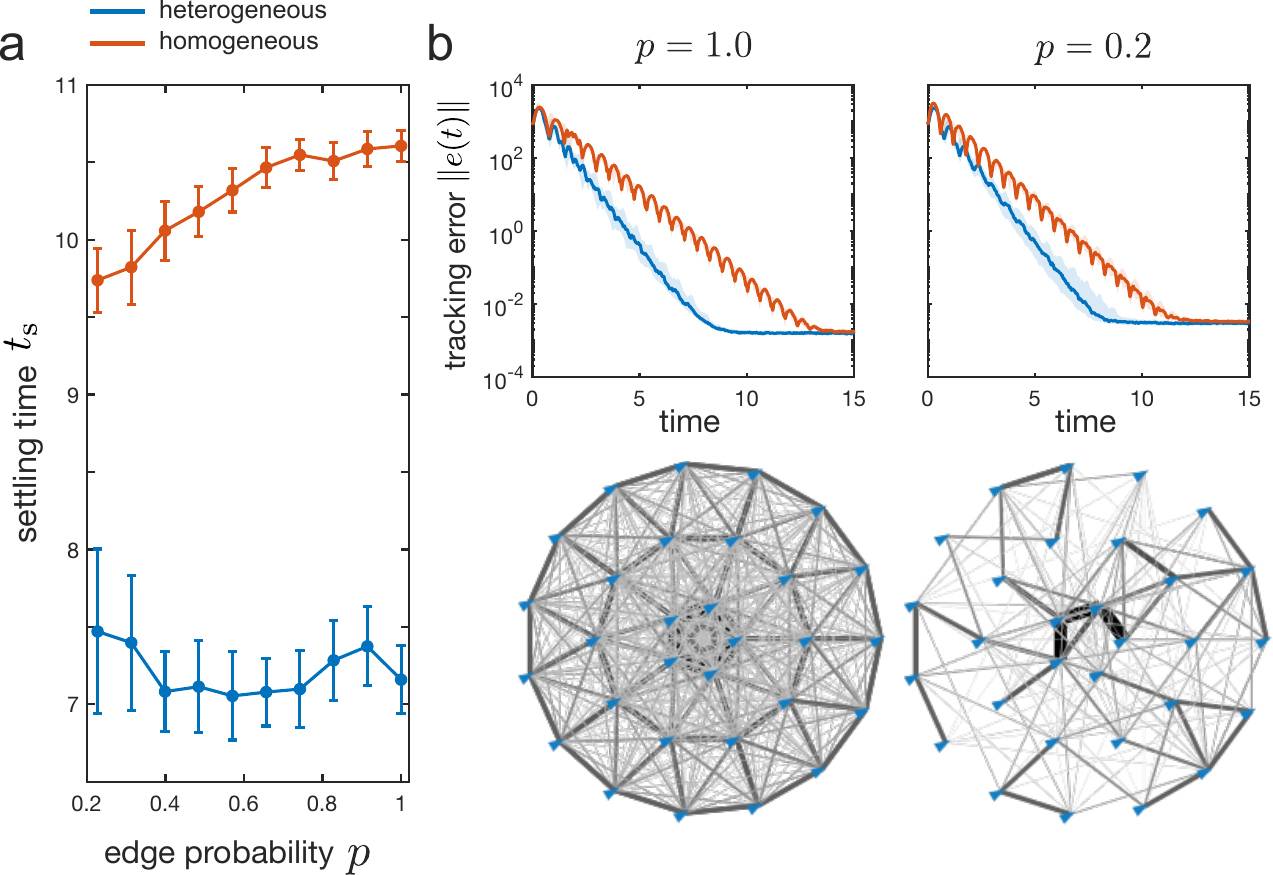}
	\caption{\rev{\textbf{Flock convergence for directed networks with varying levels of connectivity.}
		(\textbf{a})~Settling time $t_{\rm s}$ as a function of the edge probability $p$ for heterogeneous (blue) and homogeneous (orange) flocks. In this network model, each agent pair $(i,j)$ has a directed weighted edge $A_{ij} =  K(\rho^2 + \norm{\bm q_i(t) - \bm q_j(t)}^2)^{-\beta}$ from $j$ to $i$ with probability $p$, otherwise $A_{ij} = 0$. Each data point represents an average over 100 independent realizations, with the error bars indicating one standard deviation. 
		(\textbf{b})~Tracking error $\norm{\bm e(t)}$ over time (top) and representative  networks (bottom) for $p = 1$ (left) and $p = 0.2$ (right).
            In the top panels, the solid lines represent the median over 100 realizations and the shaded areas indicate the first and third quartile.
            In the bottom panels, the directionality, which is random, is omitted to facilitate visualization.
            The other parameters are set as in Fig.~\ref{fig.converror}.}
        }
	\label{fig.sm.directned}
\end{figure}

\rev{
We thus investigate the potential impact of the underlying network structure on the convergence rate of flocking dynamics. For instance, in Fig.~\ref{fig.converror}d, we showed that the performance of heterogeneous flocks strongly depends on the interaction range $\beta$, exhibiting a faster convergence at $\beta\approx 0.8$. Fig.~\ref{fig.sm.beta} further illustrates that an interaction network with $\beta = 0.8$ is more heterogeneous compared to the structures obtained with extreme values of $\beta$ (all-to-all networks for $\beta\rightarrow 0$ and disconnected networks for $\beta\rightarrow\infty$).
This result demonstrates that increased network heterogeneity can facilitate flocking among heterogeneous agents.
To further evaluate the role of sparsity in optimal flocking, we consider \textit{directed} networks with varying levels of connectivity. Specifically, we generate random graphs by assigning an edge probability $p$ such that agent $i$ interacts with agent $j$ (i.e., $A_{ij}\neq 0$). Fig.~\ref{fig.sm.directned} shows the performance of optimal flocks across different connectivity levels. While the settling time of the heterogeneous flocks is not strongly impacted by $p$, the settling time of homogeneous flocks tends to decrease for smaller $p$. This result suggests that heterogeneous systems may possess an intrinsic ability to compensate for reduced connectivity, maintaining stable flocking behavior even in sparser, highly directed networks.}

\begin{SCfigure*}[\sidecaptionrelwidth][t]
	\centering
\includegraphics[width=0.59\textwidth]{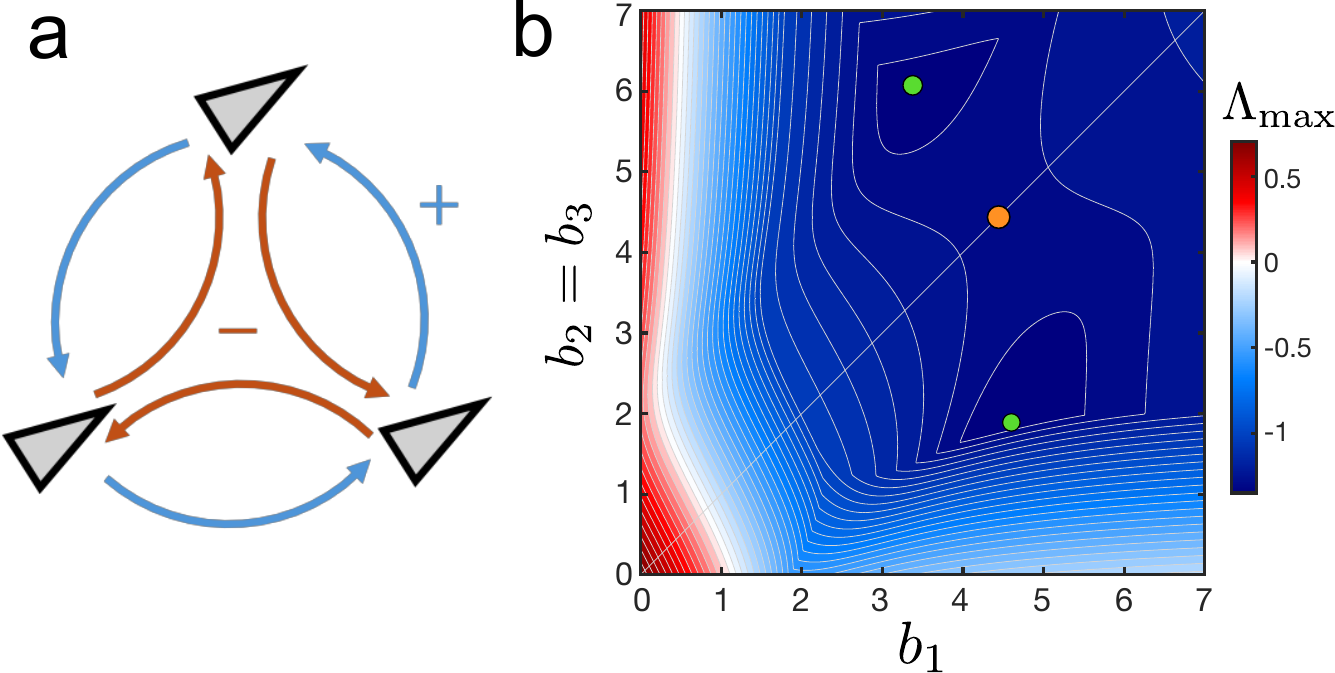}
	\caption{\rev{\textbf{Network symmetries and parameter asymmetries.}
            (\textbf{a}) Symmetric network, where blue and orange edges denote positive and negative interactions, respectively. 
            (\textbf{b}) Lyapunov exponent $\Lambda_{\rm max}$ as a function of the feedback gains $b_1$ and $b_2=b_3$. For this section of the parameter space, the optimal homogeneous gain $\bm b^*_{\rm hom}$ is indicated by the orange dot (along the diagonal line) while the  optimal heterogeneous gains $\bm b^*_{\rm het}$ are indicated by the green dots. The flock formation is stable (unstable) for $\Lambda_{\rm max}<0$ ($\Lambda_{\rm max}>0$).
        }}
	\label{fig.sm.symmetry}
\end{SCfigure*}

\rev{
Given the interplay between the flock's convergence rate (characterized by $\Lambda_{\rm max}$), the feedback matrices $B$ and $C$, and the Laplacian matrix $L$, one might expect that the degree of heterogeneity across $b_i$ and $c_i$ could depend on asymmetries in the network structure. However, even in an entirely symmetric network, the optimal gains can still be asymmetric \cite{nishikawa2016symmetric}. To show this, we consider the simplified scenario in which the multi-agent model \eqref{eq.flockmodel} consists of $N=3$ agents, the position gains and velocity gains are equal for each agent (i.e., $b_i = c_i$, $\forall i$), and the network structure is defined by the time-independent adjacency matrix:
\begin{equation}
    A = \begin{bmatrix}
        0 & 1 & -1 \\ 
        -1 & 0 & 1 \\
        1 & -1 & 0
    \end{bmatrix}.
\end{equation}
\noindent
The corresponding Jacobian matrix is given by Eq.~\eqref{eq.constrainedjacobian}.
Fig.~\ref{fig.sm.symmetry}a illustrates the underlying network structure, showing that any permutation of nodes preserves its topology. Nevertheless, the stability landscape depicted in Fig.~\ref{fig.sm.symmetry}b reveals that the optimal parameter set $\bm b^* = [b_1,b_2,b_3]$, which minimizes $\Lambda_{\rm max}$, does not lie within the homogeneous (symmetric) space where $b_1=b_2=b_3$; instead, the optimal solution is heterogeneous (asymmetric), with $b_1 \neq b_2$. (The 2D visualization in Fig.~\ref{fig.sm.symmetry}b effectively illustrates this asymmetry, but we note that the global optimal parameters are specifically given by $\bm b^* = [3.747,1.979,5.515]$.) These results suggest that the optimal gains cannot be trivially inferred directly from the flock geometry or network symmetries alone. A full analysis of the spectral properties of the Laplacian matrix $L$ and Jacobian matrix $J$ is required. In
Section \ref{sec.optimalhom}, we provide such analytical characterization of the relationship between the optimal gains and the eigenvalues of $L$ for homogeneous systems.
}

\section{Optimal feedback gain for homogeneous flocks}
\label{sec.optimalhom}

Consider the Jacobian matrix $J(t)$ presented in Eq.~\eqref{eq.constrainedjacobian}. We analytically determine the optimal feedback gain for a flock of \textit{homogeneous} agents that minimizes the largest Lyapunov exponent $\Lambda_{\rm max} (J(t))$. Given that $c_i = c$ and $b_i = b$, $\forall i$, the Jacobian matrix reduces to
\begin{equation}
    J = \begin{bmatrix}
        0_{Nm} & I_{Nm} \\
        -(bI + L) \otimes I_m & - \gamma (cI+L) \otimes I_m
    \end{bmatrix}.
\end{equation}

\noindent
We consider the problem of minimizing $\Lambda_{\rm max}(J)$ for two scenarios: 1) the special case $b=c$ and 2) the general case $b\neq c$.

\medskip\noindent
\textbf{Special case ($b = c$).} 
Let $\ell_1 < \ldots < \ell_N$ be the eigenvalues of the Laplacian matrix $L$. Following Ref.~\cite[Section 4.1.1]{ren2008distributed}, the eigenvalues of the Jacobian matrix  can be expressed as
\begin{equation}
\lambda_{i\pm}= -\frac{\gamma(b + \ell_i)}{2}\pm\frac{1}{2}\sqrt{\gamma^2(b + \ell_i)^2 - 4(b + \ell_i)}, \quad \text{for} \,\, i=1,\ldots,N.
\label{eq.eigenvaluesofJ_hom}
\end{equation}

\noindent
Note that $\Lambda_{\rm max}(J) = \max_i\Re\{\lambda_{i\pm}\}$ depends non-trivially on the feedback gain $b$ and the eigenvalues of the Laplacian matrix.
However, we can determine how $\lambda_{i+}$ changes as a function of $b$ by evaluating the derivative
\begin{equation}
\frac{{\rm d} \lambda_{i+}}{{\rm d}b}= 
\begin{cases}
    -\frac{\gamma}{2} \qquad &\mbox{if} \quad b^{(i)} < \frac{4}{\gamma^2}-\ell_i, \\ \\
		  -\frac{\gamma}{2}+\frac{1}{2}\frac{\gamma^2(b + \ell_i) -2}{\sqrt{\gamma^2(b + \ell_i)^2 - 4(b + \ell_i)}} \qquad &\mbox{if} \quad b^{(i)} > \frac{4}{\gamma^2}-\ell_i .\\
\end{cases}
\end{equation}

\noindent
Given that $\frac{{\rm d} \lambda_{i+}}{{\rm d}b}>0$ if $\gamma^2(b + \ell_i)^2 - 4(b + \ell_i) > 0$, it follows that $\gamma^2(b + \ell_i)^2 - 4(b + \ell_i) = 0$ is a point of non-differentiability where the derivative changes signal. Thus, $b^{(i)} = \frac{4}{\gamma^2}-\ell_i$ is the parameter that minimizes $\lambda_{i+}$. 
At this local minimum, $b^{(i)}$ decreases as a function of $\ell_i$ and $\frac{{\rm d} \lambda_{i+}}{{\rm d}b}>\frac{{\rm d} \lambda_{j+}}{{\rm d}b}$ if $i>j$. Thus, the optimal solution $b_{\rm hom}^* = \argmin_b \Lambda_{\rm max}(J(b))$ is obtained when $\lambda_{1+} = \lambda_{N+}$, leading to
\begin{align}
b^*_{\rm hom} &= \frac{2}{\gamma^2} - \ell_N + \sqrt{(\ell_N-\ell_1)^2+ \frac{4}{\gamma^4}},
\label{eq.hom.obtb}
\\
\Lambda_{\rm max}(b_{\rm hom}^*) &= -\frac{1}{\gamma} + \frac{\gamma \ell_N}{2} - \frac{\gamma}{2} \sqrt{(\ell_N-\ell_1)^2+ \frac{4}{\gamma^4}}.
\label{eq.hom.optlambda}
\end{align}

In Section \ref{sec.stabilitylandscape}, we provide a detailed analysis of the stability landscape (determined by $\Lambda_{\rm max}$) around the homogeneous solution \eqref{eq.hom.obtb}. This allows us to establish the conditions for which a  descending path exists from the homogeneous optimum $\bm b_{\rm hom}$ to some heterogeneous optimum $\bm b_{\rm het}$

\medskip\noindent
\textbf{General case ($b\neq c$).}
As in Eq.~\eqref{eq.eigenvaluesofJ_hom}, the eigenvalues of $J$ can be expressed as
\begin{equation}
\lambda_{i\pm}= -\frac{\gamma(c + \ell_i)}{2}\pm\frac{1}{2}\sqrt{\gamma^2(c + \ell_i)^2 - 4(b + \ell_i)}.
\end{equation}

\noindent
Here, $\gamma^2(c + \ell_i)^2 - 4(b + \ell_i) = 0$ determines a non-differentiable point where the derivative $\frac{{\rm d} \lambda_{i+}}{{\rm d}c}$ changes signal and hence $c^{(i)} = \frac{2\sqrt{b+\ell_i}}{\gamma}-\ell_i$ is the point that minimizes $\lambda_{i+}$. However, unlike Eq.~\eqref{eq.hom.obtb}, the optimal velocity feedback gain $c_{\rm hom}^* = \argmin \Lambda_{\rm max}(J)$ cannot be directly determined in terms of the smallest and largest eigenvalues of the Laplacian matrix since its solution depends nontrivially on the interplay between all eigenvalues $\ell_i$ and the position feedback gain $b$.

\section{Stability analysis of delay differential equations}
\label{sec.dde}
The DDE system~\eqref{eq.timedelayconsensus} can be expressed in matrix form as
\begin{equation}
    \dot{\bm x}(t) = (L_1 \otimes I_m)\bm x(t) + (L_2\otimes I_m)\bm x(t-\tau),
\label{eq.timedelaymatrix}
\end{equation}
\noindent
where
\begin{equation}
L_1 = \begin{bmatrix}
        0_N & I_N \\
        0_N & 0_N
    \end{bmatrix},
    \,\,
    L_2 = \begin{bmatrix}
        0_N & 0_N \\
        -KL & -KL
    \end{bmatrix},
    \,\,
    K = \operatorname{diag}(k_1,\ldots,k_N).
\end{equation}
\noindent
Eq.~\eqref{eq.timedelaymatrix} has the following characteristic equation \cite{yu2010some}:
\begin{equation}
    \det(\lambda I_{2N} - L_1 - e^{-\lambda \tau} L_2) = 0.
\label{eq.characteristiceq}
\end{equation}
\noindent
The ``eigenvalues'' (roots) $\lambda$ of Eq.~\eqref{eq.characteristiceq} determine the stability of the linear DDE system \cite{bellen2013numerical}: the equilibrium point $\bm x^* = 0$ is asymptotically stable if and only if every eigenvalue has negative real part (i.e., $\Re\{\lambda\}<0$).
Note that the number of eigenvalues is infinite for DDEs, although the system stability can be directly characterized by the largest Lyapunov exponent $\Lambda_{\rm max} = \max_\lambda \Re{\lambda}$.
Assuming a homogeneous gain $k_i = \bar k$, $\forall i$, Ref.~\cite[Theorem 2]{yu2010some} shows that consensus can be achieved if and only if
\begin{equation}
    \tau < \tau_0 := \min_{2\leq i \leq N} \frac{\theta_i}{\omega_i} ,
\label{eq.tau0bound}
\end{equation}
\noindent
where $\omega_i= \sqrt{ \frac{1}{2}(\norm{\ell_i}^2 \bar k^2 + \norm{\ell_i} \bar k \sqrt{ \norm{\ell_i}^2 \bar k^2 + 4})}$ and the angle $0\leq \theta_i\leq 2\pi$ satisfies the equalities $\cos\theta_i = \frac{\bar k}{\omega_i^2}(\Re\{\ell_i\} - \Im\{\ell_i\}\omega_i)$ and $\sin\theta_i = \frac{\bar k}{\omega_i^2}(\Re\{\ell_i\}\omega_i + \Im\{\ell_i\})$. The nonzero eigenvalues of the Laplacian matrix $L$ are defined as $\ell_i$, for $i=2,\ldots,N$.

For the simulations in Fig. \ref{fig.timedelay}, we use the same system considered in Ref. \cite[Section 5.2]{yu2010some}, composed by $N=4$ agents interacting according to the Laplacian matrix
\begin{equation}
    L = 
    \begin{bmatrix}
        1 & 0 & -1 & 0 \\
        -1 & 1 & 0 & 0 \\
        0 & -1 & 1 & 0 \\
        -1 & 0 & 0 & 1
    \end{bmatrix}.
\end{equation}

\begin{figure}[t!]
    \centering
    \includegraphics[width=0.32\textwidth]{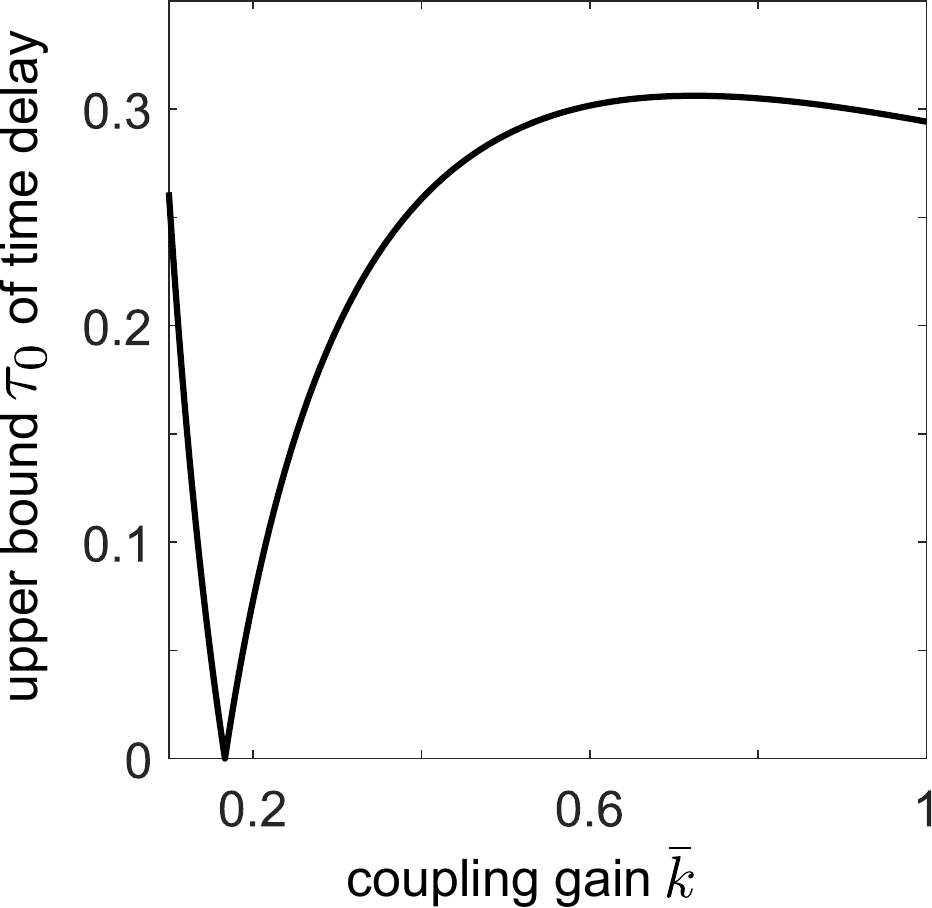}
    \caption{\textbf{Upper bound $\tau_0$ of the time delay as a function of the coupling gain $\bar k$.} This upper bound represents the maximum communication delay $\tau$ for which the consensus model \eqref{eq.timedelayconsensus} is asymptotically stable, given that a homogeneous gain $\bar k$ is applied uniformly across all agents ($k_i = \bar k$, $\forall i$).}
    \label{fig.tau0}
    \vspace{-0.3cm}
\end{figure}

\noindent
Fig. \ref{fig.tau0} depicts the upper bound $\tau_0$ as a function of $\bar k$ for this system, showing that $\tau_0$ reaches a maximum of $\tau_0^*=0.306$ at $\bar k^* = 0.724$.

To numerically compute the eigenvalues of Eq.~\eqref{eq.characteristiceq}, we use the function \texttt{ddebiftool\_stst\_stabil} available as part of the DDE-BIFTOOL toolbox for MATLAB \cite{engelborghs2002numerical,sieber2014dde}.
For each choice of time delay $\tau$ (in Fig.~\ref{fig.timedelay}), we determine the best homogeneous and heterogeneous choice of parameters $k_i$ that solves the optimization problem:
\begin{equation}
    \begin{aligned}
        \min_{k_i, \, i=1,\ldots,N} \quad & \Lambda_{\rm max}, \\
        \text{s.t.} \quad & 0< k_i \leq k_{\rm max}.
    \end{aligned}
    \label{eq.optimizationtimedelay}
\end{equation}
%
%
We solve Eq.~\eqref{eq.optimizationtimedelay} for $k_{\rm max} = 1$ employing the interior-point method \cite{nocedal1999numerical}, implemented by the MATLAB function \texttt{fmincon}.

\section{Stability analysis of the free-flocking model}
\label{sec.stabilityfreeflock}

We derive the Lyapunov function used to characterize the convergence rate of the flocking model \eqref{eq.osflock} to a steady-state formation (i.e., the lattice formation).

Consider the position $\bm q_{\rm c} = \frac{1}{N}\sum_{i}\bm q_i$ and momentum $\bm p_{\rm c} = \frac{1}{N}\sum_{i}\bm p_i$ of the center of mass of all agents. The change of coordinates $\bm e_i = [\bm e_{q,i}, \,\, \bm e_{p,i}] = [\bm q_i - \bm q_c, \,\, \bm p_i - \bm p_c]$ defines a moving frame centered at this center of mass in which relative positions among agents are invariant: $V(\bm q) = V(\bm e_q)$, $\gradient_{\bm q}V(\bm q) = \gradient_{\bm e_q} V(\bm e_q)$, and $A(\bm q) = A(\bm e_q)$. Let $\bm e_q = [\bm e_{q,1},\ldots,\bm e_{q,N}]$ and $\bm e_p = [\bm e_{p,1},\ldots,\bm e_{p,N}]$. 
As in Ref.~\cite[Lemma 2]{olfati2006flocking}, the dynamics of the flocking model \eqref{eq.osflock} can be decomposed into the set of $2Nm$ equations describing the structural dynamics (i.e., the relative motion of agents in the reference frame moving with the center of mass):
\begin{equation}
    \begin{bmatrix}
        \dot{\bm e}_{q} \\ \dot{\bm e}_{p}
    \end{bmatrix}
    =
    \underbrace{
    \begin{bmatrix}
        0_{Nm} & I_{Nm} \\
        -B \otimes  I_m & -(C + k_2^\alpha L(\bm e_q)) \otimes I_m 
    \end{bmatrix}
    }_{J(\bm e_q)}
    \underbrace{
    \begin{bmatrix}
        {\bm e}_{q} \\ {\bm e}_{p}
    \end{bmatrix}
    }_{\bm e}
    +
    \begin{bmatrix}
        0 \\ -k_1^\alpha \gradient V(\bm e_q)
    \end{bmatrix},
\label{eq.structuraldyn}
\end{equation}

\noindent
and the set of $2m$ equations describing the translational dynamics (i.e., the motion of the center of mass):
\begin{equation}
    \begin{bmatrix}
        \dot{\bm q}_{\rm c} \\ \dot{\bm p}_{\rm c}
    \end{bmatrix}
    =
    \begin{bmatrix}
        0_{m} & I_{m} \\
        - b_{\rm avg}\otimes I_m & - c_{\rm avg}\otimes I_m
    \end{bmatrix}
    \begin{bmatrix}
        {\bm q}_{\rm c} \\ {\bm p}_{\rm c}
    \end{bmatrix}
    +
    \begin{bmatrix}
        0_m & 0_m \\  b_{\rm avg}\otimes I_m &  c_{\rm avg}\otimes I_m
    \end{bmatrix}
    \begin{bmatrix}
        \bm q_{\rm t} \\ \bm p_{\rm t}
    \end{bmatrix}
    ,
\label{eq.translationaldyn}
\end{equation}

\noindent
where $L(\bm e_q)$ is the Laplacian matrix associated with the adjacency matrix $A(\bm e_q)$, $B = \operatorname{diag}(b_1,\ldots,b_N)$, $C = \operatorname{diag}(c_1,\ldots,c_N)$, $b_{\rm avg} = \frac{1}{N}\sum_i b_i$, and $c_{\rm avg} = \frac{1}{N}\sum_i c_i$. 

We focus on the stability analysis of achieving  a lattice formation among agents, which is given by Eq.~\eqref{eq.structuraldyn}. To this end, we propose the following Hamiltonian function
\begin{equation}
    H(\bm e) = k_1^\alpha V(\bm e_q) + \frac{1}{2}\bar{\bm e}^\transp\bar{\bm e},
\end{equation}

\noindent
where $\bar{\bm e} = \bm e - \bm e^*$ and $\bm e^*$ is an equilibrium point of system \eqref{eq.structuraldyn}.
Note that, if $\bm e_q^*$ is a local minimum of $V(\bm e_q)$, then $\bm e^*= [\bm e_q^*, 0]$ is an equilibrium point; moreover, every local minimum of $V(\bm e_q)$ corresponds to a lattice configuration among agents \cite[Lemma 3]{olfati2006flocking}. 
Likewise, if $\bm e_q^{*}$ is a local maximum or a saddle point of $V(\bm e_q)$, then $\bm e^{*} = [\bm e_q^{*}, 0]$  is also an equilibrium point. Let $\mathcal D_1 = \{\bm e_q^{*} : \gradient V(\bm e_q^{*}) = 0$ and $\operatorname{Hessian}(V(\bm e_q^{*}))$ has all non-negative eigenvalue$\}$ denote the set of local minima of $V(\bm e_q)$ and $\mathcal D_2 = \{\bm e_q^{*} : \gradient V(\bm e_q^{*}) = 0$ and $\operatorname{Hessian}(V(\bm e_q^{*}))$ has at least one negative eigenvalue$\}$ denote the set of local maxima and saddle points of $V(\bm e_q)$. The set of equilibrium points is thus given by $\mathcal D_1\cup \mathcal D_2$. In what follows, we assume that $\bm e_q^*\in\mathcal D_1$ given that the equilibria in $\mathcal D_2$ are unstable. 
Thus, the following statements hold:
\begin{enumerate}
    \item[i)]  $H(\bm e^*) = 0$ and $H(\bm e) > 0$, $\forall \bar{\bm e}\neq 0$;

    \item[ii)] $\alpha_1\norm{\bar{\bm e}}^2 \leq H(\bm e) \leq \alpha_2\norm{\bar{\bm e}}^2$, for some $\alpha_1,\alpha_2 > 0$.
\end{enumerate}

The derivative of the Hamiltonian function is given by
\begin{equation}
\begin{aligned}
    \dot{H}(\bm e) &= k_1^\alpha \gradient V^\transp \bm e_p + \frac{1}{2} \bar{\bm e}^\transp\left(J(\bm e_q) + J(\bm e_q)^\transp\right)\bar{\bm e} 
    - \frac{k_1^\alpha }{2}\left(\gradient V^\transp\bm e_p + \bm e_p^\transp\gradient V\right) \\
    &= \bar{\bm e}^\transp J(\bm e_q) \bar{\bm e}.
\end{aligned}
\label{eq.lyapunovfunction}
\end{equation} 

\noindent
Note that $J(\bm e_q)$ is a state-dependent matrix, which, for every state $\bm e_q \in\R^{Nm}$, is negative definite. Therefore, it holds that
\begin{enumerate}
    \item[iii)] $\dot{H}(\bm e)< 0$, $\forall \bar{\bm e}\neq 0$.
\end{enumerate}

\noindent
Following Ref.~\cite[Theorem 4.1]{Khalil2002}, statements i) and iii) imply that $H(\bm e)$ is a Lyapunov function and hence the equilibrium point $\bm e^*$ is asymptotically stable. Moreover, for any initial position $\bm e_q(0)$ (except critical points $\bm e_q^{*}\in\mathcal D_2$) and momentum $\bm e_p(0)$, the flock of agents converges asymptotically to a lattice configuration $\bm e_q^*$ and moves asymptotically with the same momentum $\bm p_i(t) = \bm p_c(t)$, $\forall i$ \cite[Theorem 2]{olfati2006flocking}.

Under the assumption that the timescale associated with changes in $J(\bm e_q)$ is much slower than the timescale of the state dynamics $\bm e(t)$, we approximate $J(\bm e_q(t))$ by a piecewise-constant function, yielding $J(\bm e_q(t))\approx J(t_k)$ within each time interval $t\in[t_k,t_k+T]$ and some $T>0$, as defined in Eq.~\eqref{eq.osjacobian}. 
From Eq. \eqref{eq.lyapunovfunction}, the following relation holds:
\begin{enumerate}
    \item[iv)] $\dot H(\bm e,t) = \bar{\bm e}^\transp J(t_k) \bar{\bm e} = \bar{\bm e}^\transp P^{-1}_k \Sigma_k P_k  \bar{\bm e} \leq \eta_k \Lambda_{\rm max}(J(t_k)) 
    \norm{\bar{\bm e}}^2$, for an interval $t\in[t_k,t_k+T]$, where $\eta_k = \norm{P_k}\norm{P_k^{-1}}$.
\end{enumerate}

\noindent
Here, we have used the fact that $J(t_k)$ is diagonalizable, i.e., $J(t_k) = P_k^{-1} \Sigma_k P_k$, $\Sigma_k = \operatorname{diag}(\lambda_1,\ldots,\lambda_N)$, and $\lambda_i$ are eigenvalues of $J(t_k)$, for $i=1,\ldots,n$. Recall that $\Lambda_{\rm max}(J(t_k)) = \max_i \Re\{\lambda_i\}$.

\begin{figure}[t]
    \centering
    \includegraphics[width=0.7\textwidth]{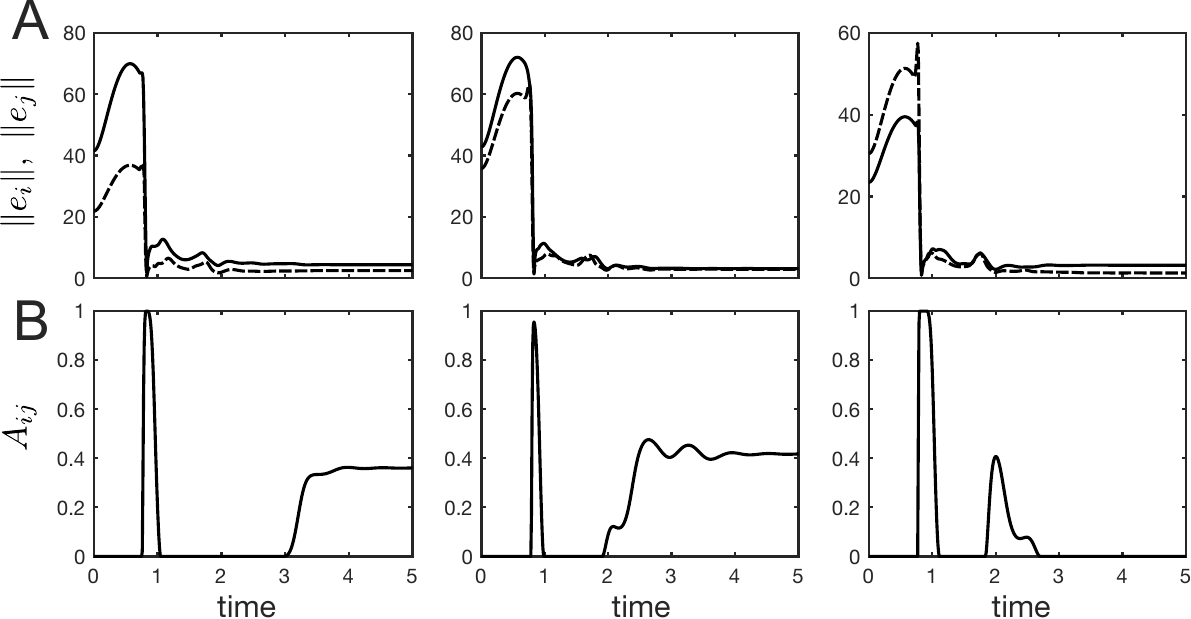}
    \caption{\textbf{Timescale separation between the agent dynamics and the time-varying structure of the communication network.} 
    (\textbf{a}) Euclidean norm of the relative state of two agents $i$ (solid line) and $j$ (dashed line) with respect to the flock's center of mass as a function of time.
    (\textbf{b}) Coupling term between agents $(i,j)$ in the adjacency matrix $A(\bm q)$.
    The simulations are shown for three representative pairs of agents using the same parameters as in Fig.~\ref{fig.freeflock}.
    }
    \label{fig.timescale}
\end{figure}

Statements ii) and iv) together imply that the trajectories of the agents converge exponentially and are upper bounded according to \cite[Theorem 4.10]{Khalil2002}:
\begin{equation}
    \norm{\bar{\bm e}(t)} \leq \eta \exp\left\{\frac{\eta_k}{2\alpha_2}\Lambda_{\rm max}(J(t_k)) T\right\}\norm{\bar{\bm e}(t_k)},
\label{eq.SMupperboundosmodel}
\end{equation}

\noindent
for the time interval $t\in[t_k,t_k+T]$ and some constant $\eta=\sqrt{\frac{\alpha_2}{\alpha_1}}>0$. This upper bound shows that minimizing $\Lambda_{\rm max}$ at every interval $[t_k,t_k+T]$ can increase the convergence rate of a flock of agents toward its equilibrium $\bm e^*$.
Considering the simulation parameters reported in Methods, Fig.~\ref{fig.timescale} shows the temporal evolution of the states of three representative pairs of agents $i$ and $j$, along with the evolution of the corresponding coupling term $A_{ij}(\bm e_q)$. It is evident that the timescale of the system state $\bm e$ is substantially faster (stiffer) compared to that of the entries of the adjacency matrix $A(\bm e_q)$ and, therefore, of the time-varying matrix $J(\bm e_q)$. This analysis suggests that the piecewise-constant assumption is appropriate for the estimation of the upper bound \eqref{eq.SMupperboundosmodel} (for a suitable choice of $T$).

\medskip\noindent
\textbf{Stability analysis with obstacles.}
For flocking applications involving obstacles, an upper bound to the convergence time can also be derived following similar steps to the derivation above. Recall that $\bm u_i^\beta$ is defined by Eq.~\eqref{eq.ubeta}. 
The flocking model \eqref{eq.osflock} can thus be decomposed as
\begin{equation}
    \begin{bmatrix}
        \dot{\bm e}_{q} \\ \dot{\bm e}_{p}
    \end{bmatrix}
    =
    \underbrace{
    \begin{bmatrix}
        0_{Nm} & I_{Nm} \\
        - B \otimes I_m & - (C + k_2^\alpha L(\bm e_q) + k_2^\beta L_\beta(\bm e_q))\otimes I_m 
    \end{bmatrix}
    }_{J'(\bm e_q)}
    \begin{bmatrix}
        {\bm e}_{q} \\ {\bm e}_{p}
    \end{bmatrix}
    +
    \begin{bmatrix}
        0 \\ -k_1^\alpha \gradient V(\bm e_q) -k_1^\beta \gradient V_\beta (\bm e_q)
    \end{bmatrix},
\end{equation}

\noindent
where $L_\beta$ is a block diagonal matrix that describes the coupling between agents and obstacles, which has entries $[L_\beta]_{ii} = - \sum_{k\in\mathcal N_i^\beta} A_{ik}^\beta(\bm q) (\mu_{i,k} P_{i,k} - I_m)$. By defining the Hamiltonian function $H'(\bm e) = k_1^\alpha V(\bm e_q) + k_2^\beta V_\beta(\bm e_q) + \frac{1}{2}\bar{\bm e}^\transp\bar{\bm e}$ and noting that $\dot H'(\bm e) = \bar{\bm e}^\transp J'(\bm e_q) \bar{\bm e}$, it follows that statements i)-iv) also hold for the extended Hamiltonian $H'$ and matrix $J'$. Thus, the error dynamics $\bar{\bm e}(t)$ is also upper bound by an exponential function of $\Lambda_{\rm max}(J')$. The optimization of the flock formation in the presence of obstacles can then be achieved by solving Eq.~\eqref{eq.optimization} using $J'$ instead of $J$.

\section{Relation to optimal control theory}
\label{sec.optimalcontrol}
Consider the multi-agent control system \eqref{eq.flockmodel}. The feedback control law is given by the proportional control signal $\bm u_i = -b_i(\bm q_i - \bm q_{\rm t} - \bm r_i) - \gamma c_i(\bm p_i - \bm p_{\rm t})$, where $[\bm q_i,\bm p_i]$ is the state  of each agent $i$ and $[\bm q_{\rm t} + \bm r_i,\bm p_{\rm t}]$ is a reference signal that each agent must track. Based on the LTV system \eqref{eq.ltv}, we can express the multi-agent system as a feedback control system
\begin{equation}
    \bm{\dot e} = \bar A(t)\bm e + \bar B\bm u,
\label{eq.ltvcontrol}
\end{equation}

\noindent
where
$\bar A(t) = \begin{bmatrix}
        0_{Nm} & I_{Nm} \\
        -L(t)\otimes I_m & -\gamma L(t) \otimes I_m
\end{bmatrix}$ is the system matrix, 
$\bar B = \begin{bmatrix}
        0_{Nm} \\ I_{Nm}
\end{bmatrix}$ is the input matrix, and
$\bm u = - \bar K(t)\bm e$ is the control signal, with the feedback matrix $\bar K(t) = \begin{bmatrix}
    \bar K_1(t) & \bar K_2(t)
\end{bmatrix}$ defined by $\bar K_1(t) = B(t)\otimes I_m$ and $\bar K_2(t) = C(t) \otimes I_m$. Note that $\bar K_1$ and $\bar K_2$ are diagonal matrices (i.e., $\bar K_1, \bar K_2 \in \mathcal {D}_n$, where $\mathcal D_n = \{ D \in \mathbb{R}^{Nm \times Nm} : D = \text{diag}(d_1, d_2, \ldots, d_n) \}$ is the set of all possible diagonal matrices). To maximize the convergence rate of the LTV system~\eqref{eq.ltvcontrol}, we propose the optimization problem \eqref{eq.optimization} in which the controller gains $b_i$ and $c_i$ are tuned in real time. As a result, the gain matrices $B(t)$ and $C(t)$, and hence the feedback matrix $\bar K(t)$, are functions of time.

The design of a the feedback matrix $\bar K\in\R^{Nm\times 2Nm}$ capable of controlling the time response of system~\eqref{eq.ltvcontrol} according to some pre-specified characteristics is a classical control-theory problem. The solution of this problem is based on the eigenvalue placement of the closed-loop system $\dot{\bm e} = (\bar A - \bar B \bar K)\bm e$. Yet, a standard approach assumes that $\bar K$ can be an unconstrained dense matrix, implying that the control signal $\bm u_i$ of each agent $i$ has access to the state of all other agents. This is often not the case in multi-agent systems due to communication constraints. In contrast, by optimizing solely parameters $b_i$ and $c_i$ (as in this paper), the control signal $\bm u_i(t)$ depends uniquely on the state of each agent $[\bm q_i,\bm p_i]$. Accordingly, to operate under the same information constraints, matrices $\bar K_1$ and $\bar K_2$ must be constrained to be diagonal matrices. Here, we improve the time response of Eq.~\eqref{eq.ltvcontrol} through the minimization of the largest Lyapunov exponent $\Lambda_{\rm max}(J(t_k))$ (via Eq.~\eqref{eq.optimization}). However, an alternative approach to achieve this control task is to formulate the optimization problem as an optimal control problem:
\begin{equation}
    \begin{aligned}
        \min_{K} \quad &\int_{t_k}^{t_k+w} \left(\bm e^\transp Q\bm e + \bm u^\transp R\bm u\right) {\rm d}t, \\
        \text{s.t.} \quad & \bar K_1, \bar K_2 \in \mathcal {D}_n.
    \end{aligned}
\label{eq.optimalcontrol}
\end{equation}

\noindent
The matrices $Q$ and $R$ are responsible for respectively tuning the convergence of the tracking error $\bm e(t)$ and the amplitude of the input signal $\bm u(t)$ according to the desired specifications. The optimization problem \eqref{eq.optimalcontrol} is a promising research direction to optimize the flocking dynamics, and may lead to even better results than those presented in this paper, but this problem is nontrivial and challenging to solve due to the constraints in the feedback matrix $\bar K$ and the finite time horizon $t_k+w$.

\section{Stability landscape in the neighborhood of the homogeneous optimum}
\label{sec.stabilitylandscape}

Here, we present an analysis of the stability landscape around the point $\boldsymbol{b}_{\rm hom}=\left(b^*_{\rm hom}, \ldots, b^*_{\rm hom}\right)$, where $b^*_{\rm hom}$ is given by Eq.~\eqref{eq.hom.obtb}. This analysis follows the framework introduced in Ref.~\cite[Supplementary Notes]{molnar2021asymmetry}, originally  derived for a power-grid model,  while extending the results to more general network systems. In the power-grid model, the Jacobian matrix is given by $\tilde J = \begin{bmatrix}
    0_N & I_N \\ \tilde C & - \tilde B
\end{bmatrix}$,
where $\tilde C$ represents the network coupling (akin to the Laplacian matrix $L$) and $\tilde B$ is a diagonal matrix with the damping coefficients of generators along its diagonal (akin to the diagonal matrix $B$ in the multi-agent model). Note that matrices $\tilde C$ and $\tilde B$ are isolated in different block matrices of $J$, whereas, in the multi-agent system considered in this work, both block matrices $J_{21}$ and $J_{22}$ are given by linear combinations of $L$ and $B$ (cf. Eq.~\eqref{eq.constrainedjacobian}).

Here, we establish a generalization of the derivation presented in Ref. \cite{molnar2021asymmetry} to accomodate this change and then draw conclusions about the landscape of the optimization problem \eqref{eq.optimization}.
First, we determine the conditions under which $\Lambda_{\rm max}$ can be further decreased along some arbitrary path starting at the homogeneous optimum $b^*_{\rm hom}$. Second, we prove that the conditions necessary to apply the Implicit Function Theorem (IFT) are locally satisfied. Third, we apply the IFT to show that the parameterization of the path is continuously differentiable. This allows us to derive an analytical expression for $\Lambda_{\max}$ as a function of a distance $\varepsilon$ from $b^*_{\rm hom}$.

\medskip\noindent
\textbf{Existence of a descending path in the optimization landscape.}
\label{subsec.CharacPol}
Consider the Jacobian matrix \eqref{eq.constrainedjacobian}, where, without loss of generality, we omit the Kronecker product to simplify the notation:
\begin{equation}
  J=\begin{bmatrix}
      0_N & I_N \\
-(B + L) & -\gamma(B + L)
  \end{bmatrix}.
\label{Jac}
\end{equation}
\\
Recall that $B = \operatorname{diag}(b_1,\ldots,b_N)$. Define 
$\bar{J}=\begin{bmatrix}
0 & I \\
-P / b_{\rm hom}^{*2} & -\gamma P / b^*_{\rm hom}
\end{bmatrix}$,
where $P = B + L$ and, by assumption, $b^*_{\rm hom}>0$. Note that $\nu$ is an eigenvalue of $\bar{J}$ if and only if $b^*_{\rm hom}\nu$ is an eigenvalue of $J$, as it can be shown that $\operatorname{det}(\bar{J}-\nu I) = \operatorname{det}\left(J-b^*_{\rm hom} \nu I\right)$.
%

Consider a (potentially curved) path $\boldsymbol{\zeta}$ that passes through $\boldsymbol{b}^*_{\rm hom}$ within the space of all $\boldsymbol{b}\in\R^N$, parameterized by $\varepsilon$ through a differentiable vector function $\boldsymbol{b}=\boldsymbol{\zeta}(\varepsilon)$ with the condition $\boldsymbol{\zeta}(0)=\boldsymbol{b}_{\rm hom}^*$. Let $\nu_{j \pm}(\varepsilon)$, for $j=1, \ldots, N$, be the eigenvalues of $\bar{J}$. Assume the eigenvalues of $L$ are all real, distinct, and ordered with $\ell_1=0$ (connected network). At $\varepsilon=0$ (corresponding to the homogeneous optimum $\bm b^*_{\rm hom}$), it follows that $\operatorname{det}(\bar{J}-\nu I)=b_{\rm hom}^{*-2 n} \operatorname{det}\left[b_{\rm hom}^{*2}\left(\nu^2+\gamma\nu+b^*_{\rm hom}\right) I+(b^*_{\rm hom}\gamma\nu + 1)L\right]$ and, hence,
\begin{equation}
\nu_{j \pm}(0)=-\frac{\gamma}{2}\left(1+\frac{\ell_j}{b^*_{\rm hom}}\right) \pm \frac{1}{2} \sqrt{\gamma^2\left(1+\frac{\ell_j}{b^*_{\rm hom}}\right)^2 - \frac{4}{b^*_{\rm hom}}\left(1+\frac{\ell_j}{b^*_{\rm hom}}\right)}.
\label{nu_j}
\end{equation}

\noindent
Accordingly, the largest Lyapunov exponent $\Lambda_{\rm max}$ is parameterized as
\begin{equation}
\Lambda_{\max }(\varepsilon)=b^*_{\rm hom} \cdot \max _{1 \leq j \leq N} \max \left\{\operatorname{Re}\left\{\nu_{j_{+}}(\varepsilon)\right\}, \operatorname{Re}\left\{\nu_{j-}(\varepsilon)\right\}\right\} ,
\label{lammaxeps}
\end{equation}

\noindent
where it follows that $\Lambda_{\max }(0)=b^*_{\rm hom} \left(\operatorname{Re}\left\{\nu_N(0)\right\}+\frac{\gamma}{2}\right)=\Lambda_{\max }(b^*_{\rm hom})$, as given by Eq.~\eqref{eq.hom.optlambda}.

The eigenvalues $\nu_{j_{ \pm}}(\varepsilon)$ change with $\varepsilon$ continuously given that $\bm\zeta$ is a continuous function and the eigenvalues are continuous functions of the matrix elements. Thus, for sufficiently small $\varepsilon \neq 0$, the eigenvalue $\nu_{N_{+}}(\varepsilon)$ determines the maximum in Eq.~\eqref{lammaxeps}:
\begin{equation}
\Lambda_{\max }(\varepsilon)=b^*_{\rm hom} \cdot \left(\operatorname{Re}\left\{\nu_{N_+}(0)\right\}+\frac{\gamma}{2}\right)=b^*_{\rm hom} \cdot \operatorname{Re}\left\{-\frac{c_N}{2}+\frac{1}{2} \sqrt{c_N^2-4 d_N}+\frac{\gamma}{2}\right\}.
\label{lammaxeps2}
\end{equation}
\\
In the expression above, the coefficients $c_i$ and $d_i$, for $i=1,\ldots,N$, are given by the characteristic polynomial of $\bar{J}$ (expressed as a product of quadratic factors): 
\begin{equation}
\operatorname{det}(\bar{J}-\nu I)=\left(\nu^2+c_1 \nu+d_1\right)\left(\nu^2+c_2 \nu+d_2\right) \cdots\left(\nu^2+c_N \nu+d_N\right).
\label{evalexpansion}
\end{equation}

Eq.~\eqref{lammaxeps2} expresses $\Lambda_{\max }$ as a function of $c_N$ and $d_N$, thereby determining a landscape over the $\left(c_N, d_N\right)$-plane in which the path $\boldsymbol{\zeta}(\varepsilon)$ lies. By defining $f\left(c_N, d_N\right) = \operatorname{Re}\left\{-c_N+\gamma+\sqrt{c_N^2-4 d_N}\right\} / 2$,  we have that $\Lambda_{\max }(\varepsilon)=b^*_{\rm hom}f\left(c_N, d_N\right)$. Thus, the condition for $\Lambda_{\max }$ to decrease along the path $\boldsymbol{\zeta}(\varepsilon)$ is given by $f\left(c_N(\varepsilon), d_N(\varepsilon)\right)<f\left(c_N(0), d_N(0)\right)$. This condition is equivalent to the existence of a path $(c_N(\epsilon),d_N(\epsilon))$ starting at $\left(c_N(0), d_N(0)\right)=\frac{b^*_{\rm hom}+\ell_j}{b^*_{\rm hom}} \left(\gamma,\frac{1}{b^*_{\rm hom}}\right)$ that immediately enters the region:
\begin{equation}
\left\{\left(c_N, d_N\right): f\left(c_N, d_N\right)<-\gamma / 2\right\}=\left\{\left(c_N, d_N\right): \frac{d_N}{\gamma}-\gamma> c_N-2\gamma>0\right\}.
\label{cndn_plane}
\end{equation}

\noindent
In this region, the following condition on the derivatives is always satisfied: $d_N^{\prime}(0) \geq$ $\gamma c_N^{\prime}(0)$, where the prime denotes the derivative of a function with respect to its argument. We assume $c_N^{\prime}(0) \geq 0$ without loss of generality. Since $d_N^{\prime}(0) = \frac{1}{\gamma b^*_{\rm hom}}$, if $\Lambda_{\max }$ decreases along the path, then one of the following conditions must hold: i) $b^*_{\rm hom}\leq \frac{1}{\gamma^2}$ or ii) $c_N^{\prime}(0)=d_N^{\prime}(0)=0$. Given that $b^*_{\rm hom} - \frac{1}{\gamma^2} = \frac{1}{\gamma^2} - \ell_N + \sqrt{\ell_N^2+\frac{4}{\gamma^4}} \geq 0$, condition i) is false and, as a result, we have that condition ii) is the condition for the existence of a descending path.

\medskip\noindent
\textbf{Conditions for the parameterization of a descending path.}
\label{subsec.Gnonsing}
To parameterize the polynomial coefficients $c_i$ and $d_i$ as functions $c_i(\varepsilon)$ and $d_i(\varepsilon)$, we express Eq.~\eqref{evalexpansion} as 
 \begin{equation}
\operatorname{det}(\bar{J}-\nu I)=\operatorname{det}\left(\nu^2 I+\nu \gamma(B+L) / b^*_{\rm hom}+(B+L) / b_{\rm hom}^{*2}\right)=\operatorname{det}\left(\nu^2 I+\nu \gamma (\bar{B} + D)+(\bar{B} + D)\right).
\label{evalsJbar3}
\end{equation}

\noindent Here, we applied the transformations $Q^{-1} L Q = D$ and  $Q^{-1}\left(B / b^*_{\rm hom}\right) Q = \bar{B}$ based on the diagonalization of $L$, where $D=\operatorname{diag}(\ell_1,\ldots,\ell_N)$. Moreover, given the path $\boldsymbol{\zeta}(\varepsilon) = \left[\zeta_1(\varepsilon), \ldots, \zeta_N(\varepsilon)\right]$, we define
\begin{equation}
   \bar{B}_{i j}(\varepsilon)=\sum_{l=1}^N u_{i l} v_{j l} \zeta_{l}(\varepsilon),
   \label{Bbarcompon}
\end{equation}
\noindent where $u_{i l}$ and $ v_{j l}$ are the $l$th component of the left and right eigenvectors of $L$, respectively.

Now, consider the Leibniz formula for the determinant: $\operatorname{det}(J) = \sum_{\bm\sigma\in S_n} \operatorname{sgn}(\bm\sigma)\prod_{i=1}^n J_{i\sigma (i)}$, where $\bm\sigma = [\sigma(1),\ldots,\sigma(n)]$ is a  permutation of set $\{1,\ldots,n\}$ onto itself, $\operatorname{sgn}(\bm\sigma)$ denotes the signature of a permutation $\bm \sigma$ ($+1$ if the number of transpositions is even and $-1$ otherwise), and $S_n$ is the symmetric group.
Following the same steps as in Ref.~\cite[Supplementary Note, Sec. 2]{molnar2021asymmetry}, by equating the coefficients in Eqs.~\eqref{evalexpansion} and \eqref{evalsJbar3} to each other, we obtain the following implicit equation in terms of $\{c_i,d_i,\varepsilon\}$, for $i=1,\ldots,N$:
\begin{equation}
\mathbf{F}\left(c_1, \ldots, c_N, d_1, \ldots, d_N, \varepsilon\right)={0}.
   \label{F_vec}
\end{equation}

\noindent The components of function $\mathbf{F}$ are given by
\begin{equation}
F_k =\sum_{\left\{k_i\right\}} \chi\left(\sum_i k_i=k\right) \cdot\left[\prod_{i=1}^N a_i^{\left(k_i\right)}-\sum_{\bm \sigma} \operatorname{sgn}(\bm \sigma) \prod_{i=1}^N E_{i \sigma(i)}^{\left(k_i\right)}\right] ,
\label{F_k2}
\end{equation}

\noindent
where  $\sum_{\{k_i\}}$  comprises all possible combinations of $k_i=0,1,2$ and $\chi$ is an indicator function defined as  $\chi\left(\sum_i k_i=k\right)=1$ if $\sum_i k_i=k$ and $\chi\left(\sum_i k_i=k\right)=0$ otherwise. Note that Eq.~\eqref{F_vec} implicitly determines functions  $c_i(\varepsilon)$ and $d_i(\varepsilon)$. For $\varepsilon=0$, it follows that $E_{i j}^{(1)}=\gamma \frac{\bar{B}_{ij}(0) + \ell_i \delta_{i j}}{ b^*_{\rm hom}}=\gamma\left(1+\frac{\ell_i}{b^*_{\rm hom}}\right)$ and, therefore, 
$a_i^{(0)}=d_i(0)=\frac{1}{b^*_{\rm hom}}\left(1+\frac{\ell_i}{b^*_{\rm hom}}\right)$, $a_i^{(1)}=c_i(0)=\gamma\left(1+\frac{\ell_i}{b^*_{\rm hom}}\right)$, $a_i^{(2)}=1$, and $E_{i j}^{(k)}=a_i^{(k)} \delta_{i j}$.

Our goal is to apply the IFT to show that  $c_i(\varepsilon)$ and $d_i(\varepsilon)$ are continuously differentiable functions when $\varepsilon$ is sufficiently small. This proof implies the existence a smooth curve in the neighborhood of the point $
\frac{b^*_{\rm hom}+\ell_j}{b^*_{\rm hom}}\left(\gamma,\frac{1}{b^*_{\rm hom}}\right)$ in the $(c_i, d_i)$-plane, for which we can locally determine the first derivatives. As a result, we can determine the direction of the path $\bm\zeta(\varepsilon)$ that minimizes $\Lambda_{\rm max}(\varepsilon)$. To apply the IFT to Eq.~\eqref{F_k2} at the point $\left(c_1(0), \ldots, c_N(0), d_1(0), \ldots, d_N(0), 0\right)$, the following $2N \times 2N$ matrix must be nonsingular:
\begin{equation}
G = \left(\begin{array}{cccccc}
\frac{\partial F_0}{\partial c_1} & \cdots & \frac{\partial F_0}{\partial c_N} & \frac{\partial F_0}{\partial d_1} & \cdots & \frac{\partial F_0}{\partial d_N} \\
\vdots & & \vdots & \vdots & & \vdots \\
\frac{\partial F_{2 N-1}}{\partial c_1} & \cdots & \frac{\partial F_{2 N-1}}{\partial c_N} & \frac{\partial F_{2 N-1}}{\partial d_1} & \cdots & \frac{\partial F_{2 N-1}}{\partial d_N}
\end{array}\right).
\label{G_matrix}
\end{equation}

\noindent
We observe that $G$ is solely determined by  the matrix $L$ since it only depends on  $\frac{1}{b^*_{\rm hom}}\left(1+\frac{\ell_2}{b^*_{\rm hom}}\right), \ldots,\frac{1}{b^*_{\rm hom}}\left(1+\frac{\ell_N}{b^*_{\rm hom}}\right)$.  We can explicitly determine the entries of $G$ by differentiating Eq.~\eqref{F_k2}.  Let $x_s=c_s$, for $s=1, \ldots, N$, and $x_s=d_{s-N}$, for $s=N+1, \ldots, 2 N$. At the point $\left(c_1(0), \ldots, c_N(0), d_1(0), \ldots, d_N(0), 0\right)$, it follows that 
\begin{equation}
\begin{split}
\label{G_ks}
G_{k s}&= \frac{\partial F_k-1}{\partial x_s} = \begin{cases}\sum_{\left\{k_i\right\}} \prod_{i \neq s} a_i^{\left(k_i\right)} \cdot \chi\left(s, k-1,\left\{k_i\right\}\right), \, \text { if } s=1, \ldots, N, \\
\sum_{\left\{k_i\right\}} \prod_{i \neq \hat{s}} a_i^{\left(k_i\right)} \cdot \chi\left(s, k-1,\left\{k_i\right\}\right), \, \text { if } s=N+1, \ldots, 2 N,\end{cases}
\end{split}
\end{equation}

\noindent where $\hat{s} = s-N$ and  $a_i^{\left(k_i\right)}$ is defined in Eq.~\eqref{F_k2}. Moreover

\begin{equation}
\label{Chi}
\chi\left(s, k,\left\{k_i\right\}\right) = \begin{cases}1 & \text { if } \sum_i k_i=k, k_s=1, \text { and } s=1, \ldots, N, \\ 1 & \text { if } \sum_i k_i=k, k_3=0, \text { and } s=N+1, \ldots, 2 N, \\ 0 & \text { otherwise. }\end{cases}
\end{equation}

The non-singularity of $G$ is guaranteed if any two columns are linearly independent. Given a distinct pair $s$ and $s'$, we can simplify Eq.~\eqref{G_ks} as follows:

\begin{itemize}
    \item For $k=1$,
$G_{1 s}=\begin{cases}0, & \text { if } s=1, \ldots, N, \\
\prod_{i \neq s} d_i, & \text { if } s=N+1, \ldots, 2 N.\end{cases}$

\item 
For $k=2$, $G_{2 s}=\begin{cases}\prod_{i \neq s} d_i, & \text { if } s=1, \ldots, N, \\
\sum_{t \neq i} \gamma b^*_{\rm hom} \prod_{i \neq s} d_i = (N-1)\gamma b^*_{\rm hom}\prod_{i \neq s} d_i, & \text { if } s=N+1, \ldots, 2 N.\end{cases}$

\item
For $k=3$, $G_{3 s}=\sum_{t \neq s} \gamma b^*_{\rm hom} \prod_{i \neq s} d_i = (N-1)\gamma b^*_{\rm hom}\prod_{i \neq s} d_i, \quad \text { if } s=1, \ldots, N$. Note that we do not explicitly define $G_{3s}$  for $s=N+1,\ldots,2N$ since this expression is not needed to prove the linear independence of two columns.

\item
For $k = 2N-1$, $G_{2 N-1, s}=\begin{cases}\sum_{i \neq s} c_i = \sum_{i \neq s} \gamma\left(1+\frac{\ell_i}{b^*_{\rm hom}}\right) = \gamma(N-1) + \frac{\gamma}{b^*_{\rm hom}}\sum_{i\neq s} \ell_i, & \text { if } s=1, \ldots, N, \\
1, & \text { if } s=N+1, \ldots, 2 N.\end{cases}$

\item
For $k = 2N$, $G_{2 N, s}=\begin{cases}\prod_{i \neq s} a_i^{\left(k_k\right)}=1, & \text { if } s=1, \ldots, N, \\
0, & \text { if } s=N+1, \ldots, 2 N.\end{cases}$

\end{itemize}

The equations above are sufficient to show that, for any pair $s$ and $s^{\prime}$, the $s$th and $s^{\prime}$th columns of $G$ are linearly independent. First, note that when $1 \leq s \leq N$ and $N+1 \leq s^{\prime} \leq 2 N-1$, the last two components of the $s$th and $s^{\prime}$th column vectors form the two-dimensional, linearly independent vectors $[\gamma(N-1) + \frac{\gamma}{b^*_{\rm hom}}\sum_{i\neq s} \ell_i, \, 1]$ and $[1, \, 0]$, respectively. This implies that the $2N$-dimensional vectors in both columns of $G$ are also linearly independent. 
Second, consider the case $1 \leq s < s^{\prime} \leq N$. It follows that both $s$th and $s^{\prime}$th column have their last components equal to $1$. Therefore, to demonstrate their linear independence, it is sufficient to show that the third component is different, which is true since
\begin{equation}
G_{3 s}-G_{3 s^{\prime}} =\frac{\gamma}{b^*_{\rm hom}}\sum_{t \neq s, s^{\prime}}\left(\prod_{i \neq t, s, s^{\prime}} d_i\right)\left(\ell_{s^{\prime}}-\ell_s\right)>0.
\end{equation}

\noindent
Finally, consider the case $1 \leq s<s^{\prime} \leq N$. Analogously to the previous case, linear independence is guaranteed given that the second component differs:
\begin{equation}
G_{2 s} - G_{2 s^{\prime}} =\frac{\gamma}{b^*_{\rm hom}}\sum_{t \neq s, s^{\prime}}\left(\prod_{i \neq t, s, s^{\prime}} d_i\right)\left(\ell_{s^{\prime}}-\ell_s\right)>0.
\label{G_2s-g2sprime}
\end{equation}

\noindent
This concludes the proof that all distinct pairs $(s,s^{\prime})$ of column vectors of $G$ are linearly independent. Thus, $G$ is non-singular.

\medskip\noindent
\textbf{Parameterization of a descending path.}
\label{subsec.AnalyticalLammax}
We show that $c_N^{\prime}(0)=0$ implies that $\boldsymbol{\zeta}$ is tangent to a hyperplane $H$ at the point $\boldsymbol{b}_{\rm hom}$. Given that the conditions of the IFT are satisfied for Eq.~\eqref{F_k2}, $c_i(\varepsilon)$ and $d_i(\varepsilon)$ are continuously differentiable functions. Thus, we have that
\begin{equation}
\frac{\rm d}{{\rm d} \varepsilon} F_k\left(c_1(\varepsilon), \ldots, c_N(\varepsilon), d_1(\varepsilon), \ldots, d_N(\varepsilon), \varepsilon\right)=0,
\label{dF_k}
\end{equation}

\noindent
which, as shown in Ref.~\cite{molnar2021asymmetry}, is equivalent to $\sum_{s=1}^{2 N} G_{k s}\left[x_s^{\prime}(0)-y_s^{\prime}(0)\right]=0$. Expressing this function in vector form yields $G(\bm{x}-\bm{y}) = {0}$, where $\bm{x} = [x_1^{\prime}(0), \ldots, x_{2N}^{\prime}(0)]$ and $\bm{y} = [y_1^{\prime}(0), \ldots, y_{2N}^{\prime}(0)]$. Given that $G$ is non-singular, it follows that $\bm{x} = \bm{y}$ and, hence, $x_i^{\prime}(0) = y_i^{\prime}(0)$. Therefore, the polynomial coefficients $c_i(\varepsilon)$ and $d_i(\varepsilon)$ at $\varepsilon = 0$ are given by
\begin{equation}
\begin{split}
    c_i^{\prime}(0)&=x_i^{\prime}(0)=y_i^{\prime}(0)=\frac{\gamma}{b^*_{\rm hom}}\frac{{\rm d} \bar{B}_{i i}(0)}{{\rm d} \varepsilon}=\gamma\left.\frac{{\rm d}}{{\rm d} \varepsilon}\left(\sum_{l=1}^N \frac{u_i v_i \zeta_{l}(\varepsilon)} {b^*_{\rm hom}}\right)\right|_{\varepsilon=0}=\gamma\sum_{l=1}^N \frac{u_{i l} v_{i e} \zeta_{l}^{\prime}(0)} {b^*_{\rm hom}},\\
    d_i^{\prime}(0)&=x_{i+N}^{\prime}(0)=y_{i+N}^{\prime}(0)=\frac{1}{b^{*2}_{\rm hom}}\frac{{\rm d} \bar{B}_{i i}(0)}{{\rm d} \varepsilon}=\sum_{l=1}^N \frac{u_{i l} v_{i e} \zeta_{l}^{\prime}(0)} {b_{\rm hom}^{*2}},
\end{split}
\label{cd_iprime}
\end{equation}
\\
for $i=1, \ldots, N$, where we applied Eq.~\eqref{Bbarcompon} and the fact that $\ell_i$ remains constant and does not vary with $\varepsilon$. Specifically, we find that
\begin{equation}
c_N^{\prime}(0)=\gamma\sum_{l=1}^N \frac{u_{n l} v_{n l} \zeta_{l}^{\prime}(0)} {b^*_{\rm hom}}\quad \text { and } \quad d_N^{\prime}(0)=\frac{1}{\gamma b^*_{\rm hom}}c_N^{\prime}(0).
\label{cn_dn}
\end{equation}

As we have shown, when $d_N^{\prime}(0) = 0$, and hence $c_N^{\prime}(0) = 0$, it follows that $\Lambda_{\text{max}}$ decreases along the path $\bm\zeta(\varepsilon)$ in the $b$-space. Under this particular condition, it is evident from Eq.~\eqref{cn_dn} that, when all eigenvalues of $L$ are distinct, any descending path $[\bm\zeta_1^{\prime}(0), \ldots, \bm\zeta_N^{\prime}(0)]$ aligns with the hyperplane $H$, where $H$ is uniquely determined by $\sum_{i=1}^N u_{N i} v_{N i}\left(b_i-b^*_{\rm hom}\right)=0$. 
On the other hand, if $c_N^{\prime}(0), d_N^{\prime}(0) \neq 0$, it follows that $\bm \zeta(\varepsilon)$ is not tangent to $H$ due to the existence of higher-order terms in
\begin{equation}
\begin{split}
& c_N(\varepsilon)=c_N(0)+c_N^{\prime}(0) \varepsilon+\mathcal O\left(\varepsilon^2\right), \\
& d_N(\varepsilon)=d_N(0)+d_N^{\prime}(0) \varepsilon+\mathcal O\left(\varepsilon^2\right).
\end{split}
\label{cn_dn_eps}
\end{equation}

\noindent
Substituting Eq.~\eqref{cn_dn_eps} into Eq.~\eqref{lammaxeps2} leads to the following approximation for $\Lambda_{\max }(\varepsilon)$:
\begin{equation}
\Lambda_{\max }(\varepsilon)= \Lambda_{\rm max}(b_{\rm hom}^*) + b^*_{\rm hom} \operatorname{Re}\left\{-\frac{c_N^{\prime}(0)}{2} + \frac{c_N(0)c_N^{\prime}(0)-2d_N^{\prime}(0)}{2\sqrt{c_N^2(0)-4d_N(0)}}\right\}\varepsilon+\mathcal O\left(\varepsilon^2\right).
\label{lammaxapprox}
\end{equation}

\noindent
We now show that the first-order term is always positive. Recall the assumption $c_2^{\prime}(0) \geq 0$. Since $d_N^{\prime}(0) =\frac{1}{\gamma b^*_{\rm hom}}c_N^{\prime}(0)$, the conditions for the first-order term to be positive are $c_N(0)-\frac{2}{\gamma b^*_{\rm hom}}>0$ and $\left(c_N(0)-\frac{2}{\gamma b^*_{\rm hom}}\right)^2 - c_N(0)^2 + 4d_N(0)\geq 0$. These conditions are satisfied since $c_N(0)-\frac{2}{\gamma b^*_{\rm hom}} =  \frac{\gamma}{b^*_{\rm hom}}\sqrt{\ell_N +\frac{4}{\gamma^4}}>0$ and $\left(c_N(0)-\frac{2}{\gamma b^*_{\rm hom}}\right)^2 - c_N(0)^2 + 4d_N(0) = \frac{4}{\gamma^2b_{\rm hom}^{*2}}\geq 0$.



Finally, this analysis concludes that $\bm b^*_{\rm hom}$ is a local minimum of $\Lambda_{\max}$ along any path $\bm\zeta$ that transversally intersects with the hyperplane $H$ at $\bm b^*_{\rm hom}$. Thus, any first-order path that crosses $\bm b^*_{\rm hom}$ will necessarily increase $\Lambda_{\rm max}$.


\begin{footnotesize}


\begin{thebibliography}{10}
\urlstyle{rm}
\expandafter\ifx\csname url\endcsname\relax
  \def\url#1{\texttt{#1}}\fi
\expandafter\ifx\csname urlprefix\endcsname\relax\def\urlprefix{URL }\fi
\expandafter\ifx\csname doiprefix\endcsname\relax\def\doiprefix{DOI: }\fi
\providecommand{\bibinfo}[2]{#2}
\providecommand{\eprint}[2][]{\url{#2}}

\bibitem{reynolds1987flocks}
\bibinfo{author}{Reynolds, C.~W.}
\newblock \bibinfo{title}{Flocks, herds and schools: A distributed behavioral model}.
\newblock In \emph{\bibinfo{booktitle}{Proceedings of the 14th Annual Conference on Computer Graphics and Interactive Techniques}}, \bibinfo{pages}{25--34} (\bibinfo{year}{1987}).

\bibitem{silva2010boids}
\bibinfo{author}{Silva, A. R.~D.}, \bibinfo{author}{Lages, W.~S.} \& \bibinfo{author}{Chaimowicz, L.}
\newblock \bibinfo{journal}{\bibinfo{title}{{Boids that see: Using self-occlusion for simulating large groups on GPUs}}}.
\newblock {\emph{{Computers in Entertainment}}} \textbf{\bibinfo{volume}{7}}, \bibinfo{pages}{1--20} (\bibinfo{year}{2010}).

\bibitem{vicsek1995novel}
\bibinfo{author}{Vicsek, T.}, \bibinfo{author}{Czir{\'o}k, A.}, \bibinfo{author}{Ben-Jacob, E.}, \bibinfo{author}{Cohen, I.} \& \bibinfo{author}{Shochet, O.}
\newblock \bibinfo{journal}{\bibinfo{title}{Novel type of phase transition in a system of self-driven particles}}.
\newblock {\emph{{Physical Review Letters}}} \textbf{\bibinfo{volume}{75}}, \bibinfo{pages}{1226} (\bibinfo{year}{1995}).

\bibitem{helbing2000simulating}
\bibinfo{author}{Helbing, D.}, \bibinfo{author}{Farkas, I.} \& \bibinfo{author}{Vicsek, T.}
\newblock \bibinfo{journal}{\bibinfo{title}{Simulating dynamical features of escape panic}}.
\newblock {\emph{{Nature}}} \textbf{\bibinfo{volume}{407}}, \bibinfo{pages}{487--490} (\bibinfo{year}{2000}).

\bibitem{gazi2004stability}
\bibinfo{author}{Gazi, V.} \& \bibinfo{author}{Passino, K.~M.}
\newblock \bibinfo{journal}{\bibinfo{title}{Stability analysis of social foraging swarms}}.
\newblock {\emph{{IEEE Transactions on Systems, Man, and Cybernetics, Part B (Cybernetics)}}} \textbf{\bibinfo{volume}{34}}, \bibinfo{pages}{539--557} (\bibinfo{year}{2004}).

\bibitem{couzin2009collective}
\bibinfo{author}{Couzin, I.~D.}
\newblock \bibinfo{journal}{\bibinfo{title}{Collective cognition in animal groups}}.
\newblock {\emph{{Trends in Cognitive Sciences}}} \textbf{\bibinfo{volume}{13}}, \bibinfo{pages}{36--43} (\bibinfo{year}{2009}).

\bibitem{katz2011inferring}
\bibinfo{author}{Katz, Y.}, \bibinfo{author}{Tunstr{\o}m, K.}, \bibinfo{author}{Ioannou, C.~C.}, \bibinfo{author}{Huepe, C.} \& \bibinfo{author}{Couzin, I.~D.}
\newblock \bibinfo{journal}{\bibinfo{title}{Inferring the structure and dynamics of interactions in schooling fish}}.
\newblock {\emph{{Proceedings of the National Academy of Sciences}}} \textbf{\bibinfo{volume}{108}}, \bibinfo{pages}{18720--18725} (\bibinfo{year}{2011}).

\bibitem{marras2012fish}
\bibinfo{author}{Marras, S.} \& \bibinfo{author}{Porfiri, M.}
\newblock \bibinfo{journal}{\bibinfo{title}{Fish and robots swimming together: attraction towards the robot demands biomimetic locomotion}}.
\newblock {\emph{{Journal of The Royal Society Interface}}} \textbf{\bibinfo{volume}{9}}, \bibinfo{pages}{1856--1868} (\bibinfo{year}{2012}).

\bibitem{pearce2014role}
\bibinfo{author}{Pearce, D.~J.}, \bibinfo{author}{Miller, A.~M.}, \bibinfo{author}{Rowlands, G.} \& \bibinfo{author}{Turner, M.~S.}
\newblock \bibinfo{journal}{\bibinfo{title}{Role of projection in the control of bird flocks}}.
\newblock {\emph{{Proceedings of the National Academy of Sciences}}} \textbf{\bibinfo{volume}{111}}, \bibinfo{pages}{10422--10426} (\bibinfo{year}{2014}).

\bibitem{gomez2022intermittent}
\rev{\bibinfo{author}{G{\'o}mez-Nava, L.}, \bibinfo{author}{Bon, R.} \& \bibinfo{author}{Peruani, F.}
\newblock \bibinfo{journal}{\bibinfo{title}{Intermittent collective motion in sheep results from alternating the role of leader and follower}}.
\newblock {\emph{{Nature Physics}}} \textbf{\bibinfo{volume}{18}}, \bibinfo{pages}{1494--1501} (\bibinfo{year}{2022}).}

\bibitem{sinha2023optimal}
\bibinfo{author}{Sinha, S.}, \bibinfo{author}{Krishnan, V.} \& \bibinfo{author}{Mahadevan, L.}
\newblock \bibinfo{journal}{\bibinfo{title}{Optimal control of interacting active particles on complex landscapes}}.
\newblock {\emph{{arXiv:2311.17039}}}  (\bibinfo{year}{2023}).

\bibitem{sar2023flocking}
\bibinfo{author}{Sar, G.~K.} \& \bibinfo{author}{Ghosh, D.}
\newblock \bibinfo{journal}{\bibinfo{title}{Flocking and swarming in a multi-agent dynamical system}}.
\newblock {\emph{{Chaos}}} \textbf{\bibinfo{volume}{33}}, \bibinfo{pages}{123126} (\bibinfo{year}{2023}).

\bibitem{xiao2024perception}
\bibinfo{author}{Xiao, Y.} \emph{et~al.}
\newblock \bibinfo{journal}{\bibinfo{title}{Perception of motion salience shapes the emergence of collective motions}}.
\newblock {\emph{{Nature Communications}}} \textbf{\bibinfo{volume}{15}}, \bibinfo{pages}{4779} (\bibinfo{year}{2024}).

\bibitem{wang2022coverage}
\bibinfo{author}{Wang, P.}, \bibinfo{author}{Song, C.} \& \bibinfo{author}{Liu, L.}
\newblock \bibinfo{journal}{\bibinfo{title}{Coverage control for mobile sensor networks with double-integrator dynamics and unknown disturbances}}.
\newblock {\emph{{IEEE Transactions on Automatic Control}}} \textbf{\bibinfo{volume}{68}}, \bibinfo{pages}{6299--6306} (\bibinfo{year}{2022}).

\bibitem{bertuccelli2009real}
\bibinfo{author}{Bertuccelli, L.}, \bibinfo{author}{Choi, H.-L.}, \bibinfo{author}{Cho, P.} \& \bibinfo{author}{How, J.}
\newblock \bibinfo{title}{{Real-time multi-UAV task assignment in dynamic and uncertain environments}}.
\newblock In \emph{\bibinfo{booktitle}{AIAA Guidance, Navigation, and Control Conference}}, \bibinfo{pages}{5776} (\bibinfo{year}{2009}).

\bibitem{balazs2024decentralized}
\bibinfo{author}{Bal{\'a}zs, B.}, \bibinfo{author}{Vicsek, T.}, \bibinfo{author}{Somorjai, G.}, \bibinfo{author}{Nepusz, T.} \& \bibinfo{author}{V{\'a}s{\'a}rhelyi, G.}
\newblock \bibinfo{journal}{\bibinfo{title}{Decentralized traffic management of autonomous drones}}.
\newblock {\emph{{Swarm Intelligence}}}  (\bibinfo{year}{2024}).

\bibitem{nguyen2021swarm}
\bibinfo{author}{Nguyen, T.-H.} \& \bibinfo{author}{Jung, J.~J.}
\newblock \bibinfo{journal}{\bibinfo{title}{Swarm intelligence-based green optimization framework for sustainable transportation}}.
\newblock {\emph{{Sustainable Cities and Society}}} \textbf{\bibinfo{volume}{71}}, \bibinfo{pages}{102947} (\bibinfo{year}{2021}).

\bibitem{chen2019control}
\bibinfo{author}{Chen, F.}, \bibinfo{author}{Ren, W.} \emph{et~al.}
\newblock \bibinfo{journal}{\bibinfo{title}{On the control of multi-agent systems: A survey}}.
\newblock {\emph{{Foundations and Trends in Systems and Control}}} \textbf{\bibinfo{volume}{6}}, \bibinfo{pages}{339--499} (\bibinfo{year}{2019}).

\bibitem{beaver2021overview}
\bibinfo{author}{Beaver, L.~E.} \& \bibinfo{author}{Malikopoulos, A.~A.}
\newblock \bibinfo{journal}{\bibinfo{title}{An overview on optimal flocking}}.
\newblock {\emph{{Annual Reviews in Control}}} \textbf{\bibinfo{volume}{51}}, \bibinfo{pages}{88--99} (\bibinfo{year}{2021}).

\bibitem{leonard2024fast}
\bibinfo{author}{Leonard, N.~E.}, \bibinfo{author}{Bizyaeva, A.} \& \bibinfo{author}{Franci, A.}
\newblock \bibinfo{journal}{\bibinfo{title}{Fast and flexible multiagent decision-making}}.
\newblock {\emph{{Annual Review of Control, Robotics, and Autonomous Systems}}} \textbf{\bibinfo{volume}{7}}, \bibinfo{pages}{19--45} (\bibinfo{year}{2024}).

\bibitem{olfati2006flocking}
\bibinfo{author}{Olfati-Saber, R.}
\newblock \bibinfo{journal}{\bibinfo{title}{Flocking for multi-agent dynamic systems: Algorithms and theory}}.
\newblock {\emph{{IEEE Transactions on Automatic Control}}} \textbf{\bibinfo{volume}{51}}, \bibinfo{pages}{401--420} (\bibinfo{year}{2006}).

\bibitem{ren2007formation}
\bibinfo{author}{Ren, W.}
\newblock \bibinfo{journal}{\bibinfo{title}{Formation keeping and attitude alignment for multiple spacecraft through local interactions}}.
\newblock {\emph{{Journal of Guidance, Control, and Dynamics}}} \textbf{\bibinfo{volume}{30}}, \bibinfo{pages}{633--638} (\bibinfo{year}{2007}).

\bibitem{nagy2010hierarchical}
\bibinfo{author}{Nagy, M.}, \bibinfo{author}{{\'A}kos, Z.}, \bibinfo{author}{Biro, D.} \& \bibinfo{author}{Vicsek, T.}
\newblock \bibinfo{journal}{\bibinfo{title}{Hierarchical group dynamics in pigeon flocks}}.
\newblock {\emph{{Nature}}} \textbf{\bibinfo{volume}{464}}, \bibinfo{pages}{890--893} (\bibinfo{year}{2010}).

\bibitem{baronchelli2012consensus}
\bibinfo{author}{Baronchelli, A.} \& \bibinfo{author}{Diaz-Guilera, A.}
\newblock \bibinfo{journal}{\bibinfo{title}{Consensus in networks of mobile communicating agents}}.
\newblock {\emph{{Physical Review E}}} \textbf{\bibinfo{volume}{85}}, \bibinfo{pages}{016113} (\bibinfo{year}{2012}).

\bibitem{griparic2022consensus}
\bibinfo{author}{Griparic, K.}, \bibinfo{author}{Polic, M.}, \bibinfo{author}{Krizmancic, M.} \& \bibinfo{author}{Bogdan, S.}
\newblock \bibinfo{journal}{\bibinfo{title}{Consensus-based distributed connectivity control in multi-agent systems}}.
\newblock {\emph{{IEEE Transactions on Network Science and Engineering}}} \textbf{\bibinfo{volume}{9}}, \bibinfo{pages}{1264--1281} (\bibinfo{year}{2022}).

\bibitem{cucker2007emergent}
\bibinfo{author}{Cucker, F.} \& \bibinfo{author}{Smale, S.}
\newblock \bibinfo{journal}{\bibinfo{title}{Emergent behavior in flocks}}.
\newblock {\emph{{IEEE Transactions on Automatic Control}}} \textbf{\bibinfo{volume}{52}}, \bibinfo{pages}{852--862} (\bibinfo{year}{2007}).

\bibitem{valcher2017consensus}
\bibinfo{author}{Valcher, M.~E.} \& \bibinfo{author}{Zorzan, I.}
\newblock \bibinfo{journal}{\bibinfo{title}{On the consensus of homogeneous multi-agent systems with arbitrarily switching topology}}.
\newblock {\emph{{Automatica}}} \textbf{\bibinfo{volume}{84}}, \bibinfo{pages}{79--85} (\bibinfo{year}{2017}).

\bibitem{mikaberidze2024consensus}
\bibinfo{author}{Mikaberidze, G.}, \bibinfo{author}{Chowdhury, S.~N.}, \bibinfo{author}{Hastings, A.} \& \bibinfo{author}{D’Souza, R.~M.}
\newblock \bibinfo{journal}{\bibinfo{title}{Consensus formation among mobile agents in networks of heterogeneous interaction venues}}.
\newblock {\emph{{Chaos, Solitons \& Fractals}}} \textbf{\bibinfo{volume}{178}}, \bibinfo{pages}{114298} (\bibinfo{year}{2024}).

\bibitem{amichay2024revealing}
\bibinfo{author}{Amichay, G.}, \bibinfo{author}{Li, L.}, \bibinfo{author}{Nagy, M.} \& \bibinfo{author}{Couzin, I.~D.}
\newblock \bibinfo{journal}{\bibinfo{title}{Revealing the mechanism and function underlying pairwise temporal coupling in collective motion}}.
\newblock {\emph{{Nature Communications}}} \textbf{\bibinfo{volume}{15}}, \bibinfo{pages}{4356} (\bibinfo{year}{2024}).

\bibitem{OlfatiSaber2004}
\bibinfo{author}{Olfati-Saber, R.} \& \bibinfo{author}{Murray, R.~M.}
\newblock \bibinfo{journal}{\bibinfo{title}{{Consensus problems in networks of agents with switching topology and time-delays}}}.
\newblock {\emph{{IEEE Transactions on Automatic Control}}} \textbf{\bibinfo{volume}{49}}, \bibinfo{pages}{1520--1533} (\bibinfo{year}{2004}).

\bibitem{blondel2005convergence}
\bibinfo{author}{Blondel, V.~D.}, \bibinfo{author}{Hendrickx, J.~M.}, \bibinfo{author}{Olshevsky, A.} \& \bibinfo{author}{Tsitsiklis, J.~N.}
\newblock \bibinfo{title}{Convergence in multiagent coordination, consensus, and flocking}.
\newblock In \emph{\bibinfo{booktitle}{Proceedings of the IEEE Conference on Decision and Control}}, \bibinfo{pages}{2996--3000} (\bibinfo{year}{2005}).

\bibitem{ren2008consensus}
\bibinfo{author}{Ren, W.}
\newblock \bibinfo{journal}{\bibinfo{title}{On consensus algorithms for double-integrator dynamics}}.
\newblock {\emph{{IEEE Transactions on Automatic Control}}} \textbf{\bibinfo{volume}{53}}, \bibinfo{pages}{1503--1509} (\bibinfo{year}{2008}).

\bibitem{yu2010some}
\bibinfo{author}{Yu, W.}, \bibinfo{author}{Chen, G.} \& \bibinfo{author}{Cao, M.}
\newblock \bibinfo{journal}{\bibinfo{title}{Some necessary and sufficient conditions for second-order consensus in multi-agent dynamical systems}}.
\newblock {\emph{{Automatica}}} \textbf{\bibinfo{volume}{46}}, \bibinfo{pages}{1089--1095} (\bibinfo{year}{2010}).

\bibitem{zhang2017sliding}
\bibinfo{author}{Zhang, J.}, \bibinfo{author}{Lyu, M.}, \bibinfo{author}{Shen, T.}, \bibinfo{author}{Liu, L.} \& \bibinfo{author}{Bo, Y.}
\newblock \bibinfo{journal}{\bibinfo{title}{Sliding mode control for a class of nonlinear multi-agent system with time delay and uncertainties}}.
\newblock {\emph{{IEEE Transactions on Industrial Electronics}}} \textbf{\bibinfo{volume}{65}}, \bibinfo{pages}{865--875} (\bibinfo{year}{2017}).

\bibitem{ogren2004cooperative}
\bibinfo{author}{Ogren, P.}, \bibinfo{author}{Fiorelli, E.} \& \bibinfo{author}{Leonard, N.~E.}
\newblock \bibinfo{journal}{\bibinfo{title}{Cooperative control of mobile sensor networks: Adaptive gradient climbing in a distributed environment}}.
\newblock {\emph{{IEEE Transactions on Automatic Control}}} \textbf{\bibinfo{volume}{49}}, \bibinfo{pages}{1292--1302} (\bibinfo{year}{2004}).

\bibitem{jolles2017consistent}
\bibinfo{author}{Jolles, J.~W.}, \bibinfo{author}{Boogert, N.~J.}, \bibinfo{author}{Sridhar, V.~H.}, \bibinfo{author}{Couzin, I.~D.} \& \bibinfo{author}{Manica, A.}
\newblock \bibinfo{journal}{\bibinfo{title}{Consistent individual differences drive collective behavior and group functioning of schooling fish}}.
\newblock {\emph{{Current Biology}}} \textbf{\bibinfo{volume}{27}}, \bibinfo{pages}{2862--2868} (\bibinfo{year}{2017}).

\bibitem{niizato2024information}
\bibinfo{author}{Niizato, T.}, \bibinfo{author}{Sakamoto, K.}, \bibinfo{author}{Mototake, Y.-i.}, \bibinfo{author}{Murakami, H.} \& \bibinfo{author}{Tomaru, T.}
\newblock \bibinfo{journal}{\bibinfo{title}{Information structure of heterogeneous criticality in a fish school}}.
\newblock {\emph{{Scientific Reports}}} \textbf{\bibinfo{volume}{14}}, \bibinfo{pages}{29758} (\bibinfo{year}{2024}).

\bibitem{doering2022noise}
\bibinfo{author}{Doering, G.~N.} \emph{et~al.}
\newblock \bibinfo{journal}{\bibinfo{title}{Noise resistant synchronization and collective rhythm switching in a model of animal group locomotion}}.
\newblock {\emph{{Royal Society Open Science}}} \textbf{\bibinfo{volume}{9}}, \bibinfo{pages}{211908} (\bibinfo{year}{2022}).

\bibitem{jolles2020role}
\bibinfo{author}{Jolles, J.~W.}, \bibinfo{author}{King, A.~J.} \& \bibinfo{author}{Killen, S.~S.}
\newblock \bibinfo{journal}{\bibinfo{title}{The role of individual heterogeneity in collective animal behaviour}}.
\newblock {\emph{{Trends in Ecology \& Evolution}}} \textbf{\bibinfo{volume}{35}}, \bibinfo{pages}{278--291} (\bibinfo{year}{2020}).

\bibitem{nishikawa2016symmetric}
\bibinfo{author}{Nishikawa, T.} \& \bibinfo{author}{Motter, A.~E.}
\newblock \bibinfo{journal}{\bibinfo{title}{Symmetric states requiring system asymmetry}}.
\newblock {\emph{{Physical Review Letters}}} \textbf{\bibinfo{volume}{117}}, \bibinfo{pages}{114101} (\bibinfo{year}{2016}).

\bibitem{molnar2020network}
\bibinfo{author}{Molnar, F.}, \bibinfo{author}{Nishikawa, T.} \& \bibinfo{author}{Motter, A.~E.}
\newblock \bibinfo{journal}{\bibinfo{title}{Network experiment demonstrates converse symmetry breaking}}.
\newblock {\emph{{Nature Physics}}} \textbf{\bibinfo{volume}{16}}, \bibinfo{pages}{351--356} (\bibinfo{year}{2020}).

\bibitem{molnar2021asymmetry}
\bibinfo{author}{Molnar, F.}, \bibinfo{author}{Nishikawa, T.} \& \bibinfo{author}{Motter, A.~E.}
\newblock \bibinfo{journal}{\bibinfo{title}{Asymmetry underlies stability in power grids}}.
\newblock {\emph{{Nature Communications}}} \textbf{\bibinfo{volume}{12}}, \bibinfo{pages}{1457} (\bibinfo{year}{2021}).

\bibitem{mallada2015distributed}
\bibinfo{author}{Mallada, E.}, \bibinfo{author}{Freeman, R.~A.} \& \bibinfo{author}{Tang, A.~K.}
\newblock \bibinfo{journal}{\bibinfo{title}{Distributed synchronization of heterogeneous oscillators on networks with arbitrary topology}}.
\newblock {\emph{{IEEE Transactions on Control of Network Systems}}} \textbf{\bibinfo{volume}{3}}, \bibinfo{pages}{12--23} (\bibinfo{year}{2015}).

\bibitem{sugitani2021synchronizing}
\bibinfo{author}{Sugitani, Y.}, \bibinfo{author}{Zhang, Y.} \& \bibinfo{author}{Motter, A.~E.}
\newblock \bibinfo{journal}{\bibinfo{title}{Synchronizing chaos with imperfections}}.
\newblock {\emph{{Physical Review Letters}}} \textbf{\bibinfo{volume}{126}}, \bibinfo{pages}{164101} (\bibinfo{year}{2021}).

\bibitem{nair2021using}
\bibinfo{author}{Nair, N.}, \bibinfo{author}{Hu, K.}, \bibinfo{author}{Berrill, M.}, \bibinfo{author}{Wiesenfeld, K.} \& \bibinfo{author}{Braiman, Y.}
\newblock \bibinfo{journal}{\bibinfo{title}{Using disorder to overcome disorder: A mechanism for frequency and phase synchronization of diode laser arrays}}.
\newblock {\emph{{Physical Review Letters}}} \textbf{\bibinfo{volume}{127}}, \bibinfo{pages}{173901} (\bibinfo{year}{2021}).

\bibitem{cao2022harnessing}
\bibinfo{author}{Cao, H.} \& \bibinfo{author}{Eliezer, Y.}
\newblock \bibinfo{journal}{\bibinfo{title}{Harnessing disorder for photonic device applications}}.
\newblock {\emph{{Applied Physics Reviews}}} \textbf{\bibinfo{volume}{9}}, \bibinfo{pages}{011309} (\bibinfo{year}{2022}).

\bibitem{gast2024neural}
\bibinfo{author}{Gast, R.}, \bibinfo{author}{Solla, S.~A.} \& \bibinfo{author}{Kennedy, A.}
\newblock \bibinfo{journal}{\bibinfo{title}{Neural heterogeneity controls computations in spiking neural networks}}.
\newblock {\emph{{Proceedings of the National Academy of Sciences}}} \textbf{\bibinfo{volume}{121}}, \bibinfo{pages}{e2311885121} (\bibinfo{year}{2024}).

\bibitem{zhang2021random}
\bibinfo{author}{Zhang, Y.}, \bibinfo{author}{Ocampo-Espindola, J.~L.}, \bibinfo{author}{Kiss, I.~Z.} \& \bibinfo{author}{Motter, A.~E.}
\newblock \bibinfo{journal}{\bibinfo{title}{Random heterogeneity outperforms design in network synchronization}}.
\newblock {\emph{{Proceedings of the National Academy of Sciences}}} \textbf{\bibinfo{volume}{118}}, \bibinfo{pages}{e2024299118} (\bibinfo{year}{2021}).

\bibitem{teng2022heterogeneity}
\bibinfo{author}{Teng, R.} \emph{et~al.}
\newblock \bibinfo{journal}{\bibinfo{title}{Heterogeneity-driven collective-motion patterns of active gels}}.
\newblock {\emph{{Cell Reports Physical Science}}} \textbf{\bibinfo{volume}{3}}, \bibinfo{pages}{100933} (\bibinfo{year}{2022}).

\bibitem{yang2022emergent}
\bibinfo{author}{Yang, J.~F.} \emph{et~al.}
\newblock \bibinfo{journal}{\bibinfo{title}{Emergent microrobotic oscillators via asymmetry-induced order}}.
\newblock {\emph{{Nature Communications}}} \textbf{\bibinfo{volume}{13}}, \bibinfo{pages}{5734} (\bibinfo{year}{2022}).

\bibitem{nicolaou2021heterogeneity}
\bibinfo{author}{Nicolaou, Z.~G.}, \bibinfo{author}{Case, D.~J.}, \bibinfo{author}{Wee, E. B. v.~d.}, \bibinfo{author}{Driscoll, M.~M.} \& \bibinfo{author}{Motter, A.~E.}
\newblock \bibinfo{journal}{\bibinfo{title}{Heterogeneity-stabilized homogeneous states in driven media}}.
\newblock {\emph{{Nature Communications}}} \textbf{\bibinfo{volume}{12}}, \bibinfo{pages}{4486} (\bibinfo{year}{2021}).

\bibitem{ceron2023programmable}
\bibinfo{author}{Ceron, S.}, \bibinfo{author}{Gardi, G.}, \bibinfo{author}{Petersen, K.} \& \bibinfo{author}{Sitti, M.}
\newblock \bibinfo{journal}{\bibinfo{title}{Programmable self-organization of heterogeneous microrobot collectives}}.
\newblock {\emph{{Proceedings of the National Academy of Sciences}}} \textbf{\bibinfo{volume}{120}}, \bibinfo{pages}{e2221913120} (\bibinfo{year}{2023}).

\bibitem{keeffe2017oscillators}
\bibinfo{author}{O’Keeffe, K.~P.}, \bibinfo{author}{Hong, H.} \& \bibinfo{author}{Strogatz, S.~H.}
\newblock \bibinfo{journal}{\bibinfo{title}{Oscillators that sync and swarm}}.
\newblock {\emph{{Nature Communications}}} \textbf{\bibinfo{volume}{8}}, \bibinfo{pages}{1504} (\bibinfo{year}{2017}).

\bibitem{ghosh2022synchronized}
\bibinfo{author}{Ghosh, D.} \emph{et~al.}
\newblock \bibinfo{journal}{\bibinfo{title}{The synchronized dynamics of time-varying networks}}.
\newblock {\emph{{Physics Reports}}} \textbf{\bibinfo{volume}{949}}, \bibinfo{pages}{1--63} (\bibinfo{year}{2022}).

\bibitem{ren2007consensus}
\bibinfo{author}{Ren, W.}
\newblock \bibinfo{journal}{\bibinfo{title}{Consensus strategies for cooperative control of vehicle formations}}.
\newblock {\emph{{IET Control Theory \& Applications}}} \textbf{\bibinfo{volume}{1}}, \bibinfo{pages}{505--512} (\bibinfo{year}{2007}).

\bibitem{ren2008distributed}
\bibinfo{author}{Ren, W.} \& \bibinfo{author}{Beard, R.~W.}
\newblock \emph{\bibinfo{title}{Distributed Consensus in Multi-vehicle Cooperative Control: Theory and Applications}}, vol.~\bibinfo{volume}{27} (\bibinfo{publisher}{Springer}, \bibinfo{year}{2008}).

\bibitem{su2011stability}
\bibinfo{author}{Su, Y.} \& \bibinfo{author}{Huang, J.}
\newblock \bibinfo{journal}{\bibinfo{title}{Stability of a class of linear switching systems with applications to two consensus problems}}.
\newblock {\emph{{IEEE Transactions on Automatic Control}}} \textbf{\bibinfo{volume}{57}}, \bibinfo{pages}{1420--1430} (\bibinfo{year}{2011}).

\bibitem{horn2012matrix}
\bibinfo{author}{Horn, R.~A.} \& \bibinfo{author}{Johnson, C.~R.}
\newblock \emph{\bibinfo{title}{Matrix Analysis}} (\bibinfo{publisher}{Cambridge University Press}, \bibinfo{year}{2012}).

\bibitem{Pecora1998}
\bibinfo{author}{Pecora, L.~M.} \& \bibinfo{author}{Carroll, T.~L.}
\newblock \bibinfo{journal}{\bibinfo{title}{{Master stability functions for synchronized coupled systems}}}.
\newblock {\emph{{Physical Review Letters}}} \textbf{\bibinfo{volume}{80}}, \bibinfo{pages}{2109--2112} (\bibinfo{year}{1998}).

\bibitem{nishikawa2006synchronization}
\bibinfo{author}{Nishikawa, T.} \& \bibinfo{author}{Motter, A.~E.}
\newblock \bibinfo{journal}{\bibinfo{title}{Synchronization is optimal in nondiagonalizable networks}}.
\newblock {\emph{{Physical Review E}}} \textbf{\bibinfo{volume}{73}}, \bibinfo{pages}{065106} (\bibinfo{year}{2006}).

\bibitem{Motter2013}
\bibinfo{author}{Motter, A.~E.}, \bibinfo{author}{Myers, S.~A.}, \bibinfo{author}{Anghel, M.} \& \bibinfo{author}{Nishikawa, T.}
\newblock \bibinfo{journal}{\bibinfo{title}{{Spontaneous synchrony in power-grid networks}}}.
\newblock {\emph{{Nature Physics}}} \textbf{\bibinfo{volume}{9}}, \bibinfo{pages}{191--197} (\bibinfo{year}{2013}).

\bibitem{Dorfler2013}
\bibinfo{author}{Dorfler, F.}, \bibinfo{author}{Chertkov, M.} \& \bibinfo{author}{Bullo, F.}
\newblock \bibinfo{journal}{\bibinfo{title}{{Synchronization in complex oscillator networks and smart grids}}}.
\newblock {\emph{{Proceedings of the National Academy of Sciences}}} \textbf{\bibinfo{volume}{110}}, \bibinfo{pages}{2005--2010} (\bibinfo{year}{2013}).

\bibitem{nocedal1999numerical}
\bibinfo{author}{Nocedal, J.} \& \bibinfo{author}{Wright, S.~J.}
\newblock \emph{\bibinfo{title}{Numerical optimization}} (\bibinfo{publisher}{Springer}, \bibinfo{year}{1999}).

\bibitem{chen2024cooperative}
\bibinfo{author}{Chen, F.}, \bibinfo{author}{Sewlia, M.} \& \bibinfo{author}{Dimarogonas, D.~V.}
\newblock \bibinfo{journal}{\bibinfo{title}{Cooperative control of heterogeneous multi-agent systems under spatiotemporal constraints}}.
\newblock {\emph{{Annual Reviews in Control}}} \textbf{\bibinfo{volume}{57}}, \bibinfo{pages}{100946} (\bibinfo{year}{2024}).

\bibitem{lee2020tool}
\bibinfo{author}{Lee, J.~G.} \& \bibinfo{author}{Shim, H.}
\newblock \bibinfo{journal}{\bibinfo{title}{A tool for analysis and synthesis of heterogeneous multi-agent systems under rank-deficient coupling}}.
\newblock {\emph{{Automatica}}} \textbf{\bibinfo{volume}{117}}, \bibinfo{pages}{108952} (\bibinfo{year}{2020}).

\bibitem{zheng2011consensus}
\bibinfo{author}{Zheng, Y.}, \bibinfo{author}{Zhu, Y.} \& \bibinfo{author}{Wang, L.}
\newblock \bibinfo{journal}{\bibinfo{title}{Consensus of heterogeneous multi-agent systems}}.
\newblock {\emph{{IET Control Theory \& Applications}}} \textbf{\bibinfo{volume}{5}}, \bibinfo{pages}{1881--1888} (\bibinfo{year}{2011}).

\bibitem{zhan2012flocking}
\bibinfo{author}{Zhan, J.} \& \bibinfo{author}{Li, X.}
\newblock \bibinfo{journal}{\bibinfo{title}{Flocking of multi-agent systems via model predictive control based on position-only measurements}}.
\newblock {\emph{{IEEE Transactions on Industrial Informatics}}} \textbf{\bibinfo{volume}{9}}, \bibinfo{pages}{377--385} (\bibinfo{year}{2012}).

\bibitem{nascimento2023nmpc}
\bibinfo{author}{Nascimento, I.~B.}, \bibinfo{author}{Rego, B.~S.}, \bibinfo{author}{Pimenta, L.~C.} \& \bibinfo{author}{Raffo, G.~V.}
\newblock \bibinfo{title}{{NMPC strategy for safe robot navigation in unknown environments using polynomial zonotopes}}.
\newblock In \emph{\bibinfo{booktitle}{Proceedings of the IEEE Conference on Decision and Control}}, \bibinfo{pages}{7100--7105} (\bibinfo{year}{2023}).

\bibitem{sar2025dynamics}
\rev{\bibinfo{author}{Sar, G.~K.} \emph{et~al.}
\newblock \bibinfo{journal}{\bibinfo{title}{Dynamics of swarmalators in the presence of a contrarian}}.
\newblock {\emph{{Physical Review E}}} \textbf{\bibinfo{volume}{111}}, \bibinfo{pages}{014209} (\bibinfo{year}{2025}).}

\bibitem{ginelli2016physics}
\bibinfo{author}{Ginelli, F.}
\newblock \bibinfo{journal}{\bibinfo{title}{{The physics of the Vicsek model}}}.
\newblock {\emph{{The European Physical Journal Special Topics}}} \textbf{\bibinfo{volume}{225}}, \bibinfo{pages}{2099--2117} (\bibinfo{year}{2016}).

\bibitem{leonard2007collective}
\bibinfo{author}{Leonard, N.~E.} \emph{et~al.}
\newblock \bibinfo{journal}{\bibinfo{title}{Collective motion, sensor networks, and ocean sampling}}.
\newblock {\emph{{Proceedings of the IEEE}}} \textbf{\bibinfo{volume}{95}}, \bibinfo{pages}{48--74} (\bibinfo{year}{2007}).

\bibitem{shi2019fast}
\bibinfo{author}{Shi, F.}, \bibinfo{author}{Tuo, X.}, \bibinfo{author}{Ran, L.}, \bibinfo{author}{Ren, Z.} \& \bibinfo{author}{Yang, S.~X.}
\newblock \bibinfo{journal}{\bibinfo{title}{Fast convergence time synchronization in wireless sensor networks based on average consensus}}.
\newblock {\emph{{IEEE Transactions on Industrial Informatics}}} \textbf{\bibinfo{volume}{16}}, \bibinfo{pages}{1120--1129} (\bibinfo{year}{2019}).

\bibitem{battistelli2016stability}
\bibinfo{author}{Battistelli, G.} \& \bibinfo{author}{Chisci, L.}
\newblock \bibinfo{journal}{\bibinfo{title}{Stability of consensus extended kalman filter for distributed state estimation}}.
\newblock {\emph{{Automatica}}} \textbf{\bibinfo{volume}{68}}, \bibinfo{pages}{169--178} (\bibinfo{year}{2016}).

\bibitem{soatti2016consensus}
\bibinfo{author}{Soatti, G.}, \bibinfo{author}{Nicoli, M.}, \bibinfo{author}{Savazzi, S.} \& \bibinfo{author}{Spagnolini, U.}
\newblock \bibinfo{journal}{\bibinfo{title}{Consensus-based algorithms for distributed network-state estimation and localization}}.
\newblock {\emph{{IEEE Transactions on Signal and Information Processing over Networks}}} \textbf{\bibinfo{volume}{3}}, \bibinfo{pages}{430--444} (\bibinfo{year}{2016}).

\bibitem{montanari2022functional}
\bibinfo{author}{Montanari, A.~N.}, \bibinfo{author}{Duan, C.}, \bibinfo{author}{Aguirre, L.~A.} \& \bibinfo{author}{Motter, A.~E.}
\newblock \bibinfo{journal}{\bibinfo{title}{Functional observability and target state estimation in large-scale networks}}.
\newblock {\emph{{Proceedings of the National Academy of Sciences}}} \textbf{\bibinfo{volume}{119}}, \bibinfo{pages}{e2113750119} (\bibinfo{year}{2022}).

\bibitem{meng2018opinion}
\bibinfo{author}{Meng, X.~F.}, \bibinfo{author}{Van~Gorder, R.~A.} \& \bibinfo{author}{Porter, M.~A.}
\newblock \bibinfo{journal}{\bibinfo{title}{Opinion formation and distribution in a bounded-confidence model on various networks}}.
\newblock {\emph{{Physical Review E}}} \textbf{\bibinfo{volume}{97}}, \bibinfo{pages}{022312} (\bibinfo{year}{2018}).

\bibitem{redner2019reality}
\bibinfo{author}{Redner, S.}
\newblock \bibinfo{journal}{\bibinfo{title}{Reality-inspired voter models: A mini-review}}.
\newblock {\emph{{Comptes Rendus Physique}}} \textbf{\bibinfo{volume}{20}}, \bibinfo{pages}{275--292} (\bibinfo{year}{2019}).

\bibitem{bernardo2021achieving}
\bibinfo{author}{Bernardo, C.} \emph{et~al.}
\newblock \bibinfo{journal}{\bibinfo{title}{{Achieving consensus in multilateral international negotiations: The case study of the 2015 Paris Agreement on climate change}}}.
\newblock {\emph{{Science Advances}}} \textbf{\bibinfo{volume}{7}}, \bibinfo{pages}{eabg8068} (\bibinfo{year}{2021}).

\bibitem{crabtree2024influential}
\bibinfo{author}{Crabtree, S.~A.}, \bibinfo{author}{Wren, C.~D.}, \bibinfo{author}{Dixit, A.} \& \bibinfo{author}{Levin, S.~A.}
\newblock \bibinfo{journal}{\bibinfo{title}{Influential individuals can promote prosocial practices in heterogeneous societies: a mathematical and agent-based model}}.
\newblock {\emph{{PNAS Nexus}}} \textbf{\bibinfo{volume}{3}}, \bibinfo{pages}{pgae224} (\bibinfo{year}{2024}).

\bibitem{wang2019distributed}
\bibinfo{author}{Wang, L.} \& \bibinfo{author}{Chen, B.}
\newblock \bibinfo{journal}{\bibinfo{title}{Distributed control for large-scale plug-in electric vehicle charging with a consensus algorithm}}.
\newblock {\emph{{International Journal of Electrical Power \& Energy Systems}}} \textbf{\bibinfo{volume}{109}}, \bibinfo{pages}{369--383} (\bibinfo{year}{2019}).

\bibitem{yi2024optimal}
\bibinfo{author}{Yi, L.} \& \bibinfo{author}{Wei, E.}
\newblock \bibinfo{journal}{\bibinfo{title}{Optimal EV charging decisions considering charging rate characteristics and congestion effects}}.
\newblock {\emph{{IEEE Transactions on Network Science and Engineering}}} \textbf{\bibinfo{volume}{11}}, \bibinfo{pages}{5045--5057} (\bibinfo{year}{2024}).

\bibitem{cao2008agreeing}
\bibinfo{author}{Cao, M.}, \bibinfo{author}{Morse, A.~S.} \& \bibinfo{author}{Anderson, B.~D.}
\newblock \bibinfo{journal}{\bibinfo{title}{Agreeing asynchronously}}.
\newblock {\emph{{IEEE Transactions on Automatic Control}}} \textbf{\bibinfo{volume}{53}}, \bibinfo{pages}{1826--1838} (\bibinfo{year}{2008}).

\bibitem{zhang2016sampled}
\bibinfo{author}{Zhang, W.}, \bibinfo{author}{Tang, Y.}, \bibinfo{author}{Huang, T.} \& \bibinfo{author}{Kurths, J.}
\newblock \bibinfo{journal}{\bibinfo{title}{Sampled-data consensus of linear multi-agent systems with packet losses}}.
\newblock {\emph{{IEEE Transactions on Neural Networks and Learning Systems}}} \textbf{\bibinfo{volume}{28}}, \bibinfo{pages}{2516--2527} (\bibinfo{year}{2016}).

\bibitem{wang2016global}
\bibinfo{author}{Wang, B.}, \bibinfo{author}{Wang, J.}, \bibinfo{author}{Zhang, B.} \& \bibinfo{author}{Li, X.}
\newblock \bibinfo{journal}{\bibinfo{title}{Global cooperative control framework for multiagent systems subject to actuator saturation with industrial applications}}.
\newblock {\emph{{IEEE Transactions on Systems, Man, and Cybernetics: Systems}}} \textbf{\bibinfo{volume}{47}}, \bibinfo{pages}{1270--1283} (\bibinfo{year}{2017}).

\bibitem{pasqualetti2011consensus}
\bibinfo{author}{Pasqualetti, F.}, \bibinfo{author}{Bicchi, A.} \& \bibinfo{author}{Bullo, F.}
\newblock \bibinfo{journal}{\bibinfo{title}{Consensus computation in unreliable networks: A system theoretic approach}}.
\newblock {\emph{{IEEE Transactions on Automatic Control}}} \textbf{\bibinfo{volume}{57}}, \bibinfo{pages}{90--104} (\bibinfo{year}{2011}).

\end{thebibliography}

\begin{thebibliography}{10}
\urlstyle{rm}
\expandafter\ifx\csname url\endcsname\relax
  \def\url#1{\texttt{#1}}\fi
\expandafter\ifx\csname urlprefix\endcsname\relax\def\urlprefix{URL }\fi
\expandafter\ifx\csname doiprefix\endcsname\relax\def\doiprefix{DOI: }\fi
\providecommand{\bibinfo}[2]{#2}
\providecommand{\eprint}[2][]{\url{#2}}
 \setcounter{enumiv}{85}


\bibitem{bellen2013numerical}
\bibinfo{author}{Bellen, A.} \& \bibinfo{author}{Zennaro, M.}
\newblock \emph{\bibinfo{title}{Numerical methods for delay differential equations}} (\bibinfo{publisher}{Oxford University Press}, \bibinfo{year}{2013}).

\bibitem{engelborghs2002numerical}
\bibinfo{author}{Engelborghs, K.}, \bibinfo{author}{Luzyanina, T.} \& \bibinfo{author}{Roose, D.}
\newblock \bibinfo{journal}{\bibinfo{title}{{Numerical bifurcation analysis of delay differential equations using DDE-BIFTOOL}}}.
\newblock {\emph{{ACM Transactions on Mathematical Software (TOMS)}}} \textbf{\bibinfo{volume}{28}}, \bibinfo{pages}{1--21} (\bibinfo{year}{2002}).

\bibitem{sieber2014dde}
\bibinfo{author}{Sieber, J.}, \bibinfo{author}{Engelborghs, K.}, \bibinfo{author}{Luzyanina, T.}, \bibinfo{author}{Samaey, G.} \& \bibinfo{author}{Roose, D.}
\newblock \bibinfo{journal}{\bibinfo{title}{{DDE-BIFTOOL v. 3.1.1 Manual -- Bifurcation analysis of delay differential equations}}}.
\newblock {\emph{{arXiv:1406.7144}}}  (\bibinfo{year}{2014}).

\bibitem{Khalil2002}
\bibinfo{author}{Khalil, H.~K.}
\newblock \emph{\bibinfo{title}{{Nonlinear Systems}}} (\bibinfo{publisher}{Prentice Hall}, \bibinfo{year}{2002}), \bibinfo{edition}{3rd} edn.

\end{thebibliography}

\end{footnotesize}

\end{document}